\documentclass[12pt,a4paper]{article}

\usepackage{float}
\usepackage[latin1]{inputenc}
\usepackage{amsmath}
\usepackage{amsfonts}
\usepackage{amssymb}
\usepackage{mathtools}
\usepackage{mathrsfs}
\usepackage{amsmath}
\usepackage{comment}
\usepackage{graphicx,a4wide}
\usepackage{epsf,epsfig}

\numberwithin{equation}{section}
\makeatletter 
\@addtoreset{equation}{section} %
\makeatother

\usepackage[english]{babel}

\usepackage{amsmath,amsfonts,mathtools,mathrsfs}
\usepackage{mdframed}

\DeclareMathAlphabet{\mathpzc}{OT1}{pzc}{m}{it}

\usepackage{color}
\usepackage{enumitem}
\usepackage[hidelinks]{hyperref}

\usepackage{tikz}
\usetikzlibrary{trees}
\usetikzlibrary{snakes}
\usetikzlibrary{arrows}
\usepackage{color}
\newcommand{\beq}{\begin{equation}}
\newcommand{\eeq}{\end{equation}}
\def \be  {\begin{equation}}
\def \ee  {\end{equation}}
\newcommand{\bea}{\begin{eqnarray}}
\newcommand{\eea}{\end{eqnarray}}

\newcommand{\nn}{\nonumber}

\begin{document}

\usetikzlibrary{arrows}
% Front page here
\thispagestyle{empty}

\null\vskip-12pt \hfill  \\
%\null\vskip-12pt \hfill   \\

%\vskip-1.2truecm
\begin{center}
\vskip 0.2truecm {\Large\bf
%\titleline
{\Large One-loop string amplitudes in AdS$_5\times$S$^5$: \\[.1cm]
Mellin space and sphere splitting 
%for all single-particle operators in 
}
}\\
\vskip 1truecm
{\bf F.~Aprile$^1$, J.~M. Drummond$^2$, R. Glew$^2$ and M.~Santagata$^2$}
\vskip 0.4truecm
% 
%{\it
% ${}^{1}\,$Instituto de Fisica Teorica, UNESP, ICTP South American Institute for Fundamental Research\\
%		Rua Dr Bento Teobaldo Ferraz 271, 01140-070, S\~ao Paulo, Brazil\\
%		%\vskip .4truecm 
%\vskip .2truecm                        }
%{\it
% $^2 \ $School of Physics and Astronomy, University of Southampton,\\
% Highfield, Southampton SO17 1BJ\\
%\vskip .2truecm                        }
%\end{center}
%\vskip 1truecm 

	\vskip 0.4truecm
	
		\centerline{\it ${}^{1}\,$Instituto de Fisica Teorica, UNESP, ICTP South American Institute for Fundamental Research}
		\centerline{\it Rua Dr Bento Teobaldo Ferraz 271, 01140-070, S\~ao Paulo, Brazil} 
		\vskip .4truecm

		\centerline{\it ${}^{2}\,$School of Physics and Astronomy, University of Southampton,} 
		\centerline{\it Highfield, Southampton SO17 1BJ} 
		\vskip .2truecm

\end{center}
\vskip 1truecm

\centerline{\bf Abstract}\normalsize
We study string corrections to one-loop amplitudes of single-particle operators ${\cal O}_p$
in $AdS_5 \times S^5$.
The tree-level correlators in supergravity enjoy an accidental 10d conformal symmetry. Consequently, 
one observes a partial degeneracy in the spectrum of anomalous dimensions of double-trace operators 
and at the same time equality of many different correlators for different external charges $p_{i=1,2,3,4}$. 
The one-loop contribution is expected to lift such bonus properties, and its precise form can be predicted 
from tree-level data and consistency with the operator product expansion. 

Here we present a closed-form Mellin space formula for $\langle {\cal O}_{p_1}{\cal O}_{p_2}{\cal O}_{p_3} {\cal O}_{p_4}\rangle$ 
at order $(\alpha')^3$, valid for arbitrary external charges $p_{i}$. Our formula makes explicit the lifting of the bonus degeneracy 
among different correlators through a feature we refer to as `sphere splitting'.  While tree-level Mellin amplitudes 
come with a single crossing symmetric kernel, which defines the pole structure of the $AdS_5\times S^5$ amplitude, 
our one-loop amplitude naturally splits the $S^5$ part into two separate contributions. 
The amplitude also exhibits a remarkable consistency with the corresponding flat space IIB amplitude through the large $p$ limit.

\medskip
\noindent                                   
\newpage
\setcounter{page}{1}\setcounter{footnote}{0}
\tableofcontents
\pagebreak

%%==============================================================================
\section{Introduction}

An ongoing challenge in theoretical physics  is to understand the dynamics of quantum gravity and its relation to string theory.  
Recent advances within the analytic bootstrap program have shown that the study of the AdS/CFT correspondence, beyond the classical bulk approximation,
leads to concrete and quantitative results for both quantum gravity and string theory in $AdS$,  see \cite{Heslop:2022qgf,Gopakumar:2022kof} for a review.  
In this context, the problem of quantum gravity is well posed and the CFT approach is able to tackle $AdS$ gravity with a variety of computational tools. 
It has the potential to 
pinpoint the existence of very general structures in the theory, and
progressively allow us to move towards a more complete picture.

The most emblematic and best understood  AdS/CFT correspondence is the one between $SU(N)$ $\mathcal{N}=4$ SYM in 4d
and type-IIB string theory on $AdS_5\times S^5$. This is the theory we will study in this paper. Our probes for investigating 
quantum gravity will be four-point amplitudes of single particle operators $\langle \mathcal{O}_{p_1} \mathcal{O}_{p_2} \mathcal{O}_{p_3} \mathcal{O}_{p_4} \rangle$. We will study these objects
at tree level and one loop in Newton's constant (or equivalently, $1/N^2$ and $1/N^4$ corrections in the CFT)
and at the leading orders in string corrections to Einstein gravity, i.e. leading orders in the $\alpha' \sim \lambda^{-\frac{1}{2}}$ expansion, where $\lambda$ is the 't Hooft coupling.

%%====================================================================================================
%%====================================================================================================
\subsection{Four-point correlators of half-BPS operators}

%Single particle operators in $AdS_5\times S^5$, i.e.~
The supergravity multiplet on $AdS_5$ and the $S^5$ Kaluza-Klein spectrum 
are dual to half-BPS single-particle multiplets, whose superconformal primary operators are
\be
\mathcal{O}_p (x,y) = y^{R_1}\ldots y^{R_p} \, {\rm tr} (\phi_{R_1} \ldots \phi_{R_p}) (x) + \ldots 
\ee
where $\phi_{R}$ are the scalar fields of the $\mathcal{N}=4$ SYM, and the $y^R$ are auxillary null $SO(6)$ vectors used to project 
onto the traceless symmetric representation. Here the dots refer to multi-trace contributions determined by insuring that the 
$\mathcal{O}_p$ are orthogonal to any multi-trace operator \cite{Aprile:2018efk,Aprile:2020uxk}. The operators $\mathcal{O}_p$ 
are orthogonal and normalised as follows\footnote{The full normalisation is $R_p = p^2(p-1)\biggl[\frac{1}{(N-p+1)_{p-1}} - \frac{1}{(N+1)_{p-1}}\biggr]^{-1} =  p_1 N^{p_1} + O(N^{p_1-2})$ \cite{Aprile:2020uxk}.},
\be
\langle \mathcal{O}_{p_1} \mathcal{O}_{p_2} \rangle = \delta_{p_1 p_2} (g_{12})^{p_1} R_{p_1}\,, \qquad R_{p_1} =  p_1 N^{p_1} + O(N^{p_1-2}) \,, \qquad g_{ij} = \frac{y_{ij}^2}{x_{ij}^2} \,.
\ee
where we have defined $y_{12}^2 = y_1 \cdot y_2$.

We will study four-point functions of the operators $\mathcal{O}_p$. These correlators can be split into a free theory 
contribution plus a coupling dependent interacting term, %$($g=\lambda=0$)
\be
\langle \mathcal{O}_{p_1} \mathcal{O}_{p_2} \mathcal{O}_{p_3} \mathcal{O}_{p_4} \rangle = \langle \mathcal{O}_{p_1} \mathcal{O}_{p_2} \mathcal{O}_{p_3} \mathcal{O}_{p_4} \rangle_{\rm free} + \langle \mathcal{O}_{p_1} \mathcal{O}_{p_2} \mathcal{O}_{p_3} \mathcal{O}_{p_4} \rangle_{\rm int}\,.
\ee
From the point of view of the Operator Product Expansion (OPE), the free theory piece includes contributions from protected and long operators in any OPE channel, 
while the interacting term only receives contributions from unprotected (long) multiplets  \cite{Doobary:2015gia}.

%===========================================================================================================
%===========================================================================================================

\subsection{The Mellin space representation}
\label{sec:mellin_rep}

It was shown by Mack in \cite{Mack:2009mi}  that a CFT correlator, when written
in Mellin space, acquires a structure of poles and residues which can be nicely put in correspondence with the OPE. 
For holographic theories in $AdS$, it was later shown by Penedones \cite{Penedones:2010ue} that a suitably defined 
Mellin amplitude shares many similarities with a scattering amplitude,  and in fact this Mellin amplitude 
can be understood as a curved space completion of a flat space scattering amplitude. The flat space scattering 
amplitude is recovered in the limit of large Mellin variables. For holographic correlators in $AdS_5\times S^5$, 
it is possible to improve further the Mellin space representation, by considering a double Mellin transform in which both 
the conformal  $AdS_5$ space and the internal $S^5$ space  are treated equally.\footnote{This idea was first introduced in 
\cite{Aprile:2020luw} and further refined in \cite{Aprile:2020mus} and \cite{Abl:2020dbx}.}
This double Mellin transform reads, 
\begin{align}
\label{Mellinrep}
\langle \mathcal{O}_{p_1} \mathcal{O}_{p_2} \mathcal{O}_{p_3} \mathcal{O}_{p_4} \rangle_{\rm int} = p_1 p_2 p_3 p_4 N^{\Sigma-2} \hat{\mathcal{I}} \int\!d\hat{s}_{ij}  \sum_{\check{s}_{ij}} \prod_{i<j} \biggl[  \frac{x_{ij}^{2\hat{s}_{ij}} y_{ij}^{2\check{s}_{ij}}  \Gamma[-\hat{s}_{ij}]}{\Gamma[\check{s}_{ij} + 1]} \biggr] 
\mathcal{M}_{\vec{p}}(\hat s_{ij},\check s_{ij} )\,,
%\hat{s},\hat{t},\check{s},\check{t}) \,.
\end{align}
where $\hat{s}_{ij}$ are $AdS$ Mellin variables and $\check{s}_{ij}$ are sphere Mellin variables subject to the constraints,
\be\label{contraints_general_M}
{\textstyle \sum}_{i} \hat{s}_{ij} = -p_j - 2\,, \qquad {\textstyle \sum}_i \check{s}_{ij} = p_j - 2\,. 
\ee
The factor $\hat{\mathcal{I}}$ in \eqref{Mellinrep}, which removes two units of conformal/internal weight 
w.r.t.~to the charges $p_i$,  is a consequence of  superconformal symmetry \cite{Eden:2000bk}.\footnote{It can 
also be justified from the fact that only long representations contribute to the OPE decomposition of 
$\langle \mathcal{O}_{p_1} \mathcal{O}_{p_2} \mathcal{O}_{p_3} \mathcal{O}_{p_4} \rangle_{\rm int}$, and these have precisely the $\tilde{\mathcal{I}}$ factor.}
It takes the form
\be
\hat{\mathcal{I}} = (x_{13}^2 x_{24}^2 y_{13}^2 y_{24}^2)^2\,\tilde{\mathcal{I}}\,,\qquad \tilde{\mathcal{I}}=(x-y)(x-\bar{y})(\bar{x}-y)(\bar{x}-\bar{y})\,, 
\ee
with $x,\bar{x},y,\bar{y}$ parametrising the cross-ratios,
\begin{align}
 x \bar{x} = U = \tfrac{x_{12}^2 x_{34}^2}{x_{13}^2 x_{24}^2}\,, \qquad   &(1-x)(1-\bar{x}) = V = \tfrac{x_{14}^2 x_{23}^2}{x_{13}^2 x_{24}^2} \,, \notag \\
 y \bar{y} = \tilde{U} = \tfrac{y_{12}^2 y_{34}^2}{y_{13}^2 y_{24}^2}\,, \qquad   &(1-y)(1-\bar{y}) = \tilde{V} = \tfrac{y_{14}^2 y_{23}^2}{y_{13}^2 y_{24}^2}\,.
\end{align}
Note that $\hat{\mathcal{I}}$ is invariant under permutations of the external operators.

%The integral and the sum in \eqref{Mellinrep} only run over unconstrained variables, and the sum will actually be finite, as we will see. 
The constraints (\ref{contraints_general_M}) on the Mellin variables can be solved as follows,
%\begin{align}
%\hat{s}_{12} + p_{(12)} &= \hat{s}_{34} + p_{(34)} = \hat{s} , \qquad  && \check{s}_{12} - p_{(12)}  = \check{s}_{34} - p_{(34)} = \check{s} \,, \notag \\
%\hat{s}_{14} + p_{(14)} &= \hat{s}_{23} + p_{(23)} = \hat{t} , \qquad  && \check{s}_{14} - p_{(14)}  = \check{s}_{23} - p_{(23)} = \check{t} \,, \notag \\
%\hat{s}_{13} + p_{(13)} &= \hat{s}_{24} + p_{(24)} = \hat{u} , \qquad  && \check{s}_{13} - p_{(13)}  = \check{s}_{24} - p_{(24)} = \check{u}\,,
%\end{align}
\begin{align}
\hat{s}_{12} + \tfrac{p_1+p_2}{2}  &= \hat{s}_{34} +  \tfrac{p_3+p_4}{2}  = \hat{s} , \qquad  && \check{s}_{12} -  \tfrac{p_1+p_2}{2}   = \check{s}_{34} -\tfrac{p_3+p_4}{2}= \check{s} \,, \notag \\
\hat{s}_{14} +  \tfrac{p_1+p_4}{2} &= \hat{s}_{23} + \tfrac{p_2+p_3}{2} = \hat{t} , \qquad  && \check{s}_{14} -\tfrac{p_1+p_4}{2} = \check{s}_{23} - \tfrac{p_2+p_3}{2}= \check{t} \,, \notag \\
\hat{s}_{13} + \tfrac{p_1+p_3}{2} &= \hat{s}_{24} +\tfrac{p_2+p_4}{2} = \hat{u} , \qquad  && \check{s}_{13} -  \tfrac{p_1+p_3}{2}  = \check{s}_{24} -\tfrac{p_2+p_4}{2}  = \check{u}\,,
\end{align}
with  
\be
\hat{s} + \hat{t} + \hat{u} = \Sigma -2\,, \qquad  \check{s} + \check{t} + \check{u} = -\Sigma - 2\,\qquad;\qquad \Sigma=\tfrac{p_1+p_2+p_3+p_4}{2}.
\ee 
We shall say that the variables $\hat s,\check s$ define the ${\tt s}$-channel and similarly for the other channels. 
The variables $\hat s,\check s$  and $\hat t,\check t$ will be taken to be independent. 
The integral and the sum in \eqref{Mellinrep} only run over the independent variables $\hat s,\check s$  and $\hat t,\check t$, and the sum will actually be finite, as we will see.
We can further accompany 
each channel in position and internal space with the following combinations of charges, 
\beq
c_s=\tfrac{p_1+p_2-p_3-p_4}{2}\qquad;\qquad c_t=\tfrac{p_1+p_4-p_2-p_3}{2}\qquad;\qquad c_u=\tfrac{p_2+p_4-p_1-p_3}{2}
\eeq
The parametrisation of $\vec{p}$ in terms of $\{\Sigma,c_s,c_t,c_u\}$ is invertible, with $\Sigma$ being an invariant under permutation, and the various triplets $\{ \hat s,\check s, c_s\}$, $\{ \hat t,\check t, c_t\}$, $\{\hat u,\check u,c_u\}$ transforming into one another under crossing. 

It is convenient to rewrite the correlator \eqref{Mellinrep} by making manifest the dependence on the cross-ratios, namely
\be
\label{MellinrepUV}
\langle \mathcal{O}_{p_1} \mathcal{O}_{p_2} \mathcal{O}_{p_3} \mathcal{O}_{p_4} \rangle_{\rm int} 
= p_1 p_2 p_3 p_4 N^{\Sigma-2} \, \tilde{\mathcal{I}}\, \frac{\prod_{i<j} g_{ij}\!\!\!\,^{ \frac{ p_i+p_j}{2} }}{(g_{13} g_{24})^{\Sigma}} \mathcal{H}_{\vec{p}}(U,V,\tilde{U},\tilde{V}),
\ee
where
\be
\mathcal{H}_{\vec{p}}(U,V,\tilde{U},\tilde{V})=\int\!\! d\hat{s} d \hat{t} \,\sum_{\check{s}, \check{t}}  U^{\hat s} V^{\hat t} {\tilde U}^{\check s} {\tilde V}^{\check t}\,\, {\bf \Gamma}\, \mathcal{M}_{\vec{p}}\,.
\ee
By definition, ${\bf \Gamma}= \Gamma_{\tt s} \Gamma_{\tt t} \Gamma_{\tt u}$ with  
\be\label{ex_gammas}
\Gamma_{\tt s} = \frac{\Gamma_{\hat s}}{\Gamma_{\check s}}\qquad;\qquad
\begin{array}{c} 
\Gamma_{\hat s}= \Gamma[\tfrac{p_1+p_2}{2} -\hat{s}]\Gamma[\tfrac{p_3+p_4}{2}-\hat{s}] =\Gamma[\frac{ \Sigma}{2}\pm \frac{c_s}{2} -\hat{s}]\ \\[.2cm]
\Gamma_{\check s} =  \Gamma[1+\tfrac{p_1+p_2}{2}+\check{s}]\Gamma[1+\tfrac{p_3+p_4}{2} +\check{s}]=\Gamma[1+\frac{\Sigma}{2} \pm \frac{c_s}{2} +\check{s}].
\end{array}
\ee
Similarly for $\Gamma_{\tt t}$ and $\Gamma_{\tt u}$. The $\pm$ abbreviation stands for taking the product 
of both $\Gamma$ with $+$  and $-$ signs. Note, the sum over $\check s$ and $\check t$ is automatically truncated by the $\Gamma$ functions in the denominator of ${\bf \Gamma}$, thus it is finite. See also section \ref{sec_Mellin_numbers}. 

Under crossing transformation, the factor ${\bf \Gamma}$ is invariant while for the Mellin amplitude we find
the following table. 
\be\label{unoduetrequattro}
\begin{array}{c}
\begin{tikzpicture}[scale=1]
\node[scale=.9] at (0,0)  {$\begin{array}{|ccll|}
\hline &  & & \\[0cm]
{\cal M}_{p_1p_2p_3p_4}(\hat s,\hat t, \check s,\check t)& =& {\cal M}_{p_2,p_1,p_3,p_4}(\hat s,\hat u, \check s,\check u)& = {\cal M}_{c_s, +c_u,+c_t}(\hat s,\hat u, \check s,\check u)\\[.2cm]
%%{\cal M}_{p_1p_2p_3p_4}(\check s,\check t, \tilde s,\tilde t)
& =& {\cal M}_{p_1,p_2,p_4,p_3}(\hat s,\hat u, \check s,\check u)&= {\cal M}_{c_s, -c_u,-c_t}(\hat s,\hat u, \check s,\check u)\\[.3cm]	
\hline & &  & \\
{\cal M}_{p_1p_2p_3p_4}(\hat s,\hat t, \check s,\check t)& =& {\cal M}_{p_4,p_2,p_3,p_1}(\hat u, \hat t, \check u,\check t)& = {\cal M}_{c_u, +c_t,+c_s}(\check u, \check t, \check u,\check t)\\[.2cm]
%%{\cal M}_{p_1p_2p_3p_4}(\check s,\check t, \tilde s,\tilde t)
& =& {\cal M}_{p_1,p_3,p_2,p_4}(\hat u, \hat t, \check u,\check t)&= {\cal M}_{-c_u, c_t,-c_s}(\hat u, \hat t, \check u,\check t)\\[.3cm]	
\hline & &  & \\
{\cal M}_{p_1p_2p_3p_4}(\hat s,\hat t, \check s,\check t)& =& {\cal M}_{p_1,p_4,p_3,p_2}(\hat t, \hat s, \check t,\check s)& = {\cal M}_{c_t, +c_s,+c_u}(\hat t, \hat s, \check t,\check s)\\[.2cm]
%%{\cal M}_{p_1p_2p_3p_4}(\check s,\check t, \tilde s,\tilde t)
& =& {\cal M}_{p_3,p_2,p_1,p_4}(\hat t, \hat s, \check t,\check s)&= {\cal M}_{-c_t, -c_s,c_u}(\hat t, \hat s, \check t,\check s)\\[.25cm]		
\hline
 \end{array}$};
\end{tikzpicture}
\end{array}
\ee
The dependence on $\Sigma$ is left implicit since $\Sigma$ does not transform under crossing. 
It is clear that crossing has simple properties w.r.t.~the triplets $\{ \hat s,\check s, c_s\}$, $\{ \hat t,\check t, c_t\}$, $\{\hat u,\check u,c_u\}$.

%========================================================================================================
\subsection{A tree level primer}\label{tree_level_primer}
%========================================================================================================

We will focus in this paper on the quantum regime of $\mathcal{N}=4$ SYM  where the theory is dual to classical
supergravity (i.e.~the regime where we first take $N$ large and then take large 't Hooft coupling $\lambda$), and we will study the tree-level and 
the one-loop contribution to the Mellin amplitude, 
\be
\mathcal{M} = \mathcal{M}^{(1)} + \frac{1}{N^2} \mathcal{M}^{(2)} + \ldots,
\ee
with both contributions themselves expanded for large $\lambda$ as follows, 
\begin{align}
\mathcal{M}^{(1)} &=  \mathcal{M}^{(1,0)} + %\rule{2.2cm}{0pt} 
\lambda^{-\frac{3}{2}} \mathcal{M}^{(1,3)} + \lambda^{-\frac{5}{2}} \mathcal{M}^{(1,5)} +  \ldots \notag \\
\mathcal{M}^{(2)} &= \lambda^{\frac{1}{2}} \mathcal{M}^{(2,-1)} + 
\mathcal{M}^{(2,0)} +\lambda^{-1} \mathcal{M}^{(2,2)}
+ \lambda^{-\frac{3}{2}} \mathcal{M}^{(2,3)}  + \ldots \label{oneloopemmeformula}
\end{align}
The large $\lambda$ expansion is more precisely an expansion in curvature/derivative corrections, 
thus it is weighted by $\alpha'= \lambda^{-\frac{1}{2}}$.\footnote{The term $\lambda^{\frac{1}{2}}{\cal M}^{(2,-1)}$ 
corresponds to the $R^4$ counterterm.  A term $\lambda^{-\frac{1}{2}}\mathcal{M}^{(2,1)}$ would correspond 
to the genus one contribution to the modular completion of $\lambda^{-\frac{5}{2}}{\cal M}^{(1,5)}$ and it vanishes. 
The term $\lambda^{-1}{\cal M}^{(2,2)}$ corresponds to the genus one contribution to the modular completion of ${\cal M}^{(1,6)}$.
}

The combination of analytic bootstrap techniques and the knowledge about the spectrum of supergravity, 
which consists of protected half-BPS single-particle states and multi-particle states (but no excited string states), 
allows one to solve the problem of computing the tree-level  contribution to the four-point correlators. 
The expression for $\mathcal{M}^{(1,0)}$  was given in \cite{Rastelli:2016nze,Rastelli:2017udc}, building 
on previous work \cite{Dolan:2006ec}.\footnote{See \cite{Arutyunov:2017dti} and refs.~therein for the gravity counterpart of this result.}
In our notation 
\be
\label{M00}
\mathcal{M}^{(1,0)} = \frac{1}{({\bf s} +1)({\bf t}+1)({\bf u}+1)}\,,
\ee
where the bold variables are given by 
\be
{\bf s} = \hat{s} + \check{s} \qquad;\qquad {\bf t} = \hat{t} + \check{t} \qquad;\qquad%\quad  {\bf u} = \hat{u} + \check{u}\,, 
{\bf s} + {\bf t} + {\bf u} = -4\,.
\ee
The appearance of these bold-font variables has a precise meaning \cite{Aprile:2020luw}, which we will revisit later. 
The first tree-level $(\alpha')^3$ string correction is given simply by \cite{Drummond:2019odu}
\be
\label{M03}
\mathcal{M}^{(1,3)} = 2\zeta_3 (\Sigma-1)_3\,.
\ee
Higher order $\alpha'$ corrections come with non trivial polynomials in the Mellin variables and have been studied 
systematically via the bootstrap programme \cite{Goncalves:2014ffa,Alday:2018pdi,Drummond:2019odu}.  To various orders, fully explicit results have been 
computed in \cite{Drummond:2020dwr,Aprile:2020mus,Abl:2020dbx}.

At tree level the ultimate goal would be to resum all the $\alpha'$ corrections  to $\mathcal{M}^{(1,0)}$. 
The resummed $\mathcal{M}^{(1)}$, as function of $\alpha'$, would then give the Virasoro-Shapiro amplitude in $AdS_5\times S^5$, 
i.e.~the generalisation of the well known type IIB flat space amplitude. It is a non-trivial problem because the bootstrap program 
leaves unfixed a number of ambiguities at each order in $\alpha'$.  Additional input,\footnote{See \cite{Alday:2022uxp,Caron-Huot:2022sdy} for a dispersive sum rule approach.}  from supersymmetric localisation 
\cite{Binder:2019jwn,Chester:2019pvm,Chester:2020dja}, is already crucial to fix the ambiguities that appear at $(\alpha')^5$ \cite{Drummond:2020dwr}. %\cite{Drummond:2019odu}.
%\\[.15cm]

Both ${\cal M}^{(1,0)}$ and  ${\cal M}^{(1,3)}$ come with interesting properties which are simple to see in our formalism:  
We would expected ${\cal M}^{(1,0)}$ to be function of all variables $\{\hat{s},\hat{t},\check{s},\check{t},p_1,p_2,p_3,p_4\}$ 
but it happens to depend only on the specific combinations ${\bf s} = \hat{s} + \check{s}$ and 
${\bf t}=\hat{t}+\check{t}$, ${\bf u} = -{\bf s} - {\bf t} - 4$. Moreover, it does not depend explicitly on the charges $p_i$ at all.  
In a similar way, ${\cal M}^{(1,3)}$ is just a constant, but for the factor of $(\Sigma-1)_3$ (which is singlet under crossing).  The crucial observation is 
that both ${\cal M}^{(1,0)}$ and ${\cal M}^{(1,3)}$ can be understood as 10d objects. In the case of supergravity, it was shown in \cite{Caron-Huot:2018kta} that ${\cal M}^{(1,0)}$ enjoys a 10d conformal symmetry. In the case of tree level  $\alpha'$ corrections, the authors of 
\cite{Abl:2020dbx} showed that  ${\cal M}^{(1,3)}$ (and in fact  ${\cal M}^{(1,i\ge 3)}$)  is the dimensional reduction of a 
10d contact diagram in $AdS_5\times S^5$.  Since the operators $\mathcal{O}_p (x,y)$ are Kaluza-Klein modes,
the 10d structure of the tree-level correlators implicitly constrains the way the Mellin amplitude depends on the charges. 
%

%=================================================================================================================
\subsection{Summary of results: One-loop string corrected $AdS_5\times S^5$ }
\label{summary_sec}

This paper is dedicated to the study of one-loop string  contributions beyond the one-loop supergravity correlators 
$\mathcal{M}^{(2,0)}$ studied in \cite{Aprile:2019rep}.  We will focus on $\mathcal{M}^{(2,3)}$ at $(\alpha')^3$, 
 generalising previous work done in \cite{Alday:2018kkw,Alday:2018pdi,Drummond:2019hel,Drummond:2020uni,Drummond:2019odu,Drummond:2020dwr}. 
 Along the way  we will discuss the general picture for higher orders in $\alpha'$. 
 We will explain how the bootstrap program works in the next section. Here we would like to  
 summarise our main results and novelties, compared to the existing literature. 

The OPE determines the gravity amplitude from CFT data of exchanged two-particle operators at tree 
level.\footnote{up to a handful of ambiguities which cannot be fixed by the current bootstrap program.}
In the case of $\mathcal{M}^{(2,3)}$, the CFT data comes from $\mathcal{M}^{(1,0)}$ and $\mathcal{M}^{(1,3)}$, 
which we reviewed above in \eqref{M00} and \eqref{M03}. 
Within certain ranges of twist which we refer to as Above Window, Window, and Below Window,
the tree level OPE carries information about the maximal $\log^2 U$ discontinuity, the $\log U$ discontinuity 
and the analytic contribution, respectively.  
The Above Window region contains the $\log^2 U$ discontinuity and is fully determined by the OPE. 
The Window and the Below Window are finite ranges in the twist and give additional information 
on the structure of the amplitude. Indeed, as for $\mathcal{M}^{(2,0)}$ in supergravity \cite{Aprile:2019rep}, the $\log^2 U$ discontinuity 
is not enough to bootstrap the full one-loop amplitude, and the information coming 
from Window and Below Window is crucial to obtain the final result.

In order to appreciate the various novelties of one-loop physics, let us begin by noting that the ${\bf \Gamma}$ factor 
used to define the Mellin transform in \eqref{MellinrepUV} has itself a \emph{bonus property}. It
is invariant under \emph{variations} of the charges $p_i$ which swap pairs of values $p_{i}+p_{j}$ 
for $i,j$ in the ${\tt s}$, or ${\tt t}$, or ${\tt u}$ channel. For instance,  if we highlight the ${\tt s}$ channel in, 
\beq\label{example_Gamma_s}
{\bf \Gamma}=\frac{ \Gamma[\frac{p_1+p_2}{2} -\hat{s}]\Gamma[\frac{p_3+p_4}{2}-\hat{s}]}{  \Gamma[1+\frac{p_1+p_2}{2}+\check{s}]\Gamma[1+\frac{p_3+p_4}{2}+\check{s}]}\Gamma_{\tt t}\Gamma_{\tt u}
\eeq
this is invariant under variations of the charges such that the values of ${p_1+p_2}{}$ and ${p_3+p_4}{}$ 
swap, but the other combinations  $p_i+p_j$  remain unchanged. Let us emphasise that this 
property is not crossing symmetry, hence the use of the term \emph{variations}. In terms of the $c_s,c_t,c_u$ 
parametrisation of the charges,  see e.g.~\eqref{ex_gammas}, the variations we are discussing 
just amount to exchange the values $\pm c_s$.  In \eqref{example_Gamma_s} we looked at the ${\tt s}$ channel, but 
of course the same reasoning applies to the other channels.

To have a concrete and simple example of the bonus property in mind, consider the case of correlators 
$\vec{p}=(4424)$ and $\vec{p}=(3335)$. These correlators have indeed the same ${\bf \Gamma}$, 
but more importantly, since the Mellin amplitudes $\mathcal{M}^{(1,0)}$ and $\mathcal{M}^{(1,3)}$ 
are themselves invariant under the aforementioned variations, the interacting correlators 
are equal at the corresponding orders in the expansion,
\be
\label{degenex}
\mathcal{H}_{4424}(U,V,\tilde U,\tilde V)=\mathcal{H}_{3335}(U,V,\tilde U,\tilde V).
\ee
Recall that ${\cal H}_{\vec{p}}$ is introduced in \eqref{MellinrepUV} and, up to numerical factors normalisations 
dictated by \eqref{MellinrepUV} and free propagators removed, is just the interacting part of $\langle \mathcal{O}_{p_1} \mathcal{O}_{p_2} 
\mathcal{O}_{p_3} \mathcal{O}_{p_4} \rangle$.

More generally, we shall say that two correlators are degenerate when they have 
the same values of $\Sigma$ and $|c_s|,|c_t|,|c_u|$.  
We will show that whenever two correlators are degenerate in the tree level $\alpha'$ expansion, this degeneracy 
is lifted at one-loop  at the corresponding order in $\alpha'$. This lift was first discussed in supergravity \cite{Aprile:2019rep} 
but its expression at the level of the Mellin amplitude was not yet investigated.  
In this paper we provide very concrete formulae which exhibit the degeneracty lifting in the case
of $\mathcal{M}^{(2,3)}$, and we believe that analogous formulae will hold  
at higher orders in $\alpha'$.

The Mellin amplitude will be written in the following way,
\begin{align}
\label{M13structure}
\mathcal{M}^{(2,3)}_{\vec{p}} = \Bigl[ \mathcal{W}^{(AW)}_{\vec{p}}(\hat{s},\check{s}) + \mathcal{R}^{(W)}_{\vec{p}}(\hat{s},\check{s})+ \mathcal{B}^{(BW)}_{\vec{p}}(\hat{s},\check{s}) \Bigr] + {\tt crossing }\,,
\end{align}
where the superscripts indicate which region of OPE data was used to fix the function, 
i.e.~Above Window (AW), Window (W) and Below Window (BW).

\underline{The first term} in \eqref{M13structure} is given by
\be
\mathcal{W}_{\vec{p}}^{(AW)} = \mathpzc{w}_{\vec{p}}(\hat{s},\check{s})\, \tilde{\psi}^{(0)}(-{\bf s})\,,
\ee
where $ \mathpzc{w}_{\vec{p}}(\hat{s},\check{s})$ is a polynomial and $\tilde{\psi}^{(0)}(-{\bf s}) \equiv \psi^{(0)}(-{\bf s}) + \gamma_E$ 
is the digamma function shifted by the Euler constant.
This term is entirely determined by the OPE prediction for the $\log^2 U$ discontinuity, 
corresponding to exchange of two-particle operators Above Window (AW).

\underline{The second term} in \eqref{M13structure} represents the novelty of the one-loop function. It takes the form, 
\be
\label{window_splitting1int}
\mathcal{R}^{(W)}_{\vec{p}}(\hat{s},\check{s})= \sum_{z=0}^6 \frac{1}{{\bf s} -z}\biggl[
(\check{s}+\tfrac{p_{1}+p_2}{2}-z)_{z+1} \,
\mathpzc{r}^{+}_{\vec{p};z} 
(\hat{s},\check{s}) 
+
(\check{s}+\tfrac{p_{3}+p_4}{2}-z)_{z+1}\, \mathpzc{r}^{-}_{\vec{p};z}(\hat{s},\check{s})
\biggr]\,,
\ee
where crossing implies that $\mathpzc{r}^{\pm}$ are related to each other, and given in terms of a single function with certain residual symmetry in the charges $\vec{p}$:
\beq\label{solu_crossing_intro}
\left\{\begin{array}{rl}   
%
%\mathpzc{r}^+_{\vec{p}}  = \mathpzc{r}_{\{-c_s,-c_t,c_u\}}\qquad;\qquad \mathpzc{r}^-_{\vec{p}} = \mathpzc{r}_{\{+c_s,+c_t,c_u\}},
%
\mathpzc{r}^{+}_{\vec{p}}  &= \mathpzc{r}_{\{-c_s,-c_t,c_u\}}\\[.25cm]
\mathpzc{r}^{-}_{\vec{p}}  &= \mathpzc{r}_{\{+c_s,+c_t,c_u\}}
\end{array}\right.
\qquad;\qquad
\mathpzc{r}_{\{c_s,c_t,c_u\}}= \mathpzc{r}_{\{c_s,-c_t,-c_u\}}\quad;\quad\mathpzc{r}_{\{c_s,c_t,c_u\}}=\mathpzc{r}_{\{c_s,c_u,c_t\}}.
\eeq
The function $\mathpzc{r}_{\vec{p};z}(\hat s,\check s)$ is a polynomial, for each $z$,
determined by OPE predictions for the $\log^1 U$  discontinuity in the Window. The poles in $z$ come with 
the bold font variables, and the structure of poles in $\hat s, \check s,$ follows. We will explain this in the next sections.  

Let us comment on the reason why ${\cal R}^{(W)}_{}$ represents a novelty: 
When we look at ${\cal R}^{(W)}_{}$ together with the ${\bf \Gamma}$ functions,  
say  we focus on $\Gamma_{\tt s}$ in the ${\tt s}$-channel, the total amplitude undergoes the following split,
 \beq\label{sph_split_intro}
\!\!\!\!\!\!\begin{array}{c}
\begin{tikzpicture}
\draw (-.5,.5) node[scale=.9] {$ \ \ \ \ \ \ \ 
\displaystyle %\frac{1}{{\bf s} -z}
\frac{ {\cal R}^{(W)}_{\vec{p}} }{  \Gamma[1+\frac{p_{1}+p_2}{2}+\check{s}]\Gamma[1+\frac{p_{3}+p_4}{2}+\check{s}]} $};  %\frac{1}{s+\tilde s-z}\Bigg[ \frac{1}{\Gamma[\tilde{s} +1] \Gamma[\tilde{s} +c_s+1]} \Bigg]$};
\draw[-latex] (-0.1,-0.1)--(-1,-1.1);
\draw[-latex] (-0.1,-0.1)--(1.2,-1.1);

\draw (-.5,-1.5) node[scale=.9] {$
\displaystyle
=\frac{1}{ {\bf s} -z}\Bigg[ \frac{\mathpzc{r}_{\vec{p};z}^{+}  }{ \Gamma[-z+\frac{p_{1}+p_2}{2}+\check{s}]\Gamma[1+\frac{p_3+p_4}{2}+\check{s}] }    \,  
+ \frac{\mathpzc{r}_{\vec{p};z}^{-}  }{ \Gamma[1+\frac{p_1+p_2}{2}+\check{s}]\Gamma[-z+\frac{p_3+p_4}{2}+\check{s}] } \Bigg]$};
\end{tikzpicture}
\end{array}
\eeq
In particular the sphere $\Gamma$ functions (in $\bf{\Gamma}$) split into two $z$-dependent gamma functions, with residues
$\mathpzc{r}^{\pm}$ respectively. %related by crossing through  \eqref{solu_crossing_intro}. do not swap under 
%$c_s\leftrightarrow -c_s$ when there is non trivial $c_t,c_u$  dependence. 
 %
Since we will find that 
$\mathpzc{r}^{\pm}$ have generic charge dependence, i.e.~they depend on $c_t$ and $c_u$ non trivially,
it follows that $\mathpzc{r}^{\pm}$ do not map to each other under variations of charges that leave ${\bf \Gamma}$ invariant, and therefore
the bonus property, exemplified for instance in (\ref{degenex}), is lifted. 
%In the {\tt s}-channel the breaking is due to non trivial $c_t,c_u$  dependence. 
All together we refer to this phenomenon as {\it sphere splitting}.

\underline{The third term} in \eqref{M13structure} is found to take the form
\be
\label{belwindow}
\mathcal{B}^{(BW)}_{\vec{p}}(\hat{s},\check{s})=\sum_{z=0}^2 \frac{(\check{s}+\frac{p_{1}+p_2}{2} -z)_{z+1} (\check{s}+\frac{p_{3}+p_4}{2}-z)_{z+1}}{{\bf s} -z} \mathpzc{b}^{}_{\vec{p} ;z}(\hat{s},\check{s}) \,,
\ee
where again $\mathpzc{b}_{\vec{p}}(\hat{s},\check{s};z)$ are polynomials. 
This contribution is determined by OPE predictions from the $\log^0 U$ term in the Below Window region.

Remarkably, we find that $\sum_{z}$ in \eqref{window_splitting1int} and \eqref{belwindow} 
are finite and moreover the number of poles indexed by $z$ is independent of the charges!
This feature was already observed in \cite{Drummond:2020uni} by studying 
$\langle \mathcal{O}_2 \mathcal{O}_2 \mathcal{O}_p \mathcal{O}_p \rangle$ in the Window. It is not manifest from 
the form of the OPE, which instead depends on charges by construction. In fact, 
in order for this truncation to happen, there is a delicate interplay between $\mathcal{W}_{}^{(AW)} $ 
and $\mathcal{R}^{(W)}_{}$, and  $\mathcal{B}_{}^{(BW)}$.

The function $\mathcal{W}^{(AW)}$ contributes to $\log^2 U$ discontinuity by construction, but also contributes to the $\log^1 U$ 
discontinuity in the Window and the analytic term $\log^0 U$ in the Below Window regions. Similarly, $\mathcal{R}^{(W)}_{}$ 
contributes to $\log^1 U$ in the Window region by construction but also to the analytic term $\log^0 U$ in the Below Window region. 
This cascading behaviour results from our choice to use the bold font variables, ${\bf s,t,u,}$ in the parametrisation of the poles 
of $\mathcal{M}$, say $\tilde{\psi}^{(0)}(-\bf{s})$ and ${\bf s}-z$ in the ${\tt s}$ channel. 
This choice of parametrisation then reveals an additional simplicity: the truncation of the number of poles in $z$. In particular, 
we find that $\mathcal{R}^{(W)}_{}$ only contains seven poles $z=0,\ldots 6$ at order $(\alpha')^3$.

We can argue that the use of the bold font variables, ${\bf s,t,u}$ in the parametrisation of the poles of $\mathcal{M}$ is natural 
from the perspective of large $p$ limit \cite{Aprile:2020luw}, i.e.~from the expected behaviour of the amplitude when the charges $\vec{p}$ 
are taken to be large.  The large $p$ limit is well established in various $AdS\times S$ backgrounds \cite{Aprile:2021mvq,Drummond:2022dxd}. 
In the case we are interested in, the crucial observation is that, in the large $\vec{p}$ limit, the $AdS_5\times S^5$ 
Mellin amplitude asymptotes the flat space amplitude of IIB supergravity,  where ${\bf s}$ is identified with the corresponding ten-dimensional 
Mandelstam invariant of the flat space amplitude. Now, the flat space amplitude of the one-loop IIB amplitude has various $\log$ 
contributions, e.g. $\log(-{\bf s})$. It is natural that such logs should arise from the digamma in the limit of large ${\bf s}$ 
as $\tilde{\psi}^{(0)}(-{\bf s})\rightarrow \log(-{\bf s})$. It follows that ${\bf s}$ is 
the natural variable entering $\tilde{\psi}^{(0)}(-{\bf s})$ in the $AdS_5\times S^5$ Mellin amplitude. 
More evidence supporting the use of  ${\bf s,t,u,}$ in parametrising the poles of $\mathcal{M}$ also comes from 
our preliminary investigations on the $(\alpha')^{n+3}$ terms for $n>0$,  which show that the number of poles 
grows with $n$, but remains finite, and is independent of tree level ambiguities \cite{unpublished}.

Following similar logic, the poles of the function $\mathcal{B}^{(BW)}_{}$, are parametrised by ${\bf s,t,u}$. This function was 
previously studied for the single correlator $\langle \mathcal{O}_3 \mathcal{O}_3 \mathcal{O}_3 \mathcal{O}_3 \rangle$ in \cite{Drummond:2020uni}.
We find that only three poles at $z=0,1,2$ are needed to match the OPE data. Again, this truncation 
depends on the delicate interplay with both  $\mathcal{W}_{}^{(AW)} $ and $ \mathcal{R}^{(W)}_{}$. Finally, combining all 
contributions we are able to explicitly verify consistency with the ten-dimensional flat space limit. Since this property was not 
used in the detailed construction of the individual contributions, this provides a strong consistency check on the form of our final results.

The rest of the paper is organised as follows: In section \ref{sec:OPE} we review  
the bootstrap program for one-loop amplitudes, explaining in particular how the OPE of two-particle operators 
works.  In section \ref{one_loopM23_sec} we discuss the construction of the one-loop amplitude ${\cal M}^{(2,3)}$.
Emphasis is put on the `sphere splitting', which is a new effect at one loop.
Finally, in section \ref{sec_largeflatlimit} we show consistency of the one-loop amplitude with the large $p$ limit and 
the flat space limit in type IIB string theory.

%%=====================================================================================================================

\section{The AdS$_5 \times$S$^5$ OPE}
\label{sec:OPE}

This section provides preparatory material for the construction of one-loop amplitudes. 
It draws mainly from \cite{Aprile:2019rep} where it was explained how the CFT data 
collected from tree-level correlators needs to be organised at one-loop, in order to initiate the bootstrap 
program for correlators with arbitrary external charges $\vec{p}$. 
There are various important subtleties to be taken into account, that have to do with the relation among 
1) the spectrum of two-particle operators, 2) the (super)block decomposition of 
the correlator, and 3) the arrangement of poles in the Mellin amplitude.\footnote{We will study 
arbitrary Kaluza-Klein amplitudes, this discussion is far beyond a similar analysis for say the 
stress-tensor correlator in the $\mathcal{O}_2$ supermultiplet.} Explaining this will be our primary task in this section.

Some readers already familiar with the CFT construction of one-loop amplitudes in $AdS_5\times S^5$, might wish to go directly to section \ref{one_loopM23_sec}.

%===========================================================================================================
%===========================================================================================================

\subsection{Tree Level OPE}

Given four operators ${\cal O}_{p_i}(x_i)$ there are three independent ways to make them approach 
each other in pairs, and use the OPE.  To fix conventions, we will always consider the OPE channel 
where points $x_1\rightarrow x_2$ and  $x_4\rightarrow x_3$ approach each other, ordered as illustrated below
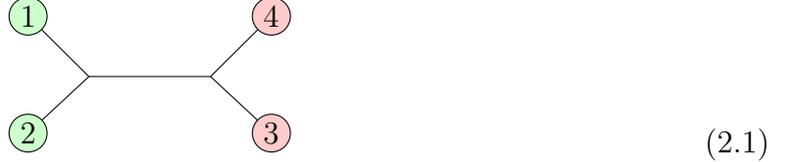
\beq
\begin{tikzpicture}

%\draw (0.5, 2.225)  node {$c_s< 0$};

\draw (0,1.3) -- (.8,.5); 
\draw (0,-.25) -- (.8,.5); 
\draw (3.2,1.3) -- (3.2-.8,.5); 
\draw (3.2,-.25) -- (3.2-.8,.5); 

\draw (3.2-.8,.5) -- (.8,.5); 

%\filldraw[fill=black!50,draw=black!50,rotate=-90]  (+.05,1) rectangle (-.05,-.25); 
%\filldraw[fill=black!50,draw=black!50,rotate=-45] (-.05,-.25) rectangle (+.05,1);
%
%\filldraw[fill=black!50,draw=black!50,rotate=-15] (0,0+.05)	 rectangle (1.2,-.05)	; 
%\filldraw[fill=black!50,draw=black!50] (-.05+1.2,-.25) rectangle (+.05+1.2,1.3); 
%\filldraw[fill=black!50,draw=black!50,rotate=+15] (0,1+.05)	 rectangle (1.3,1-.05)	; 

%\filldraw[fill=black!50,draw=black!50] (-.05+1.2,-.25) rectangle (+.05+1.2,1.3); 

\path		 (0,1.3) 	node [shape=circle,inner sep=0pt,minimum size=5mm,draw, fill=green!20] {1}
		 (0,-.25)		node [shape=circle,inner sep=0pt,minimum size=5mm,draw,fill=green!20] {2}
		 (3.2,-.25)		node [shape=circle,inner sep=0pt,minimum size=5mm,draw,fill=red!20] {3}
		 (3.2,1.3) 	node [shape=circle,inner sep=0pt,minimum size=5mm,draw,fill=red!20] {4};

%\path (0,1) node [shape=circle,inner sep=0pt,minimum size=5mm,draw] {1};
%\path (0,1) node [shape=circle,inner sep=0pt,minimum size=5mm,draw] {1};

%;    \draw (1.2,1.25)  circle (5pt);
%\draw (0,0)  circle (5pt);    \draw (1.2,-.25)  circle (5pt);

\end{tikzpicture}
\eeq
For a given set of charges $\vec{p}$, we will need to consider all three OPE channels. Thus we will consider
the three possible orderings of $\vec{p}$ (mod $x_1\leftrightarrow x_2$ and $x_3\leftrightarrow x_4$ exchange, 
since this is a symmetry of the OPE).

%=========================================================================
\subsubsection{Mellin space vs Quantum Numbers}
%=========================================================================
\label{sec_Mellin_numbers}

The interacting part of the correlator \eqref{MellinrepUV} 
is written as an expansion in monomials ${\tilde U}^{\check s} {\tilde V}^{\check t}$ over a triangle
in the $(\check{s} ,\check{t})$ plane given by 
\beq\label{triangle_h}
\check s \geq  - {\min}(\tfrac{p_{1}+p_2}{2} ,\tfrac{p_{3}+p_4}{2} )\quad;\quad
\check t \geq  - {\min}(\tfrac{p_{1}+p_4}{2},\tfrac{p_{2}+p_3}{2})\quad;\quad
\check u \geq  - {\min}(\tfrac{p_{1}+p_3}{2},\tfrac{p_{2}+p_4}{2})
\eeq
where $ \check{u} = -\Sigma - 2-\check{s} - \check{t} $. 
Finiteness of this sum is due to the sphere $\Gamma$ functions in the denominator 
of ${\bf \Gamma}$ factor, i.e.~outside of \eqref{triangle_h} the factor $1/\Gamma_{\check s}\Gamma_{\check t}\Gamma_{\check u}$ vanishes.
The amplitude is thus polynomial in ${\tilde U}$ and ${\tilde V}$. The triangle \eqref{triangle_h} 
is also in correspondence with  the $su(4)$ representations $[aba]$ flowing between the 
OPEs $\mathcal{O}_{p_1}(x_1)\times \mathcal{O}_{p_2}(x_2)$ and $\mathcal{O}_{p_3}(x_3)\times \mathcal{O}_{p_4}(x_4)$. 
In particular, we will think of $\check s$ as the conjugate variable to `twist' for the sphere, 
i.e. $b$, and $\check t$ as the conjugate variable for 'spin' on sphere, i.e.~$a$.

The number of long $su(4)$ channels a correlator contributes to depends only on the charges, 
and can be accounted by introducing the degree of extremality $\kappa$. 
A nice and fully symmetric expression for $\kappa$ can be given as follows,
\beq
\kappa= {\rm min}(\tfrac{p_1+p_2}{2} ,\tfrac{p_{3}+p_4}{2} ) + {\rm min}(\tfrac{p_1+p_4}{2},\tfrac{p_{2}+p_3}{2} ) + {\rm min}(\tfrac{p_{1}+p_3}{2},\tfrac{p_{2}+p_4}{2}) - \Sigma - 2\,
\eeq
The number of long $su(4)$ channels is then $\frac{(\kappa+1)(\kappa+2)}{2}$. The reps $[aba]$ flowing 
in the OPE instead depend on the orientation of the charges and they are %\footnote{The value of $b_{min}$ is also 
%such that the internal blocks have well defined entries in the $F^{+}_{-{b}/{2} }(x)$.}
%
\beq
a=0,1,\ldots, \kappa\qquad;\qquad \begin{array}{l} b=b_{min},b_{min}+2,\ldots, b_{min}+2\kappa \\[.2cm] b_{min}=\max({|p_{1}-p_2|}{},{|p_{4}-p_{3}|}{}) \end{array}
\eeq
Let us now translate the triangle in $su(4)$ labels into the one in $\check s$ and $\check t$. 
We shall focus on $\check s$ first, since this is relevant for the $(\alpha')^3$ amplitude. 

It is almost immediate to see that 
\beq
\begin{array}{l}
\check s=\check s_{min},\check s_{min}+1,\ldots ,\check s_{max} \\[.2cm]
b=b_{max},b_{max}-2,\ldots, b_{min}
\end{array} \qquad;\qquad 
\left\{ \begin{array}{rl} \check s_{min}=&-\min(\tfrac{p_{1}+p_2}{2},\tfrac{p_{3}+p_4}{2}) \\[.2cm]  
\check s_{max}=&-\max(\tfrac{|p_{1}-p_2|}{2},\tfrac{|p_{4}-p_{3}|}{2})-2 \end{array}\right.
\eeq
where $\check s$ is max when $\check t$ and $\check u$ are minimum and $\check s_{max}-\check s_{min}=\kappa$. 
In sum 
\beq
\frac{b}{2}=-\check s -2.
\eeq
The change from $su(4)$ harmonics $Y_{[0b0]}$ and monomials ${\tilde U}^{\check s}$ is a triangular matrix of the form
\beq\label{change_aba_monomials}
\begin{tikzpicture}

\draw (-.2,0) node[scale=.9] {$\left[\begin{array}{c} Y_{[0,b_{min},0] } \\ \vdots \\ Y_{[0,b_{max},0] }\end{array}\right]=$};
\draw (4,0) node[scale=.9] {$\left[\begin{array}{c} {\tilde U }^{\check s_{max}}\\ \vdots \\ {\tilde U}^{\check s_{min}} \end{array}\right]$};

\filldraw[ blue!20] (1.3,.65) -- (3.1,.65) -- (3.1,-.66)-- cycle;
\draw[line width=.6pt] (1.3,.65) rectangle (3.1,-.66);
\draw (1.8,-.2) node[scale=1.5] {$0$};

\end{tikzpicture}
\eeq
In particular, the monomial ${\tilde U}^{\check s_{min}}$ is the one and only one contributing to $[0b_{max}0]$, 
but as we lower $b\leq b_{max}$ sequentially a new monomial each time starts contributing. 
The inclusion of $\check t$ and $a$ for each $\check s$ and $b$ is straightforward at this point. 
The total range of $\check t$ simply covers 
$\check t=\check t_{min},\check t_{min}+1,\ldots ,\check t_{max}=\check t_{min}+\kappa$ 
with $\check t_{min}=- {\min}(\tfrac{p_{1}+p_4}{2},\tfrac{p_{2}+p_3}{2})$.
In the same way the total range of $a$ covers $a=0,1,\ldots, \kappa$.

Now that we have understood how the 4pt Mellin amplitude contributes to the various $su(4)$ channels, 
we would like to explain how the poles of 
\be
\label{M00_again}
\mathcal{M}^{(1,0)} = \frac{1}{({\bf s} +1)({\bf t}+1)({\bf u}+1)}\,,
\ee
are in correspondence with contributions in twist for each $su(4)$ channel. This will then lead us to our classification of the three regions Below Window, Window, and Above Window.

The simple pole at ${\bf s} +1=0$ is equivalently described by $\hat{s}=-1-\check{s}$.
It follows from \eqref{change_aba_monomials} that if we look at $[0b0]$ channels we find 
\beq\label{altra_tab}
\begin{array}{c|ll}
&{\rm lowest\ pole\ from\ }{\bf s}+1=0 & \\
\hline\\
{[}0b_{max}0{]} &\ \  \hat{s} = -1 - \check{s}_{min} = -1+\min(\tfrac{p_{1}+p_2}{2},\tfrac{p_{3}+p_4}{2})&= \frac{b_{max}}{2}+1\\
\vdots  &\rule{3.05cm}{0pt} \vdots &\ \vdots \\
{[}0b_{min}0{]} &\ \   \hat{s} = -1 - \check{s}_{max} = +1+\max(\tfrac{|p_{1}-p_2|}{2},\tfrac{|p_{4}-p_{3}|}{2})& = \frac{b_{min}}{2}+1 \\ \\
\hline
& %\ \   \hat{s}_{min} 
&= {\rm unitarity\ bound }\rule{0pt}{.5cm}
\end{array}
\eeq
Since there is a triangular transformation between monomials and $su(4)$ harmonics 
the value of $\hat{s}$ written in the above table is actually the minimum value.  Thus for labels $[aba]$ simple poles are given by
\beq\label{BW_uno}
 \hat{s}=[ a+\tfrac{b}{2}+1, \ldots ,\min(\tfrac{p_{1}+p_2}{2},\tfrac{p_{3}+p_{4}}{2})-1]
\eeq
Each simple pole contributes with a $U^{\hat{s}}$. Now, because a long block of twist $\tau$ has 
a leading power in $U$ given by $U^{\tau/2}$, we find that poles in \eqref{altra_tab}  imply the presence of 
a contribution in twist starting at the unitarity bound $\tau = 2a +b +2$.  Then, other simple poles coming 
from $({\bf s}+1)=0$ add a \emph{new} contribution in twist up to $\tau = {\rm min}(p_1+p_2,p_3+p_4)-2$. We refer to this region as `Below Window'.

Another sequence of simple poles contributing to the correlator comes from the ${\bf \Gamma}$ factor. 
We define the Window Region by the sequence of simple poles at 
\beq\label{W_uno}
\hat{s} = p_{min},\ldots, p_{\text{max}}-1\qquad;\qquad  {\rm Window\ Region}, 
\eeq
where $p_{\text{min}}={\rm min}(\tfrac{p_{1}+p_2}{2},\tfrac{p_{3}+p_{4}}{2})$ and $p_{\text{max}}={\rm max}(\tfrac{p_{1}+p_2}{2},\tfrac{p_{3}+p_{4}}{2})$. 

Finally, also from ${\bf \Gamma}$, we have an infinite sequence of double poles: 
\beq\label{AW_uno}
\rule{2.2cm}{0pt}\hat{s} \geq p_{\text{max}}\qquad;\qquad {\rm Above\ Window}.
\eeq
These double poles give rise to $\log U$ terms at tree level, upon performing the Mellin integral over $\hat{s}$.
We should notice that for $p_{1}+p_2=p_{3}+p_4$ the Window is empty. In this case, Above 
Window simply means above the threshold for exchange of long double traces, as defined in \eqref{AW_uno}.

%=========================================================================
\subsubsection{Two-particle operators and OPE data organisation}
%=========================================================================

The operators exchanged in the regions described in \eqref{BW_uno}-\eqref{W_uno}-\eqref{AW_uno}, 
whose OPE data we are interested in, are two-particle operators of the form 
\beq
{\cal O}_{\vec{\tau},(pq)}={\cal O}_{p} \partial^l \Box^{\frac{1}{2}(\tau-p-q)}{\cal O}_{q} \qquad (p\leq q),
\eeq
These operators are degenerate in free theory and will mix in the interacting theory. In the free theory, 
for fixed free quantum numbers, $\vec{\tau}$,  there are as many operators ${\cal O}_{(pq)}$ as integer 
pairs $(pq)$ in a certain rectangle $R_{\vec{\tau}}$ described in \cite{Aprile:2018efk}, 
\begin{align}\label{ir}
	p&=i+a+1+r\,, \qquad &q&=i+a+1+b-r\,, \notag\\
	i&=1,\ldots,(t-1)\,, &r&=0,\ldots,(\mu -1)\,,
\end{align}
so that $|R_{\vec{\tau}}|=\mu(t-1)$ with
\beq\label{multiplicity}
t\equiv \frac{(\tau-b)}{2}-a\,\quad;\quad 
\mu \equiv   \left\{\begin{array}{ll}
\bigl\lfloor{\frac{b+2}2}\bigr\rfloor \quad &a+l \text{ even,}\\[.2cm]
\bigl\lfloor{\frac{b+1}2}\bigr\rfloor \quad &a+l \text{ odd.}
\end{array}\right.
\eeq
Note that long operators have a minimum twist $\tau\ge2a+b+4$, i.e.~one unit above the unitarity bound.
Protected two-particle operators are those with twist at the unitarity bound. 

\begin{figure}
	\begin{center}
	%\hypertarget{fig:test}{}
		\begin{tikzpicture}[scale=.54]
		\draw[thin] (0,0) -- (0,6);
		\draw[ultra thick,->] (0,6) -- (0,9);
		\draw[thin] (4,0) -- (4,4);
		\draw[ultra thick,->] (4,4) -- (4,9);
		\draw[thin,dashed] (-.1,4) -- (4.1,4);
		\draw[thin,dashed] (-.1,6) -- (4.1,6);
		\draw[thin,dashed] (-.1,0) -- (4.1,0);
		\foreach \x in {0,1,2,3,4,5,6,7,8}
		\draw [fill] (0,\x) circle [radius=0.1]; 
		\foreach \x in {0,1,2,3,4,5,6,7,8}
		\draw [fill] (4,\x) circle [radius=0.1]; 
		\node at (2,8) {Above};
		\node at (2,7) {Window};%{Window};
		\node at (2,5) {Window};
		\node at (2,3) {Below};
		\node at (2,2) {Window};
		\node[left] at (0,0) {$\scriptstyle \tau=2a+b+2$};
		\node[left] at (0,1) {$\scriptstyle \tau=2a+b+4$};
		\node[right] at (4,4) {$\scriptstyle \tau=p_3+p_4$};
		\node[left] at (0,6) {$\scriptstyle \tau=p_1+p_2$};
		\node[left] at (0,4) {$\scriptstyle \tau = \tau^\text{min}$};
		\node[right] at (4,6) {$\scriptstyle \tau = \tau^\text{max}$};
		\draw[red,<->] (-3,6) -- (-3,9); 
		\node[red,left] at (-3,7.5) {$O(N^0)$};
		\draw[red,<->] (-3,0) -- (-3,5.7); 
		\node[red,left] at (-3,3) {$O(1/N^2)$};
		\node[red] at (-8,5) {$C_{p_1 p_2}(\mathcal{O}_\tau)=$};
		\draw[blue,<->] (7,4) -- (7,9); 
		\node[blue,right] at (7,6.5) {$O(N^0)$};
		\draw[blue,<->] (7,0) -- (7,3.7); 
		\node[blue,right] at (7,2) {$O(1/N^2)$};
		\node[blue] at (12,5) {$=C_{p_3 p_4}(\mathcal{O}_\tau)$};
		\end{tikzpicture}
			\end{center}
		\caption{The large $N$ structure of $C_{p_1p_2,\vec{\tau}}\,C_{p_3p_4,\vec{\tau}}$ for two particle operators $\mathcal{O}_{\tau}$ in an $su(4)$ representation $[aba]$, and varying twist. \label{figCCC}}
		%the SCPW expansion of the four-point function.}
	\end{figure}
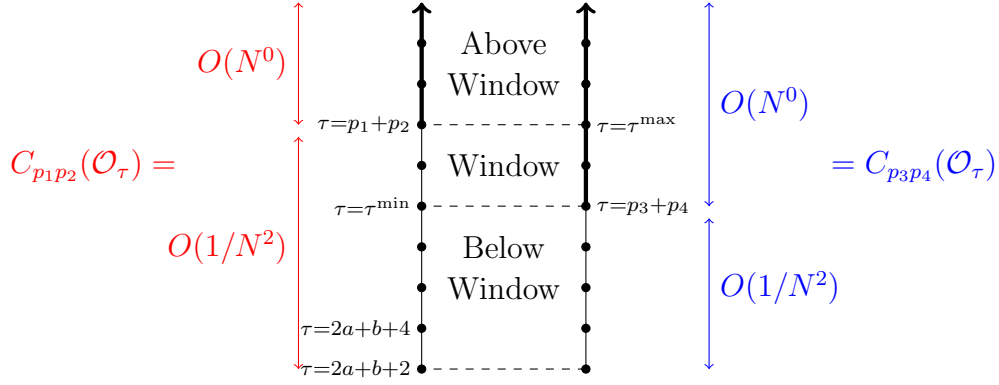

We will not discuss further the unmixing problem and we refer to the series of papers
\cite{Aprile:2017bgs,Aprile:2017qoy,Aprile:2017xsp,Aprile:2018efk,Aprile:2019rep} for details. 
The upshot is that, upon resolving the mixing, one is left with a set of true scaling eigenstates 
$\mathcal{K}_{\vec{\tau}}$ (as many of them as there are pairs $(pq)$ in $R_{\vec{\tau}}$) with dimensions,
\be
\Delta_{\mathcal{K}} = \tau + l + \frac{2}{N^2} \eta_{\mathcal{K}}^{(1)}  + O (\tfrac{1}{N^4})\,.
\ee
We can similarly expand the three-point couplings $C_{  {p_i}{p_j} {\cal K}_{\vec{\tau}} }$ of the two-particle 
operators with the external single-particle operators ${\cal O}_{p_i}$ and ${\cal O}_{p_j}$,
\be
C_{p_i p_j{\cal K}} = N^{\frac{p_{i}+p_j}{2}} \Bigl[ C^{(0)}_{p_ip_j{\cal K}}+\frac{1}{N^2} C^{(1)}_{p_ip_j{\cal K}} +\ldots \Bigr]
\ee
Note that $C^{(0)}_{p_ip_j{\cal K_{\vec{\tau}}}} = 0$ for $\tau < p_i + p_j$. This follows from the form of the 
long contribution to disconnected free theory, which is only non-zero in the Above Window region. 
Because of this, the $\frac{1}{N^2}$ below window region is empty, and it first gets contributions at one-loop. 
This is better illustrated by Figure \ref{figCCC}.\footnote{Since connected free theory contributes Below Window, the 
cancellation of  Below Window contributions at tree level can be used to fix the tree level correlator \cite{Dolan:2006ec}. 
It is actually convenient to always cancel connected free theory contributions, i.e.~at all orders in $N$, and define a 
minimal interacting correlator whose OPE decomposition is one-to-one with the OPE predictions \cite{Aprile:2019rep}.}

Having identified the spectrum we can go ahead and write down the OPE predictions in the large $N$ 
expansion. The long multiplet contribution to the \emph{full} correlator (free$+$interacting) up to one-loop is given by\footnote{We have omitted terms with derivatives acting on the blocks as they do not affect the leading logs for each region.}
\begin{align}
 \label{eq:SCPW}
\langle \mathcal{O}_{p_1} \mathcal{O}_{p_2} \mathcal{O}_{p_3} \mathcal{O}_{p_4} \rangle_{\rm long}  = 
& \  \sum_{\vec{\tau}\in AW}  L^{(0)}_{\vec{p};\vec{\tau}}\mathbb{L}^{\vec{p}}_{\vec{\tau}} \\
&+ \frac{1}{N^2} \left[ \sum_{\vec{\tau}\in W} N^{(1)}_{\vec{p};\vec{\tau}} + \sum_{\vec{\tau}\in AW} V^{(1)}_{\vec{p};\vec{\tau}} \log(u) \right]\mathbb{L}^{\vec{p}}_{\vec{\tau}} + \ldots \notag \\
 &+  \frac{1}{N^4}\left[ \sum_{\vec{\tau}\in BW} K^{(2)}_{\vec{p};\vec{\tau}} + \sum_{\vec{\tau}\in W} H^{(2)}_{\vec{p};\vec{\tau}} \log(u) +  \sum_{\vec{\tau}\in AW}  M^{(2)}_{\vec{p};\vec{\tau}}\log^2(u) \right]\mathbb{L}^{\vec{p}}_{\vec{\tau}}\ + \ldots  \notag
\end{align}
Due to the operator mixing described above, the OPE data is better organised into matrices. 
Let us package the anomalous dimensions $\eta_{\mathcal{K}}^{(1)}$ into a diagonal 
$|R_\tau| \times |R_\tau|$ matrix ${\pmb \eta}^{(1)}$. Similarly we can arrange the leading 
order three-point functions into an $|R_\tau| \times |R_\tau|$ matrix ${\bf C}^{(0)}$. 
Then the OPE implies 
\begin{align}
{\bf L}^{(0)}_{\vec{\tau}} &= {\bf C}^{(0)} \cdot {{\bf C}^{(0)}}^T \,; \qquad &&({\bf L}_{\vec{\tau}}^{(0)} )_{(p_1p_2)(p_3p_4)} =  L_{\vec{p},\vec{\tau}}^{(0)} =\sum_{\mathcal{K}} C^{(0)}_{p_1p_2{\cal K}}C^{(0)}_{p_3p_4{\cal K}}\,,\notag\\
{\bf V}^{(1)}_{\vec{\tau}} &= {\bf C}^{(0)} \cdot {\pmb \eta}^{(1)} \cdot {{\bf C}^{(0)}}^T\,; \qquad &&({\bf V}_{\vec{\tau}}^{(1)})_{(p_1p_2)(p_3p_4)} = V_{\vec{p},\vec{\tau}}^{(1)}  = \sum_{\mathcal{K}} C^{(0)}_{p_1p_2{\cal K}}{\eta}^{(1)}_{\cal K} C^{(0)}_{p_3p_4{\cal K}}\,,
\label{mixing}
\end{align}
where both $(p_1p_2)$ and $ (p_3p_4)$ belong to $R_{\vec{\tau}}$. The matrices ${\bf L}^{(0)}$ and ${\bf V}^{(1)}$ 
are symmetric and are obtained by collecting the CPW expansion, for fixed $\vec{\tau}$, of the 
correlators $\langle \mathcal{O}_{p_1} \mathcal{O}_{p_2} \mathcal{O}_{p_3} \mathcal{O}_{p_4} \rangle$ 
with varying external charges. The matrix ${\bf L}^{(0)}_{}$ gives the (diagonal) CPW 
coefficients from disconnected free theory, while ${\bf V}^{(1)}_{}$ comes from the $\log U$ contribution   
in the tree-level interacting part. For given $\vec{\tau}$ both matrices have size $|R_{\vec{\tau}}|\times |R_{\vec{\tau}}|$ and are full rank.

A similar organisation principle holds for the subleading three-point couplings $C^{(1)}$ which 
arise from the window region. This time it is more natural to arrange a vector 
$({\bf C}^{(1)}_{\vec{\tau}})_{q_1 q_2}$, labelled by $\vec{\tau}$ and a fixed pair $q_1q_2$, 
such that $q_1+q_2> \tau\ge 2a+b+4$, with the vector index running 
over the operators $\mathcal{K}_{\vec{\tau}}$. From the OPE,
\beq\label{matrix_def_sugra2}
{\bf N}_{q_1 q_2}^{(1)} = {\bf C}_{q_1 q_2}^{(1)} \cdot {{\bf C}^{(0)}}^T \qquad ; 
\qquad \bigl[{\bf N}^{(1)}_{\vec{\tau},(q_1q_2)}\bigr]_{(p_3p_4)} = \sum_{\cal K} C^{(1)}_{q_1q_2{\cal K}}C^{(0)}_{p_3p_4{\cal K}}\,,  \qquad (p_3p_4)\in R_{\vec{\tau}}\,. 
\eeq
The various entries $\bigl[{\bf N}^{(1)}_{\vec{\tau},(q_1q_2)}\bigr]_{(p_3p_4)} $ are found from the 
CPW coefficients from $\langle \mathcal{O}_{q_1} \mathcal{O}_{q_2} \mathcal{O}_{p_3} \mathcal{O}_{p_4} \rangle$ 
in the Window.  In this case it is crucial to consider both the tree-level contribution to 
$\langle \mathcal{O}_{q_1} \mathcal{O}_{q_2} \mathcal{O}_{p_3} \mathcal{O}_{p_4} \rangle_{\rm int}$ 
and connected free theory in the long sector.  

The above discussion holds for any value of the 't Hooft coupling $\lambda$, but let us recall 
that we are interested in the regime of large $\lambda$ and expand the anomalous dimensions and three-point functions accordingly,
\begin{align}
\begin{array}{rlll }
{\eta}_{\cal K}^{(1)} &=\ {\eta}^{(1,0)}_{\cal K} &+\  \lambda^{-\frac{3}{2}} {\eta}^{(1,3)}_{\cal K} \ &+\  \lambda^{-\frac{5}{2}} {\eta}^{(1,5)}_{\cal K} + \ldots\, ,\notag  \\[.2cm]
C^{(0)}_{p_ip_j{\cal K}} &= {C}^{(0,0)}_{p_ip_j{\cal K}}& +  \ \lambda^{-\frac{3}{2}} {C}^{(0,3)}_{p_ip_j{\cal K}} &+ \lambda^{-\frac{5}{2}} {C}^{(0,5)}_{p_ip_j{\cal K}} + \ldots \notag \\[.2cm]
C^{(1)}_{p_ip_j{\cal K}} &= {C}^{(1,0)}_{p_ip_j{\cal K}} & +  \ \lambda^{-\frac{3}{2}} {C}^{(1,3)}_{p_ip_j{\cal K}} &+ \lambda^{-\frac{5}{2}} {C}^{(1,5)}_{p_ip_j{\cal K}} + \ldots\,.
\end{array}
\end{align}
The supergravity contributions to the anomalous dimensions of the operators 
are given by a very simple formula \cite{Aprile:2018efk},
\be
\label{supergravityanomdims}
\eta^{(1,0)}_{\mathcal{K}_{pq}} = - \frac{2 M^{(4)}_{t} M^{(4)}_{t+l+1}}{\left(\ell_{10}+1\right)_6}
\ee
where
\begin{align}
& M^{(4)}_t\equiv (t-1) (t+a) (t+a+b+1)(t+2a+b+2)\, \\
& \ell_{10} = l+2(p-2)-a + \frac{1-(-)^{a+l}}{2}.
\end{align}
The notation for $\ell_{10}$ reflects the fact that this quantity can be interpreted as a 
ten-dimensional spin \cite{Caron-Huot:2018kta}. For $\mu>1$ and $t> 2$ there is a residual degeneracy 
because $\ell_{10}$ depends only on $p$ and not on $q$. This degeneracy is illustrated in Figure~\ref{degeneracies} 
which gives a sketch of the rectangle $R_{\vec{\tau}}$ with operators of equal anomalous dimension 
connected by vertical lines. The residual degeneracy is a reflection of the ten-dimensional conformal symmetry described in \cite{Caron-Huot:2018kta}.

\begin{figure}
\begin{center}
\begin{tikzpicture}[scale=.54]
%
%\draw[step=2cm,gray,very thin] (-4,2) grid (8,8);
%
\def\prop{.5}
\def\shifthor{\prop*2}
\def\ptuno{(\prop*2-\shifthor,\prop*8)}
\def\ptdue{(\prop*5-\shifthor,\prop*5)}
\def\pttree{(\prop*9-\shifthor,\prop*15)}
\def\ptquattro{(\prop*12-\shifthor,\prop*12)}
%
%axis horizontal
\draw[-latex, line width=.6pt]		(\prop*1   -\shifthor-4,         \prop*14          -0.5*\shifthor)    --  (\prop*1  -\shifthor-2.5  ,   \prop*14-      0.5*\shifthor) ;
\node[scale=.8] (oxxy) at 			(\prop*1   -\shifthor-2.5,  \prop*16.5     -0.5*\shifthor)  {};
\node[scale=.9] [below of=oxxy] {$p$};
%
%axis vertical
\draw[-latex, line width=.6pt] 		(\prop*1   -\shifthor-4,     \prop*14       -0.5*\shifthor)     --  (\prop*1   -\shifthor-4,        \prop*17-      0.5*\shifthor);
\node[scale=.8] (oxyy) at 			(\prop*1   -\shifthor-2,   \prop*16.8   -0.5*\shifthor) {};
\node[scale=.9] [left of= oxyy] {$q$};
%
%rectangle
\draw[] 								\ptuno -- \ptdue;
\draw[black]							\ptuno --\pttree;
%\draw[green!50!black,thin]						(\prop*3-\shifthor,\prop*7) --(\prop*10-\shifthor,\prop*14);
%\draw[orange!70!white,thin]						(\prop*4-\shifthor,\prop*6) --(\prop*11-\shifthor,\prop*13);
\draw[black]							\ptdue --\ptquattro;
\draw[]								\pttree--\ptquattro;
\draw[-latex,gray, dashed]					(\prop*0-\shifthor,\prop*10) --(\prop*8-\shifthor,\prop*2);
\draw[-latex,gray, dashed]					(\prop*3-\shifthor,\prop*3) --(\prop*16-\shifthor,\prop*16);
%		
%dots
%
\foreach \indeyc in {0,1,2,3}
\foreach \indexc  in {2,...,9}
\filldraw   					 (\prop*\indexc+\prop*\indeyc-\shifthor, \prop*6+\prop*\indexc-\prop*\indeyc)   	circle (.07);
%
%letters
%
\node[scale=.8] (puntouno) at (\prop*4-\shifthor,\prop*8) {};
\node[scale=.8]  [left of=puntouno] {$A$};   
\node[scale=.8] (puntodue) at (\prop*5-\shifthor,\prop*6+.5) {};
\node[scale=.8] [below of=puntodue]  {$B$}; 
\node[scale=.8] (puntoquattro) at (\prop*13-\shifthor,\prop*15) {};
\node[scale=.8] [below of=puntoquattro] {$C$};
\node[scale=.8] (puntotre) at (\prop*9-\shifthor,\prop*13) {};
\node[scale=.8] [above of=puntotre] {$D$}; 
%
%
%legend
\node[scale=.84] (legend) at (16,5) {$\begin{array}{l}  
													\displaystyle A=(a+2,a+b+2); \\[.1cm]
													\displaystyle B=(a+1+\mu,a+b+3-\mu); \\[.1cm]
													\displaystyle C=(a+\mu+t-1,a+b+1+t-\mu); \\[.1cm]
													\displaystyle D=(a+t,a+b+t); \\[.1 cm] \end{array}$  };
\end{tikzpicture}
\caption{The set $\mathcal{R}_{\vec{\tau}}$ pictured in the $(p,q)$ plane. With vertical lines indicating operators with the same anomalous dimension see \cite{Aprile:2018efk} for full details.\label{degeneracies}}
\end{center}
\end{figure}

At order $\lambda^{-\frac{3}{2}}$ the anomalous dimensions are even simpler \cite{Drummond:2019odu}. 
Only operators with $\ell_{10}=0$, i.e. with $a=l=0$, $i=1$, $r=0$ receive an anomalous dimension 
at this order. In relation to the diagram Figure \ref{degeneracies}, these operators sit at the left-most 
corner of the rectangle $R_{\vec{\tau}}$ where $(p,q) = (2,2+b)$. Their anomalous dimensions read,
\be
\eta^{(1,3)} = -\frac{\zeta_3}{840}  \delta_{l,0} \delta_{a,0}  M^{(4)}_{t} M^{(4)}_{t+1} (t-1)_3 (t+b+1)_3 \,.
\ee

The matrix ${\bf L}^{(0)}$  is independent of $\lambda$, being derived from the disconnected part of the 
free theory correlator, hence we can also write ${\bf L}^{(0)} = {\bf L}^{(0,0)}$. The matrix ${\bf V}^{(1)}$ has an expansion for large $\lambda$,
\be
{\bf V}^{(1)} = {\bf V}^{(1,0)} + \lambda^{-\frac{3}{2}} {\bf V}^{(1,3)} + \lambda^{-\frac{5}{2}} {\bf V}^{(1,5)} + \ldots\,.
\ee
When we expand the mixing equations (\ref{mixing}) order by order in $\lambda^{-\frac{1}{2}}$ we find the following equations at leading order,
\begin{align}
\mathbf{C}^{(0,0)}{\mathbf{C}^{(0,0)}}^T &= {\bf L}^{(0,0)} \,, \notag \\
\mathbf{C}^{(0,0)}{\pmb \eta}^{(1,0)}{\mathbf{C}^{(0,0)}}^T &= {\bf V}^{(1,0)}\,.
\end{align}
This eigenvalue problem was solved in \cite{Aprile:2017xsp,Aprile:2018efk}, yielding 
the double-trace spectrum of anomalous dimensions in supergravity, which exhibit the partial residual degeneracy 
associated with the hidden conformal symmetry \cite{Caron-Huot:2018kta}. At order $\lambda^{-\frac{3}{2}}$ we find,
\begin{align}\label{equaOPEC03}
\mathbf{C}^{(0,3)}{\mathbf{C}^{(0,0)}}^T + \mathbf{C}^{(0,0)}{\mathbf{C}^{(0,3)}}^T &= 0 \,, \notag \\
\mathbf{C}^{(0,0)}{\pmb \eta}^{(1,3)}{\mathbf{C}^{(0,0)}}^T+  \mathbf{C}^{(0,3)}{\pmb \eta}^{(1,0)}{\mathbf{C}^{(0,0)}}^T\ +  \mathbf{C}^{(0,0)}{\pmb \eta}^{(1,0)}{\mathbf{C}^{(0,3)}}^T &= {\bf V}^{(1,3)}\,,
\end{align}
and at order $\lambda^{-\frac{5}{2}}$,
\begin{align}
\mathbf{C}^{(0,5)}{\mathbf{C}^{(0,0)}}^T + \mathbf{C}^{(0,0)}{\mathbf{C}^{(0,5)}}^T &= 0 \,, \notag \\
\mathbf{C}^{(0,0)}{\pmb \eta}^{(1,5)}{\mathbf{C}^{(0,0)}}^T+  \mathbf{C}^{(0,5)}{\pmb \eta}^{(1,0)}{\mathbf{C}^{(0,0)}}^T\ +  \mathbf{C}^{(0,0)}{\pmb \eta}^{(1,0)}{\mathbf{C}^{(0,5)}}^T &= {\bf V}^{(1,5)}\,.
\end{align}
In this paper we are mostly concerned with the order $\lambda^{-\frac{3}{2}}$ equations, which in fact yield ${\bf C}^{(0,3)}=0$ \cite{Drummond:2019odu},
since it has to be a one-dimensional matrix satisfying \eqref{equaOPEC03}. 
This simplification will not hold at the next order, i.e. ${\bf C}^{(0,5)} \neq 0$ \cite{Drummond:2020dwr}. By expanding the subleading couplings $C^{(1)}$  in $\lambda^{-\frac{1}{2}}$ we have,
\begin{align}
{\bf C}_{q_1 q_2}^{(1,0)} \cdot {{\bf C}^{(0,0)}}^T &= {\bf N}_{q_1 q_2}^{(1,0)}\,, \notag \\
{\bf C}_{q_1 q_2}^{(1,3)} \cdot {{\bf C}^{(0,0)}}^T + {\bf C}_{q_1 q_2}^{(1,0)} \cdot {{\bf C}^{(0,3)}}^T &= {\bf N}_{q_1 q_2}^{(1,3)}\,, \notag \\
{\bf C}_{q_1 q_2}^{(1,5)} \cdot {{\bf C}^{(0,0)}}^T + {\bf C}_{q_1 q_2}^{(1,0)} \cdot {{\bf C}^{(0,5)}}^T &= {\bf N}_{q_1 q_2}^{(1,5)}\,.
\end{align}
Again note that the order $\lambda^{-\frac{3}{2}}$ equation simplifies because ${\bf C}^{(0,3)} = 0$.

%===========================================================================================================
%===========================================================================================================
\subsection{One loop OPE}

Having reviewed tree level data, we are ready to use it to study one-loop string theory in $AdS_5\times S^5$.
This amounts to bootstrap the one-loop correlators by knowing
\beq
 %\label{eq:SCPW}
\!\!\langle \mathcal{O}_{p_1} \mathcal{O}_{p_2} \mathcal{O}_{p_3} \mathcal{O}_{p_4} \rangle_{\rm long}  = \ 
 \ldots \ +  \frac{1}{N^4}\left[ \sum_{\vec{\tau}\in BW} K^{(2)}_{\vec{p};\vec{\tau}} 
 + \sum_{\vec{\tau}\in W} H^{(2)}_{\vec{p};\vec{\tau}} \log(u) +  \sum_{\vec{\tau}\in AW}  M^{(2)}_{\vec{p};\vec{\tau}}\log^2(u) \right]\mathbb{L}^{\vec{p}}_{\vec{\tau}}\ + \ldots  \notag
\eeq
i.e.~by knowing the values of $M^{(2)}_{\vec{p};\vec{\tau}}$, $ H^{(2)}_{\vec{p};\vec{\tau}}$, and $K^{(2)}_{\vec{p};\vec{\tau}}$. Let us address each of these in turn according to power of $\log U$:

\paragraph{$\bullet\ {\pmb \log^2 U}$ Above Window.}
For fixed quantum numbers $\vec{\tau}$ and charges $\vec{p}$ the OPE prediction can be read off the matrix ${\bf M}^{(2)}$ given by
\be
{\bf M}^{(2)} =\tfrac{1}{2}\ {\bf C}^{(0)} \cdot \bigl({\pmb \eta}^{(1)}\bigr)^2 \cdot {{\bf C}^{(0)}}^T =\tfrac{1}{2}\ {\bf V}^{(1)} \cdot \bigl({\bf L}^{(0)}\bigr)^{-1} \cdot {\bf V}^{(1)}\,.
\ee
Thus the $\log^2 U$ terms are entirely predicted from the disconnected and tree-level CPW coefficients ${\bf L}$ and ${\bf V}$.

Expanding the above relation in $\lambda^{-\frac{1}{2}}$ we have
\begin{align}
\label{logsqsupergravity}
{\bf M}^{(2,0)}  &=\tfrac{1}{2} {\bf C}^{(0,0)} \cdot \bigl({\pmb \eta}^{(1,0)}\bigr)^2 \cdot {{\bf C}^{(0,0)}}^T = \tfrac{1}{2}  {\bf V}^{(1,0)} \cdot \bigl({\bf L}^{(0)}\bigr)^{-1} \cdot {\bf V}^{(1,0)}\,, \\[.2cm]
{\bf M}^{(2,3)}  &= {\bf C}^{(0,0)} \cdot {\pmb \eta}^{(1,0)} \cdot {\pmb \eta}^{(1,3)}  \cdot {{\bf C}^{(0,0)}}^T  +\tfrac{1}{2} 
\left[ {\bf C}^{(0,3)} \cdot {\pmb \eta}^{(1,0)}\cdot {\pmb \eta}^{(1,0)} \cdot { {\bf C}^{(0,0)}}^T +{\rm tr.} \right] \notag \\
&= \tfrac{1}{2} \left( {\bf V}^{(1,3)} \cdot \bigl({\bf L}^{(0)}\bigr)^{-1} \cdot {\bf V}^{(1,0)} + {\bf V}^{(1,0)} \cdot \bigl({\bf L}^{(0)}\bigr)^{-1} \cdot {\bf V}^{(1,3)}\right)\,.
\label{logsqstring}
\end{align}
In the second line above we have used $\mathbf{C}^{(0,3)}{\mathbf{C}^{(0,0)}}^T + \mathbf{C}^{(0,0)}{\mathbf{C}^{(0,3)}}^T = 0$, 
and that ${\pmb \eta}^{(1,0)}$ and ${\pmb \eta}^{(1,3)}$ are diagonal and hence commute. 
The first condition at $(\alpha')^3$ is obvious, since $C^{(0,3)}=0$.  A formula like \eqref{logsqstring} 
holds at $(\alpha')^5$ upon replacing  ${\bf V}^{(1,5)}$.

\paragraph{$\bullet\ {\pmb \log^1 U}$ in the Window.}
In a similar way as above, we arrange the vector ${\bf H}^{(2)}_{q_1 q_2}$ labelled by a fixed pair of charges $(q_1 q_2)$,
\begin{align}\label{general_OPE_W}
{\bf H}^{(2)}_{q_1 q_2} = {\bf C}_{q_1 q_2}^{(1)} \cdot {\pmb \eta}^{(1)} \cdot {{\bf C}^{(0)}}^T = {\bf N}^{(1)}_{q_1 q_2} \cdot \bigl({\bf L}^{(0)}\bigr)^{-1} \cdot {\bf V}^{(1)}\,.
\end{align}
Recall that $ {\bf C}_{q_1 q_2}^{(1)}$ exists only for operators $\mathcal{K}_{\vec{\tau}}$ such that $\tau<q_1+q_2$. 
Recall also that for a correlator $\langle {\cal O}_p{\cal O}_q {\cal O}_p{\cal O}_q\rangle$ there is no Window. 

If we now expand in $\lambda^{-\frac{1}{2}}$ we find
\begin{align}
\label{Wsupergravity}
{\bf H}^{(2,0)}_{q_1 q_2}  &= {\bf N}^{(1,0)}_{q_1 q_2} \cdot \bigl({\bf L}^{(0,0)}\bigr)^{-1} \cdot {\bf V}^{(1,0)}\,, \\[.2cm]
{\bf H}^{(2,3)}_{q_1 q_2}  &= {\bf N}^{(1,3)}_{q_1 q_2} \cdot \bigl({\bf L}^{(0,0)}\bigr)^{-1} \cdot {\bf V}^{(1,0)} + {\bf N}^{(1,0)}_{q_1 q_2} \cdot \bigl({\bf L}^{(0,0)}\bigr)^{-1} \cdot {\bf V}^{(1,3)}\,. 
\label{Wstring}
\end{align}
which give the OPE predictions in terms of tree level data. 

\paragraph{$\bullet\ {\pmb \log^0 U}$ Below Window.}
Finally, in the below-window region,
\be\label{general_OPE_BW}
{\bf K}^{(2)}_{q_1q_2q_3q_4} = {\bf C}^{(1)}_{q_1 q_2} \cdot {{\bf C}^{(1)}_{q_3 q_4}}^T = {\bf N}^{(1)}_{q_1q_2} \cdot \bigl({\bf L}^{(0)} \bigr)^{-1} \cdot {{\bf N}^{(1)}_{q_3q_4}}^T\,.
\ee
Once again we can expand in $\lambda^{-\frac{1}{2}}$ to obtain,
\begin{align}
\label{BWsupergravity}
{\bf K}^{(2,0)}_{q_1q_2q_3q_4} &= {\bf N}^{(1,0)}_{q_1q_2} \cdot \bigl({\bf L}^{(0,0)} \bigr)^{-1} \cdot {{\bf N}^{(1,0)}_{q_3q_4}}^T \,, \\
{\bf K}^{(2,3)}_{q_1q_2q_3q_4} &= {\bf N}^{(1,3)}_{q_1q_2} \cdot \bigl({\bf L}^{(0,0)} \bigr)^{-1} \cdot {{\bf N}^{(1,0)}_{q_3q_4}}^T + {\bf N}^{(1,0)}_{q_1q_2} \cdot \bigl({\bf L}^{(0,0)} \bigr)^{-1} \cdot {{\bf N}^{(1,3)}_{q_3q_4}}^T \,.
\label{BWstring}
\end{align}
which give the OPE predictions in terms of tree level data. The Below Window predictions are non trivial for the first time only at one-loop! \\[.15cm]

The double logarithmic behaviour at one-loop in supergravity has been studied extensively 
\cite{Alday:2017xua,Aprile:2017bgs,Drummond:2020uni,Aprile:2019rep,Alday:2019nin}. 
It has been used to make predictions for the form of the one-loop correlator both in position 
\cite{Aprile:2017bgs,Aprile:2019rep,Drummond:2020uni} and Mellin space \cite{Alday:2019nin,Aprile:2020luw}. 
The relations (\ref{Wsupergravity}) and (\ref{BWsupergravity}) only become relevant for correlators 
with multiple $su(4)$ channels. These have been studied in \cite{Aprile:2019rep} with several explicit 
examples constructed in position space. The complication in supergravity comes from the fact that 
both Window and Below Window  predictions are non-trivial at all spins, and it remains difficult to find the 
one-loop Mellin amplitude explicitly for generic external charges.  %\\%[.15cm]

Our focus in this paper will be on the $(\alpha')^3$ one-loop correlator and thus equations (\ref{logsqstring}), (\ref{Wstring}) and (\ref{BWstring}). 
The advantage of studying the ${\cal M}^{(2,3)}$ amplitude, compared to supergravity, stems from the 
truncation in spin of the spectrum of two-particle operators exchanged. 
This simplifies the structure of the OPE predictions in the Window and Below Window regions, and will indeed 
allow us to find an expression for the Mellin amplitude $\mathcal{M}^{(2,3)}_{\vec{p}}$ 
for general $\vec{p}$. Given our understanding here, we believe that the same pattern applies at all orders in $\alpha'$, even though computationally it will become a bit more involved.

%===========================================================================================================
%===========================================================================================================

\subsubsection{Window splitting}\label{window_split_sec}

We already mentioned in section \ref{summary_sec} the existence of degenerate correlators, i.e.~correlators 
that have the same values of $\Sigma$ and $|c_s|,|c_t|,|c_u|$, and therefore such  
that ${\bf \Gamma}$ is unchanged. For example, two correlators whose values of 
$p_{1}+p_2$ and $p_{3}+p_4$ swap, and the other $p_{i}+p_j$ are unchanged. At order ${\cal M}^{(1,0)}$ and ${\cal M}^{(1,3)}$ 
these correlators are necessarily proportional to each other  since both these Mellin amplitudes 
do not distinguish them. The situation at one loop is quite different. The crucial point is that the OPE predictions now involve a 
sum over operators which mixes degenerate and non-degenerate data at tree level. 

The example 
of $\langle {\cal O}_3{\cal O}_3{\cal O}_3{\cal O}_5 \rangle$ and $\langle {\cal O}_4{\cal O}_4{\cal O}_2{\cal O}_4 \rangle$ discussed in \eqref{degenex} is quite useful to see what is going on. 
The one-loop predictions in the Window will involve
%we will need to consider
\begin{align}
\left[   {\bf N}^{(1,0)}_{35}\right]_{(rs)= (24),{\color{purple} (33)}} \qquad;\qquad
\left[\, {\bf N}^{(1,0)}_{44}\right]_{(rs)={\color{purple} (24)},(33)}, \notag\\
\left[  {\bf N}^{(1,3)}_{35}\right]_{(rs)= (24),{\color{purple} (33)}}\,\qquad;\qquad
\left[  {\bf N}^{(1,3)}_{44}\right]_{(rs)={\color{purple} (24)},(33)},
\end{align}
with indices $(rs)$ running over the rectangle at $\tau=6$ in $[020]$, i.e.~$\{(24),(33)\}$. So even though the 
{\color{purple}purple} colored coefficients come from correlators degenerate at tree level, and which are therefore proportional to each other, the remaining data is genuinely distinct.  Thus, after matrix multiplication 
(see \eqref{general_OPE_W} and \eqref{general_OPE_BW}), the one-loop result will distinguish 
these correlators. The general statement will be that one-loop OPE predictions 
in the window lift the tree-level degeneracy of correlators. This was first noticed in supergravity in \cite{Aprile:2019rep}, and the same mechanism is at work here.

%===========================================================================================================
%===========================================================================================================

 \section{The one loop Mellin amplitude $\mathcal{M}^{(2,3)}$}
 \label{one_loopM23_sec}

 In this section we translate the structure of the one-loop OPE data into the Mellin space amplitude. 
We claim that 
\begin{align}
\label{M13structure_2}
\mathcal{M}^{(2,3)}_{\vec{p}} = \Bigl[ \mathcal{W}^{(AW)}_{\vec{p}}(\hat{s},\check{s}) + \mathcal{R}^{(W)}_{\vec{p}}(\hat{s},\check{s})+ \mathcal{B}^{(BW)}_{\vec{p}}(\hat{s},\check{s}) \Bigr] + {\tt crossing }\,.
\end{align}
i.e.~$\mathcal{M}^{(2,3)}_{\vec{p}}$ naturally splits into three pieces, described below.

Our starting point will be to use the data from the OPE, at tree level Above Window, to completely fix $ \mathcal{W}^{(AW)}$. 
We will find that 
\be
\mathcal{W}^{(AW)}_{\vec{p}}(\hat s, \check s)  =  \tilde{\psi}^{(0)}(-{\bf s})\ \mathpzc{w}_{}(\hat{s},\check{s},c_s;\Sigma)
\ee
where $\mathpzc{w}$ is a determined polynomial, and $\tilde{\psi}^{(0)}=\psi^{(0)} + \gamma_E$ 
is a digamma function. 
The latter accounts for the fact that ${\cal W}^{AW}$ has to contribute to triple poles 
in order to generate a $\log^2 U$ discontinuity,  and the ${\bf \Gamma}$ factor only gives 
at most double poles Above Window.\footnote{Note the formula 
$\Gamma[-{\hat s}+\frac{p_1+p_2}{2}]\Gamma[-{\hat s}+\frac{p_3+p_4}{2}]\psi(-{\bf s}){\cal W}^{(AW)} U^{\hat s} \rightarrow \tfrac{1}{2} \left[  \partial^2_{\hat s} \frac{1}{(\hat{s}-\mathfrak{n})}\right]{\cal W}^{(AW)} U^{\hat s}$.}
The restricted dependence of $\mathpzc{w}$ 
on the Mellin variables comes from the fact that the OPE  has support only on $a=l=0$, 
and therefore  $\mathpzc{w}$ is function of {\tt s}-type variables only, thus $\mathpzc{w}(\hat{s},\check{s},c_s;\Sigma)$.

Note: The argument of $\tilde{\psi}^{(0)}$ being $-{\bf s}$ implies that contributions from $\mathcal{W}^{(AW)}$ 
are not restricted to Above Window,  but actually start at ${\bf s}=0$. As we explained in the previous section, 
this is one unit above the unitarity bound ${\bf s}+1=0$, therefore $\mathcal{W}_{}^{(AW)}$ contributes 
also to the $\log^1U$ and $\log^0U$ discontinuities, respectively, in the Window and Below 
Window. This fact will play an important role as observed in \cite{Drummond:2020uni}, and explained below.

Next, we will use window OPE data, and also contributions coming from 
$\mathcal{W}^{(AW)}_{}$, to fix the remainder function in the Window
\be
\label{window_splitting1}
\mathcal{R}^{(W)}_{\vec{p}}(\hat{s},\check{s})= \sum_{z= 0}^6 
\frac{(\check{s}+\frac{p_{1}+p_2}{2} -z)_{z+1}\ \mathpzc{r}^{+}_{\vec{p};z}(\hat{s},\check{s})  +(\check{s}+\frac{p_{3}+p_4}{2}-z)_{z+1}\ \mathpzc{r}^{-}_{\vec{p};z}(\hat{s},\check{s})}{{\bf s} -z} \,,
\ee
with $\mathpzc{r}^{\pm}$ are polynomials. We shall call ${\cal R}$ a remainder function since 
it gives the part of the OPE Window predictions not captured by ${\cal W}^{(AW)}$.  Remarkably, the interplay 
with $\mathcal{W}^{(AW)}_{}$ will truncate the sum over $z$  to a maximum of seven poles! It will be clear 
that this function should also be extended to contribute to the Below Window region.

The determination of the function $\mathcal{R}^{(W)}_{}$ is a central result of our investigation, since 
it characterises the way the {\it window splitting} described in the section \ref{window_split_sec} 
is implemented in Mellin space. We called this phenomenon "sphere splitting", and it will be discussed in more detail in section \ref{window_dynamics}.

Finally, using Below Window data, as well as contributions  now coming from both 
$\mathcal{W}^{(AW)}(\hat s, \check s)$ and $\mathcal{R}^{(W)}_{}(\hat s, \check s)$, 
we fix the last piece of our ansatz\footnote{The division between $\mathcal{R}^{(W)}_{}$ and $\mathcal{B}^{(BW)}_{}$ is our choice.  
We do so because the explicit solution for ${\cal R}^{(W)}$ will have nice analytic properties Below Window, see section \ref{window_dynamics}.} 
\be
\label{window_splitting1}
\mathcal{B}^{(BW)}_{\vec{p}}(\hat{s},\check{s})=\sum_{z= 0}^2 \frac{(\check{s}+\frac{p_{1}+p_2}{2}-z)_{z+1} (\check{s}+\frac{p_{3}+p_4}{2}-z)_{z+1}}{{\bf s} -z} \mathpzc{b}^{}_{\vec{p};z }(\hat{s},\check{s}) \,.
\ee
This is also a remainder function, capturing the predictions of the OPE in the Below Window region after contributions from both $\mathcal{W}^{(AW)}(\hat s, \check s)$ and $\mathcal{R}^{(W)}_{}(\hat s, \check s)$ are taken into account. We find that $\mathcal{B}^{(BW)}({\hat s},{\check s})$ also truncates, this time to just three poles in $z$. Again $\mathpzc{b}^{}_{\vec{p};z}(\hat{s},\check{s})$ is polynomial.

%===========================================================================================================
%===========================================================================================================

\subsection{Double log discontinuity}
\label{above_window_sec}

The $\log^2 U$ terms at order $\lambda^{-\frac{3}{2}}$ have a very simple form \cite{Drummond:2020uni}. 
The reason behind this is that only operators with $\ell_{10}=0$ receive an order $\lambda^{-\frac{3}{2}}$ 
contribution to their anomalous dimensions. Therefore, in the following expression for ${\bf M}_{\vec{\tau}}^{(2,3)}$,
\be\label{double_M23}
\bigl[{\bf M}_{\vec{\tau}}^{(2,3)}\bigr]_{(p_1p_2),(p_3p_4)}  = \sum_{{\cal K}\in R_{\vec{\tau}}}  {C}^{(0,0)}_{p_1p_2 \mathcal{K}} \, \eta^{(1,0)}_{\mathcal{K}} \eta^{(1,3)}_{\mathcal{K}} C^{(0,0)}_{p_3p_4\mathcal{K}}\,.
\ee
only a single operator $\mathcal{K}$ contributes to the sum for a given $\tau$ and $b$! In Figure \ref{degeneracies} 
this is the operator labelled by the leftmost corner of the rectangle. If we now insert this expression into the sum over 
long superconformal blocks to obtain the expansion of the $\log^2 U$  discontinuity, we find an expression 
that is almost identical to the expansion of the $\log U$ discontinuity at tree level,\footnote{i.e.~$\sum_{{\cal K}\in R_{\vec{\tau}}}  
{C}^{(0,0)}_{(p_1p_2) \mathcal{K}} \eta^{(1,3)}_{\mathcal{K}} C^{(0,0)}_{p_3p_4\mathcal{K}}$.} but for the insertion 
of a factor of $\eta^{(1,0)}$ restricted to $\ell_{10}=0$. By construction this factor is a number (which includes the 
value of the denominator of $\eta^{(1,0)}$) multiplying the eigenvalue of a certain eight-order Casimir 
operator $\Delta^{(8)}$ introduced in \cite{Caron-Huot:2018kta} in position space, see in particular \cite{Drummond:2020uni}. %also  \cite{Aprile:2019rep}. 

In Mellin space, the $\log^2 U$ contribution comes  from  ${\bf \Gamma}\times\tilde{\psi}^{(0)}(-{\bf s})$, 
which is the source of all triple poles. The knowledge of the $\log^2 U$ coefficient
will therefore fully fix 
\be
\mathcal{W}^{(AW)}_{\vec{p}}(\hat s, \check s)  =  \tilde{\psi}^{(0)}(-{\bf s})\ \mathpzc{w}_{}(\hat{s},\check{s},c_s;\Sigma).
\ee
From the discussion about the special form of \eqref{double_M23}  we infer that, apart for an overall prefactor, the polynomial
$\mathpzc{w}$  is given by applying $\Delta^{(8)}$ (rewritten in Mellin space) to $1$, where the latter is (up 
to a numerical factor) the tree level amplitude in the Virasoro Shapiro amplitude at order $(\alpha')^3$.

We have written the full expression of $\mathpzc{w}_{}(\hat{s},\check{s},c_s;\Sigma)$ 
in the {\color{red} \tt ancillary file}. Expanded in all variables it has the form
\beq
\mathpzc{w}_{}(\hat{s},\check{s},c_s;\Sigma)={+} \frac{(\Sigma-1)_3}{180}\Big( -36 c_s^2+9 c_s^4+ 36 c_s^2 {\check s} \,+\, \ldots \,+\, 8 {\check s} {\hat s}^3 \Sigma^4 + 2 {\hat s}^4 \Sigma^4\Big).
\eeq
A more compact representation  can be given by the following double integral, 
\be\label{integral_tr_James}
\mathpzc{w}_{}(\hat{s},\check{s}) = \frac{i}{2 \pi} \int_0^\infty  \frac{d\alpha}{\alpha} \int_\mathcal{C} d\beta\ e^{-\alpha-\beta} \alpha^{\Sigma+2}  (-\beta)^{-\Sigma+1} \, \widetilde{\mathpzc{w}}(\alpha,\beta)
\ee
where $\mathcal{C}$ is the Hankel contour. 
The $\alpha$ integral is the $\Gamma$ function integral, and the $\beta$ integral the reciprocal 
$\Gamma$ function integral. Note that  the integral in $\alpha$ is nothing but the integral 
used by Penedones to study flat space limit \cite{Penedones:2010ue}, 
and the $\beta$-integral generalises that to the compact space.

Then, $\widetilde{\mathpzc{w}}_{\vec{p}}(\alpha,\beta)$ is defined in terms of the following variables,
\be
S = \alpha \hat{s} - \beta \check{s},\qquad \tilde{S} = \alpha \hat{s} + \beta \check{s},
\ee
by the expression
 \begin{align}\label{Ppoly}
&  \widetilde{\mathpzc{w}}(\alpha,\beta,c_s;\Sigma) =\\
& \qquad\frac{1}{90} \left( S-3\Sigma \right)\left( S-2\Sigma \right)\left( S-\Sigma \right)S+\tfrac{\Sigma^2-c_s^2}{20}\bigl( 2S^2-6S \Sigma+5\Sigma^2 - c_s^2 \bigr) +\notag\\
%+2(p_{(12)}+p_{(34)}+3 p_{(12)}p_{(34)}) \bigr)+\\
&\qquad- \frac{1}{30}\tilde{S} \bigl(2S^2-9S\Sigma+11 \Sigma^2-3 c_s^2 \bigr)+ \frac{1}{180} \bigl(S^2-36 S\Sigma+36 (\Sigma^2-c_s^2)  +7\tilde{S}^2 \bigr).\notag
 \end{align}
% {\color{red} REWRITE with $c_s$.}
 It is immediate to see that only $c_s^{2\mathbb{N}}$ appear. 
In fact, since $\mathpzc{w}_{}$ is a function of ${\tt s}$ variables only, the crossing relations,
\beq
\begin{array}{ccc}
{\cal M}_{p_1p_2p_3p_4}(\hat s,\hat t, \check s,\check t) &
= &{\cal M}_{p_4,p_3,p_2,p_1}(\hat s, \hat t, \check s,\check t)\\[.2cm]
\begin{tikzpicture} \node[rotate=90] (0,0) {$=$};\end{tikzpicture} & & \begin{tikzpicture} \node[rotate=90] (0,0) {$=$};\end{tikzpicture}  \\
 {\cal M}_{c_s, c_t,c_u}(\hat s, \hat t, \check s,\check t) & &
 {\cal M}_{-c_s, c_t,-c_u}(\hat s, \hat t, \check s,\check t)
\end{array}
\eeq
implies that $\mathpzc{w}_{}$  is function of $c_s^2$, i.e.~since there is no $c_u$ dependence
the invariance under $c_s\leftrightarrow -c_s$ follows. 
In particular, as for ${\bf \Gamma}$ factor, the polynomial
 $\mathpzc{w}_{}(\hat{s},\check{s},c_s;\Sigma)$ has the same bonus property.

Let us emphasise that the OPE does not immediately predict $ \tilde{\psi}^{(0)}(-{\bf s})$. 
The Above Window region only requires  $\Gamma[\tfrac{p_{1}+p_2}{2}-\hat{s}]\Gamma[\tfrac{p_{3}+p_4}{2}-\hat{s}] \tilde{\psi}^{(0)}(-\hat{s}+p_{\text{max}})$, 
since  $\hat{s} \geq p_{\text{max}}$  is where the triple poles are.  However, the presence of $ \tilde{\psi}^{(0)}(-{\bf s})$ 
is strongly motivated by the limit in which the charges $p_i$ are large, and therefore ${\bf s}$ 
is large \cite{Aprile:2020luw}. As explained already in the Introduction, the key observation is 
that in this limit ${\bf s}$ becomes a 10d Mandelstam invariant and the  Mellin amplitude 
asymptotes the 10d flat space scattering amplitude. In the present case we must recover 
the $\log(-{\bf s})$ of the type IIB flat space amplitude, and the latter comes from $ \tilde{\psi}^{(0)}(-{\bf s})$ 
in the limit of large ${\bf s}$. We will later show in section \ref{sec_largeflatlimit} that in the 
large $p$ limit we recover exactly the type-IIB flat space amplitude!

It follows from the presence of $\tilde{\psi}^{(0)}(-{\bf s})$ that the range of twists where ${\cal W}^{(AW)}$ contributes 
is not restricted to the Above Window region, $\hat{s} \geq p_{\text{max}}$, where  ${\cal W}^{(AW)}$  naturally lives, 
but embraces the bigger region  bounded from below by  the locus ${\bf s}=0$. We now understand 
that ${\cal W}^{(AW)}$ contributes to the one-loop correlator starting from the first two-particle operator 
in the Below Window region at $\tau=2a+b+4$. Consequently ${\cal W}^{(AW)}$ gives contributions that will 
add to those of ${\cal R}_{}^{(W)}$ function and ${\cal B}_{}^{(BW)}$ in- and below- Window, respectively.
The choice to write the Above Window contribution as described above has the consequence of revealing
the truncation of sum over $z$ in the Window contribution to $z=0,\ldots,6$, rather than a range growing with the charges.

\subsubsection{Flat space relation of the $AdS_5\times S^5$ amplitude}

The fact that the polynomial $\mathpzc{w}_{}(\hat{s},\check{s},c_s;\Sigma)$ can be written in terms of 
a pre-polynomial (\ref{Ppoly}), as in the case of the Virasoro-Shapiro amplitude studied in \cite{Aprile:2020mus}, 
suggests the following little game:  In flat space the part of the $(\alpha')^3$ amplitude at one loop 
which accompanies $\log(-s)$ is $s^4$, and the tree-level $(\alpha')^7$ amplitude is proportional to $s^4+t^4+u^4$. 
Therefore, if we assemble
\begin{equation}\label{curiosity_Michele}
\mathcal{M}_{7} \equiv \mathpzc{w}(\hat{s},\check{s},c_s;\Sigma)+\mathpzc{w}(\hat{t},\check{t},c_t;\Sigma)+\mathpzc{w}(\hat{u},\check{u},c_u;\Sigma)
\end{equation}
we expect by construction that this quantity has the correct 10d flat space limit. Is there something more? Upon inspection,   
it turns out that $\mathcal{M}_{7}$ so constructed is actually  the tree-level $(\alpha')^7$ amplitude constructed in \cite{Aprile:2020mus,Abl:2020dbx},
up to an overall coefficient and a certain choice of ambiguities!  This observation stands at the moment as a curiosity, though quite intriguing.
It would be interesting to understand its origin further, and whether or not it generalises beyond this case. We will leave this for a future investigation.

%===========================================================================================================
%===========================================================================================================
\subsection{Window dynamics}
\label{window_dynamics} 

The $\log^1 U$ projection of the $(\alpha')^3$ correlator in the Window has the following one-loop OPE expansion, for fixed quantum numbers $\vec{\tau}$,\\ % the following general form
\begin{align}
\label{win_ope}
%H^{(2,3)}_{p_1p_2p_3p_4;\vec{\tau}}  = & 
%\underbrace{
\sum_{{\cal K} \in \mathcal{R}_{\vec{\tau}}} C_{p_1p_2,\mathcal{K}_{}}^{(0,0)}\left( 
\eta^{(1,3)}_{\mathcal{K}_{}} C_{p_3p_4,\mathcal{K}_{}}^{(1,0)} + \eta^{(1,0)}_{\mathcal{K}_{}} C_{p_3p_4,\mathcal{K}_{}}^{(1,3)}  \right) +
%}_{p_{1}+p_2<p_3+p_4} 
%\\
%+ &% \underbrace{
\sum_{ {\cal K} \in \mathcal{R}_{\vec{\tau}} } C_{p_3p_4,\mathcal{K}_{}}^{(0,0)}\left( \eta^{(1,3)}_{\mathcal{K}_{}} C_{p_1p_2,\mathcal{K}_{}}^{(1,0)} + \eta^{(1,0)}_{\mathcal{K}_{}} C_{p_1p_2,\mathcal{K}_{}}^{(1,3)}  \right)%\label{line_due}
%}_{p_{1}+p_2>p_{3}+p_4}
\end{align}
Recall that there is no Window when $p_1+p_2=p_3+p_4$, see section \ref{sec_Mellin_numbers}.

Formula \eqref{win_ope} contains two terms symmetric under $(p_1,p_2)\leftrightarrow (p_3,p_4)$. 
However, the two terms never contribute together. 
This is because the (free theory value of the) twist of the two-particle operator ${\cal K}$ 
is greater equal than $\max(p_1+p_2,p_3+p_4)$, and therefore in the Window either 
$C_{p_1p_2,\mathcal{K}}^{(0,0)}=0$ or $C_{p_3p_4,\mathcal{K}}^{(0,0)}=0$. 
%
%The underbrace notation used in \eqref{win_ope} wants to highlights  %when each term is non zero. 
%
It follows that when we compute the OPE predictions we only have access to one of the two terms at once. 
%However, both \eqref{win_ope} still have a chance of existing together if certain analytic properties of the OPE coefficients are satisfied. 
Nevertheless, we do expect a final formula for the coefficients which is at the same time symmetric 
under $(p_1,p_2)\leftrightarrow (p_3,p_4)$ and also analytic in the charges. We can imagine several scenarios 
of which the simplest one is perhaps the one where a single contribution, 
say for concreteness the one for $p_{1}+p_2<p_3+p_4$, is such that upon a natural analytic continuation 
in the charges, it automatically vanishes when $p_{1}+p_2>p_{3}+p_4$, and vice-versa. The final symmetric result will 
then be the sum of the two contributions. This simple scenario is indeed the one realised by the amplitude.

In the analysis that follows it will prove useful to move between a Mellin space formula of the type,
 \be
 \label{mellin_reminder}
\mathcal{H}_{\vec{p}}(U,V,\tilde{U},\tilde{V})=\int\!\!d\hat{s} d \hat{t} \ \sum_{\check{s}, \check{t}} U^{\hat s} V^{\hat t} {\tilde U}^{\check s} {\tilde V}^{\check t}\,\, {\bf \Gamma}\ \mathcal{M}_{\vec{p}}(\check s, \check t),
\ee
and a formula where we perform the sum over $\check s$ and $\check t$ and decompose into the basis of spherical harmonics $Y_{[aba]}$. 
We can start in the monomial basis, with an expression of the form 
 \be
\int \!\!d\hat{s} d \hat{t} \ U^{\hat s} V^{\hat t} \  \Gamma_{\hat s}\Gamma_{\hat t}\Gamma_{\hat u}\ 
\sum_{\check{s}, \check{t}} \frac{{\tilde U}^{\check s} {\tilde V}^{\check t}}{ \Gamma_{\check s}\Gamma_{\check t}\Gamma_{\check u} }\,\, \mathcal{F}^{}_{\vec{p}}(\hat s,\check s),
\ee
valid for ${\cal F}=\{\mathcal{W}^{(AW)},\mathcal{R}^{(W)},\mathcal{B}^{(BW)}\}$, i.e.~each  
one of the three functions that builds up our amplitude. The sum over $\check{s}$ and $\check{t}$ is finite, 
as we explained already. Then, upon decomposing into spherical harmonics we can alternatively write,
\be
\int\!\!d\hat{s} d \hat{t} \ U^{\hat s} V^{\hat t} \   \Gamma_{\hat s}\Gamma_{\hat t}\Gamma_{\hat u}\  \sum_{[0b0]} Y_{[0b0]}\,\, \mathcal{F}^{}_{\vec{p}}(\hat s,b).
\ee
The notation $\mathcal{F}^{}_{}(\hat s,\check s)$ and $\mathcal{F}^{}_{}(\hat s,b)$ will 
then refer to $\mathcal{F}$ as being written in the monomial basis and spherical harmonic basis, respectively. 

As an example, we reproduce an interesting formula for the Above Window function
\begin{align}
    \mathpzc{w}_{\vec{p}}(\hat s, b)=& \frac{ (\Sigma-1)_3 }{ {\Gamma}_{[0b0]} } \times \frac{1 }{15\times 3!} \sum _{i=0}^4 (-1)^i \binom{4}{i} ({\hat s} (\hat{s}+2-i)- \tfrac{b}{2}  (\tfrac{b}{2} +2)) \notag \\
    &  \times \left({\hat s}-\frac{b+2}{2}\right)_{3-i} \left({\hat s}+\frac{b+2}{2}\right)_{3-i} \left(\frac{ p_{1}+p_2}{2}-{\hat s}\right)_i \left(\frac{ p_{3}+p_4}{2}-{\hat s}\right)_i %Y_{[aba]}.
    \label{eq:gen_AW}
\end{align}
where the analogous factor of $\Gamma_{[aba]}$ function was found in \cite{Drummond:2020dwr}.\footnote{
\be\nn
\!\!\!\!\!\!\!\!\!\!\!\!\!\!\!\!\!\Gamma_{[aba]}^{-1}=
\frac{(\Sigma-2)! b!(b+1)!(2+a+b)  }{  \Gamma[ \pm \frac{p_1-p_2}{2} + \frac{b+2}{2}]  
\Gamma[\frac{p_1+p_2}{2} + \frac{b+2}{2}]   \Gamma[\frac{p_1+p_2}{2} - \frac{b+2a+2}{2}]  
\Gamma[ \pm \frac{p_3-p_4}{2} + \frac{b+2}{2}] \Gamma[\frac{p_3+p_4}{2} + \frac{b+2}{2} ] \Gamma[\frac{p_3+p_4}{2} - \frac{b+2a+2}{2} ] }\notag
\ee} 
%We can add to this observation the fact, noticed already  in \eqref{curiosity_Michele}, that we can
%use the one-loop $\log^2 U$ discontinuity, therefore \eqref{eq:gen_AW} as well, to generate the tree level amplitude at $(\alpha')^7$ up to a choice of 
%ambiguities.\footnote{Inspired by this, we saw \cite{unpublished} that with minor modifications of 
%\eqref{eq:gen_AW}, we can generate the tree level amplitudes $(\alpha')^{n=5,6,7,8}$ of the AdS Virasoro-Shapiro amplitude in \cite{Aprile:2020mus}.}

With either representation, i.e.~$\mathcal{F}^{}_{}(\hat s,\check s)$ and $\mathcal{F}^{}_{}(\hat s,b)$, 
we can perform the $\hat s$ and $\hat t$ integration, to arrive at an expression in position space which 
we can then decompose into conformal blocks and match against OPE predictions.

%===========================================================================================================
\subsubsection{Spherical harmonic basis}
%===========================================================================================================

%By matching the OPE predictions w
We wish to compute first $\mathcal{R}^{W}_{\vec{p}}(\hat s,b)$ in the spherical harmonic basis.
This is derived from matching the $\log^1U$ projection of the amplitude, i.e.~
\be
\label{window_start}
\int\! d\hat{s} d \hat{t} \ U^{\hat s} V^{\hat t}\ \Gamma_{\hat s}\Gamma_{\hat t}\Gamma_{\hat u}\ \left(\sum_{b} Y_{[0b0]} \left[ \mathcal{W}^{(AW)}_{\vec{p}}(\hat s,b)+\mathcal{R}^{(W)}_{\vec{p}}(\hat s,b) \right] + {\tt crossing}\right)\Bigg|_{\log^1 U}
\ee
with the prediction from the OPE in the Window.
%where $\mathcal{W}^{(AW)}(\hat s,b)$ is also written in the spherical harmonics basis. 
Schematically, we expect
\beq
\mathcal{R}^{(W)}_{\vec{p}}(\hat s,b)= \sum_{n} \frac{\#}{{\hat s}-n} \qquad;\qquad n\ge \min(\tfrac{p_1+p_2}{2} ,\tfrac{p_3+p_4}{2} ).
\eeq 
The task will be to find the residues $\#$ for the various values of the twist in the Window Region, here labelled by values of $\tau=2n$.  
In \eqref{window_start} we have already restricted the summation to $a=0$, as a consequence of the truncation to spin zero valid at order $(\alpha')^3$. 

In practice, say we focus on the ${\tt s}$ channel, we will pick double poles in $\hat s$ in the Window 
(this will select the $\log^1 U$ contribution), perform the $\hat t$ integral, and obtain a 
function in position space. This function contains up to $\log^1 V$ and $\log^0V$ contributions
(since there is no $\tilde \psi(-{\bf t})$ in the ${\tt s}$ channel). Note that for given power of $U$, both $\log^1 V$ 
and $\log^0 V$ come with a non trivial rational function of $V$. These contributions are analytic in the small 
$x,\bar{x}$ expansion, which is the expansion of the blocks we want to match.\footnote{It is actually convenient to 
resum the OPE predictions to exhibit the $\log V$ contribution explicitly, then match. This type of resummation was called a one-variable resummation in \cite{Aprile:2019rep} see section 4.3.}

As in the case of $\langle \mathcal{O}_2 \mathcal{O}_2 \mathcal{O}_p \mathcal{O}_p \rangle$ considered 
in \cite{Drummond:2020uni}, we find that {\it only five poles} are necessary to fit the OPE predictions .\footnote{In 
$\langle ppqq\rangle$, fix the $su(4)$ channel to start with, and look at the residue of the first pole as function of 
$p$ and $q$. The range of $p,q$ is infinite and this gives a $p,q$ dependent polynomial in the numerator and three $\Gamma$ functions in the denominator. 
We then vary the $su(4)$ channel, introducing $b$. By studying in the same way the second pole, 
the third pole, etc\ldots, we recognise $\Gamma[\frac{p_{3}+p_4}{2}-n]\Gamma[n-\frac{b+2a+2}{2}]\Gamma[n+\frac{b+2}{2}]$.
Thus, even though we can access five values of the twist, i.e.~five poles, 
we can single out $\Gamma[5+\frac{p_{1}+p_2}{2}-n]$ from looking at $B^3$ where $B:=\tfrac{b(b+4)}{4}$.}  
In the sector $p_1+p_2\leq p_3+p_4$, we find
\beq\label{remainder_func}
\mathcal{R}^{-}_{\vec{p}}(\hat s,b)=  \sum_{n\ge \frac{p_1+p_2}{2} }^{}\frac{\gamma_{\vec{p}}(n,b)}{\Gamma[5-n+\frac{p_1+p_2}{2} ] \Gamma[\frac{p_{3}+p_4}{2}-n]}\frac{{\tt R}^-_{\vec{p}}(n,b)}{ (\hat s-n) } ,
\eeq
where the minus superscript stands for $p_1+p_2\leq p_3+p_4$, then $n=\frac{p_1+p_2}{2},\frac{p_1+p_2}{2}+1,\ldots$ runs over half-twists, and
\beq\label{remainder_func_solu1}
\gamma_{\vec{p}}(n,b)=\frac{ b! (b+1)!(b+2) }{\Gamma\left[  \pm \frac{p_{12}}{2} +\frac{b+2}{2} \right] \Gamma\left[ \pm \frac{p_{43}}{2}+\frac{b+2}{2}  \right] }
\frac{ \left(\frac{p_1+p_2}{2}-\frac{b+2}{2}\right)_3 \left(\frac{p_1+p_2}{2}+\frac{b+2}{2}\right)_3 }{
\Gamma\left[n -\frac{b+2}{2}\right]\Gamma\left[n+\frac{b+2}{2} \right]} 
\eeq
and
\beq\label{remainder_func_solu2}
\!\!\!\!\!\!\begin{array}{l}
{\tt R}^-_{\vec{p}}(n,b)= 
 \frac{B^3}{15} + \frac{(\Sigma^2+14\Sigma-(c_s)^2-10c_s+28-2(c_{tu})^2)B^2}{60} + \ldots  \\[.2cm]
\rule{1cm}{0pt{}}-2n \left[ \frac{ (\Sigma+3)B^2}{15}+\frac{(\Sigma^2+7\Sigma -(c_s)^2 -c_s+12-(c_{tu})^2(\Sigma+3))B}{30}+\ldots \right] \qquad ;\qquad B:=\tfrac{b(b+4)}{4}.
\end{array}
\eeq
The full expression for ${\tt R}^-_{}$ can be found in the  {\color{red} {\tt ancillary file}}, it is made of simple polynomials. 
Note that above we have used the notation $c_{tu} =c_t+c_u$.  %is a peculiar property of the spin truncation at $(\alpha')^3$. 

The fact that only five poles in $\mathcal{R}^-_{}$ are needed to fit the OPE data is reflected by the factor 
$1\big/\Gamma[5+\tfrac{p_{1}+p_2}{2}-n]$, which automatically truncates when $n\ge \frac{p_1+p_2}{2}+5$.
It implies that beyond the fifth pole, the OPE predictions are fully captured already by the function ${\cal W}^{AW}$! This is quite remarkable given that we are evaluating ${\cal W}^{AW}$ in the Window. 
Let us also emphasise that ${\tt Res}_{\vec{p}}$ turns out to 
be only a linear polynomial in $n$. Considering that we are fitting five poles, this is a non-trivial consistency check of our formula. 
A closely related formula holds for ${\cal R}^+_{}$ in the sector $p_1+p_2\ge p_3+p_4$. Let us see why:

Above we considered the case $p_1+p_2< p_3+p_4$ and found $\mathcal{R}^{-}$, but note that it automatically 
vanishes when $p_1+p_2> p_3+p_4$. This is because of $1\big/\Gamma[\frac{p_{3}+p_4}{2}-n]$ and the 
fact that $n\ge \frac{p_1+p_2}{2}$ in the sum. Therefore, we are free to add both contributions in one formula 
and write the following symmetric and analytic expression
\begin{align}
\mathcal{R}^{(W)}_{\vec{p}}(\hat s,b)=\mathcal{R}^{-}_{\vec{p}}({\hat s},b)\,+&\ \mathcal{R}^{+}_{\vec{p}}({\hat s},b) %\\[.2cm]
\qquad;\qquad \mathcal{R}^{+}_{p_1p_2p_3p_4}({\hat s},b)=\mathcal{R}^{-}_{p_4p_3p_2p_1}({\hat s},b)
\end{align}
where ${\cal R}^+$ is related to ${\cal R}^-$ by swapping charges in the appropriate way. Using 
the $c_s,c_t,c_u$ parametrisation, $\mathcal{R}^+_{\{c_s,c_t,c_u\}}({\hat s},b)=\mathcal{R}^-_{\{-c_s,c_t,-c_u\}}({\hat s},b)$.

Let us now come back to the Window splitting mentioned in section \ref{window_split_sec}.
There, we explained that degenerate correlators at tree level are those correlators with the same values 
of $\Sigma$ and $|c_s|,|c_t|,|c_u|$ which therefore are proportional to each other  at order $\mathcal{M}^{(1,0)}$ 
and $\mathcal{M}^{(1,3)}$, because the bonus property is preserved.  We can see from the explicit expressions 
for ${\tt Res}_{\vec{p}}(n,b)$ that at one-loop this is not the case anymore. Coming back to our guiding example of $\vec{p}=(3335)$ and $\vec{p}=(4424)$, we can see that
\beq
{\tt R}^-_{\vec{p}=3335}(n=3,b=2)=-\frac{336}{5}\qquad;\qquad{\tt R}^+_{\vec{p}=4424}(n=3,b=2)=-\frac{348}{5}
\eeq
where the LHS is simply the evaluation of \eqref{remainder_func_solu2}, while the RHS is obtained from the \eqref{remainder_func_solu2} 
upon $c_s\rightarrow-c_s$ and $c_u\rightarrow -c_u$.  As promised, the one-loop OPE distinguishes these two correlators 
in the Window. Note instead that if we project the correlators onto the $\log^2 U$ discontinuity 
Above Window, the corresponding contributions are still degenerate.

%\begin{mdframed}
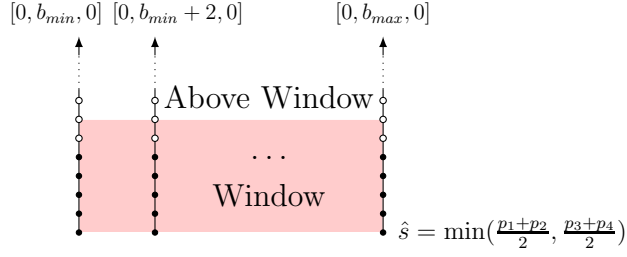
\begin{figure}[t]
\begin{center}
\begin{tikzpicture}

\filldraw[fill=red!20,draw=white] (0,1.5)  rectangle (4,0);
%\draw (4,0) -- (4.5,0);  
\draw (5.7,0) node[scale=.8] {$\hat{s}={\rm min}(\tfrac{p_{1}+p_2}{2},\tfrac{p_{3}+p_4}{2})$};
%\filldraw[fill=red!20]     (0,1.1)  rectangle (4,1.8);

\draw[-]  (0,0) --(0,1.9);   \draw[dotted]  (0,1.9) --(0,2.4);  \draw[-latex]  (0,2.35) --(0,2.6);
\draw[-]  (1,0) --(1,1.9); \draw[dotted]  (1,1.9) --(1,2.4);  \draw[-latex]  (1,2.35) --(1,2.6);

					  		 \draw (-.3,.5+2.4) node[scale=.7] {${[0,b_{min},0]}$};
							 \draw (1.3,.5+2.4) node[scale=.7] {${[0,b_{min}+2,0]}$};
							  \draw (2.5,1) node[scale=1] {$\ldots$};
							  \draw (2.5,.5) node[scale=1] {Window};
							  \draw (2.5,1.8) node[scale=1] {Above Window};
\draw[-]  (4,0) --(4,1.9);  		  \draw (4,.5+2.4) node[scale=.7] {${[0,b_{max},0]}$};
\draw[dotted]  (4,1.9) --(4,2.4);  \draw[-latex]  (4,2.35) --(4,2.6);

\foreach \x in {0,1,4}
\foreach \y in {0,.25,.5,.75,1}
    \draw[fill=black] (\x,\y) circle (1pt);
    
\foreach \x in {0,1,4}  
\foreach \y in {1.25,1.5,1.75}
    \draw[fill=white] (\x,\y) circle (1.2pt);

\end{tikzpicture}
\end{center}
\caption{Pole structure of the remainder function in the basis of spherical harmonics. Black dots indicate a 
non-zero residue. Un-filled dots only receive contributions from $\mathcal{W}_{}^{(AW)}$.}
\label{fig:win_harmonics_poles}
\end{figure}
%\end{mdframed}

\begin{figure}[t]
\begin{center}
\begin{tikzpicture}
\filldraw[fill=red!20,draw=white] (0,1.5)  rectangle (4,0);

\filldraw[ green!20] (0,-.25)  --  (0,-1.25)  -- (.5,-1.25) -- (.5,-1) -- (1.5,-1)  -- (1.5,-.75) --  (3.5,-.35) -- (3.5,-.25)  -- cycle;
%\draw (4,0) -- (4.5,0);  
%\draw (5.05,+0.3) node[scale=.8,color=blue!90] {${\bf s}-1=0$};
\draw (6.4,0.05) node[scale=.8] {$\phantom{spacing}$};
\draw (4.8,0.05) node[scale=.8,color=blue] {${\bf s}=0$};
\draw (5.08,-0.35) node[scale=.8,] {${\bf s}+1=0$};
%\filldraw[fill=red!20]     (0,1.1)  rectangle (4,1.8);

\draw[-latex]  (0,0) --(0,.5+2);  		  \draw (-.3,.5+2.2) node[scale=.7] {${[0,b_{min},0]}$};
\draw[-latex]  (1,0) --(1,.5+2); 		  \draw (1.3,.5+2.2) node[scale=.7] {${[0,b_{min}+2,0]}$};
							  \draw (2.5,1) node[scale=1.2] {$\ldots$};
							  \draw (2.5,-.4) node[scale=1.2] {$\ldots$};
							  \draw (2.5,.5) node[scale=1] {Window};
							  \draw (2.5,1.8) node[scale=1] {Above Window};
\draw[-latex]  (4,0) --(4,.5+2);  		  \draw (4,.5+2.2) node[scale=.7] {${[0,b_{max},0]}$};

\foreach \x in {0,1,4}
\foreach \y in {,0,.25,.5,.75,1}
    \draw[fill=black] (\x,\y) circle (1pt);

\foreach \y in {-1,-.75,-.5,-.25,0,.25,.5,.75,1}
    \draw[fill=black] (0,\y) circle (1pt); 
    
     \filldraw[blue] (0,-1.25) circle (1pt); 
     \draw[] (0,-1.5) circle (1.2pt);  

\foreach \y in {-.75,-.5,-.25,0,.25,.5,.75,1}
    \draw[fill=black] (1,\y) circle (1pt);  
    
    \filldraw[blue] (1,-1.) circle (1pt); 
     \draw[] (1,-1.25) circle (1.2pt);  
     
     	 \filldraw[blue] (4,0) circle (1pt);  
          \draw[] (4,-.25) circle (1.2pt);

\end{tikzpicture}
\end{center}
\caption{Pole structure of the remainder function after continuation Below Window.}
\label{fig:win_poles_cont}
\end{figure}
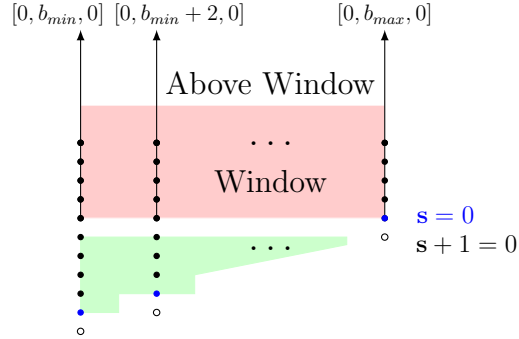

Figure \ref{fig:win_harmonics_poles} illustrates the general structure of poles of the remainder function, as we have obtained it from OPE data. 
Notice now that $\mathcal{R}^{(W)}_{}(\hat s,b)$ can be analytically continued Below Window. Quite nicely, 
the factor $1\big/\Gamma\left[n -\frac{b+2}{2}\right]$ ensures vanishing at the unitarity bound,  thus giving the correct 
physical behaviour not just in the Window, where the function was fitted, but also Below Window! 
We infer that the remainder function can be understood more properly to descend onto the range of twists above 
the unitarity bound, which means we are free to start the sum in \eqref{remainder_func} from $n \geq \frac{b+4}{2}$. 
At this point it is clear that the pole structure of ${\cal R}^{(W)}$ will be naturally labelled by bold font variables, 
in complete analogy with the way $\mathcal{W}^{(AW)}$ descends in- and below- Window. Figure \ref{fig:win_poles_cont} 
illustrates the poles from the latter viewpoint. 

To see more clearly the structure of poles in ${\bf s}$, from the Window down to the Below Window, 
let us point out that when we look at the $[0b_{max}0]$ channel, the locus ${\bf s}=0$ pinpoints the 
bottom of the window, below there is only the unitarity bound at ${\bf s}+1=0$. Lowering $b< b_{max}$, 
and looking at the $[0b0]$ channels, the same locus ${\bf s}=0$ enters the below window region depicted in green.

%===========================================================================================================
%===========================================================================================================
\subsubsection{Monomial basis, Crossing, and Sphere splitting}

In this section we will turn the expression $\mathcal{R}^{(W)}_{}(\hat s,b)$, written in the basis of harmonics $Y_{[0b0]}$,  
into the monomials basis $U^{\check s}V^{\check t}$, yielding the final Mellin amplitude corresponding to the Window region. The result right away is
\be\label{window_result_R}
\mathcal{R}^{(W)}_{\vec{p}}(\hat{s},\check{s})= \sum_{z=0}^6 \frac{1}{{\bf s} -z}\biggl[
(\check{s}+\tfrac{p_{1}+p_2}{2}-z)_{z+1} \ \mathpzc{r}^{+}_{\vec{p};z} 
(\hat{s},\check{s}) 
+
(\check{s}+\tfrac{p_{3}+p_4}{2}-z)_{z+1}\ \mathpzc{r}^{-}_{\vec{p};z}(\hat{s},\check{s})
\biggr]\,,%\,+\,\,{\tt crossing}
\ee
where the notation $\mathpzc{r}^{\pm}$ refers to the fact that $p_{1}+p_{2}=\Sigma+c_s$ and $p_{3}+p_4=\Sigma-c_s$. 
The polynomials $\mathpzc{r}^{\pm}$ are related to each other, 
\beq\label{solu_crossing}
\mathpzc{r}^+_{\vec{p}} = \mathpzc{r}_{\{-c_s,-c_t,c_u\}}
\qquad;\qquad 
\mathpzc{r}^-_{\vec{p}}  = \mathpzc{r}_{\{+c_s,+c_t,c_u\}},
\eeq
and given in the {\tt \color{red} ancillary file}.  There are seven of them,  since $z=0,\ldots,6$. The first few are are simple to write down, see the list in \eqref{example_rpolys}.

The result for $\mathcal{R}^{(W)}$ in \eqref{window_result_R} has two important properties, which are tied to the window splitting described in the previous section. 
These properties go together and are 1) the split of ${\bf \Gamma}$, and  2) the dependence of $\mathpzc{r}$ on the charges $\vec{p}$. 
Both these properties provide the way to distinguish between $p_{1}+p_2< p_{3}+p_4$ and $p_{1}+p_2> p_{3}+p_4$ at the level of the Mellin amplitude. 
All together this is the sphere splitting we mentioned in Section \ref{summary_sec}.

In more details: Assume $\hat s$ belongs to the Window, 
then $\hat{s}={\rm min}(\tfrac{p_{1}+p_2}{2},\tfrac{p_{3}+p_4}{2})+n$ for some positive integer, $n\ge 0$.
Now, from ${\bf s}-z=0$ we obtain the value of $\check{s}$, through the relation $-z+{\rm min}(\tfrac{p_{1}+p_2}{2},\tfrac{p_{3}+p_4}{2})+\check{s}=-n$. 
Considering that
\begin{align}
&\frac{ \mathcal{R}^{(W)}_{\vec{p}}(\hat{s},\check{s})}{\Gamma[1+\frac{p_{1}+p_2}{2}+\check s]\Gamma[1+\frac{p_{3}+p_{4}}{2}+\check s]}\rightarrow\notag\\
&
\sum_{z=0}^6\frac{1}{ {\bf s} -z}\Bigg[ \frac{\mathpzc{r}^+_{\vec{p};z}  }{ \Gamma[-z+\frac{p_{1}+p_{2}}{2}+\check{s}]\Gamma[1+\frac{p_{3}+p_{4}}{2}+\check{s}] }    \,  
+ \frac{\mathpzc{r}^{-}_{\vec{p};z}  }{ \Gamma[1+\frac{p_{1}+p_{2}}{2}+\check{s}]\Gamma[-z+\tfrac{p_{3}+p_{4}}{2}+\check{s}] } \Bigg]
\end{align}
we see that only one of the two terms contributes, because
\beq\label{argm_sphere_split}
\begin{array}{ccc}
\ \ \begin{tikzpicture}
\draw (.15,0) node[scale=.9]  {$p_{1}+p_2 > p_{3}+p_4$};

\draw (.1,-.75) node[scale=.9] {$-z+\tfrac{p_{3}+p_{4}}{2}+\check{s}=-n$};
%\draw (.1,-.75) node[scale=.9] {$-c_s\leq \tilde{s}\leq p_3-2$};
%\draw (+.3,-1.25) node[scale=.9] {$0\leq \tilde{s}+c_s\leq p_2-2+c_s$};

\draw (+.1,-1.4) node[scale=.9] {only\ \ $\mathpzc{r}^{+}$\ \ contributes};
\end{tikzpicture} 
& \rule{2cm}{0pt} &
\begin{tikzpicture}
\draw (0,0) node[scale=.9]  {$p_{1}+p_2 < p_{3}+p_4$};

\draw (+.05,-.75) node[scale=.9] {$-z+\tfrac{p_{1}+p_{2}}{2}+\check{s}=-n$};
%\draw  (0,-1.25) node[scale=.9] {$0\leq \tilde{s}\leq p_3-2$};
%\draw (0,-1.25) node[scale=.9] {$c_s\leq \tilde{s}+c_s\leq p_2-2+c_s$};

\draw (+0,-1.4) node[scale=.9] {only\ \ $\mathpzc{r}^{-}$\ \ contributes};
\end{tikzpicture}
\end{array}
\eeq
%We emphasise that 
This splitting is analytic in the arguments of the $\Gamma$ functions, 
and therefore the function $\mathpzc{r}^{\pm}$, will be polynomial, i.e. there are no absolute value discontinuities. 
When we change basis, the sum over $z$ truncates to  seven poles $z=0,\ldots, 6$!  and we find
\beq\label{solu_crossing22}
\mathpzc{r}^+_{\vec{p}}  = \mathpzc{r}_{\{-c_s,-c_t,c_u\}}\qquad;\qquad \mathpzc{r}^-_{\vec{p}} = \mathpzc{r}_{\{+c_s,+c_t,c_u\}},
\eeq
where $ \mathpzc{r}$ is given in the {\color{red} \tt ancillary file}. Here we will quote for illustration 
\beq\label{example_rpolys}
\!\!\begin{array}{l}
\mathpzc{r}^{}_{\vec{p};6}=\frac{(\check s + \frac{\Sigma+c_s}{2}-1)(\check s +  \frac{\Sigma+c_s}{2} )}{15}\,,  \\[.2cm]
\mathpzc{r}_{\vec{p};5}=\frac{(\check s +  \frac{\Sigma+c_s}{2} )

(-30+11 c_s+2 c_s^2   +9 \Sigma +3 c_s \Sigma  +\Sigma^2 +c_{tu}^2 + 30 {\check s}-2{\check s} \Sigma-12 {\check s}^2)}{30}\,,\\[.2cm]
\mathpzc{r}_{\vec{p};4}=\frac{ 30 \check s^4+10 \check s^3 (\Sigma - 9) -5 \check s (
-30+11 c_s+2 c_s^2+11 \Sigma+3 c_s\Sigma+\Sigma^2+c_{tu}^2)+\ldots}{15}\,,\\[.2cm]
\ \,\vdots
\end{array}
\eeq
%The rest can be found in the ancillary {\color{red} \tt ancillary file}.
%
It is now clear how the sphere splitting on the Mellin amplitude achieves the window splitting of section \ref{window_split_sec} coming from the OPE.
As mentioned already, to break the bonus property of the ${\bf \Gamma}$
we need $\mathpzc{r}_{\{c_s,c_t,c_u\}}$ to depend not only on $c_s$, but generically on all $c_s,c_t,c_u$. 
Considering \eqref{example_rpolys} for the {\tt s}-channel,
this indeed has non trivial dependence on $c_t$ and $c_u$ (actually here only on $c_{tu} =c_t+c_u$)\footnote{otherwise crossing would imply invariance under $c_s\leftrightarrow -c_s$.}
and therefore breaks the degeneracy of correlators  ${\cal M}^{(1,0)}$ and ${\cal M}^{(1,3)}$.

Let us comment further on crossing symmetry, verifying that \eqref{solu_crossing22} is consistent with crossing, and checking additional symmetries of $\mathpzc{r}$. 
By starting from the following (subset of crossing) relations, 
\begin{align}
{\cal R}^{(W)}_{p_1,p_2,p_3,p_4}(\hat s,\hat t,\check s,\check t)&= {\cal R}^{(W)}_{p_2,p_1,p_4,p_3}(\hat s,\hat t, \check s,\check t)= {\cal R}^{(W)}_{c_s, -c_t,-c_u}(\hat s,\hat t, \check s,\check t) \label{cr1} \\[.15cm]
&={\cal R}^{(W)}_{p_4,p_3,p_2,p_1}(\hat s, \hat t, \check s,\check t)= {\cal R}^{(W)}_{-c_s, c_t,-c_u}(\hat s, \hat t, \check s,\check t) \label{cr2}  \\[.15cm]
 &= {\cal R}^{(W)}_{p_3,p_4,p_1,p_2}(\hat s, \hat t, \check t,\check s)= {\cal R}^{(W)}_{-c_s, -c_t,c_u}(\hat s, \hat t, \check s,\check t)  \label{cr3} 
\end{align}
we find that $\mathpzc{r}^{\pm}$ are related to each other, and in fact are given in terms of a single function, 
\beq\label{solu_crossing2}
\qquad\mathpzc{r}^+_{\vec{p}}  = \mathpzc{r}_{\{-c_s,-c_t,c_u\}}\qquad;\qquad \mathpzc{r}^-_{\vec{p}} = \mathpzc{r}_{\{+c_s,+c_t,c_u\}}.
\eeq
Moreover, crossing shows that $\mathpzc{r}^{}$ has the following residual symmetry\footnote{The second property on the 
l.h.s.~does not follow from \eqref{cr1}-\eqref{cr3}. It has to do with exchanging $c_t\leftrightarrow c_u$ 
and it comes from imposing \eqref{unoduetrequattro} etc., 
on the full Mellin amplitude
\beq\notag
\mathcal{R}^{(W)}_{c_s,c_t,c_u}(\hat{s},\check{s})+\mathcal{R}^{(W)}_{c_t,c_u,c_s}(\hat{t},\check{t})+\mathcal{R}^{(W)}_{c_u,c_s,c_t}(\hat{u},\check{u}).
\eeq}
\beq
\mathpzc{r}_{\{c_s,c_t,c_u\}}= \mathpzc{r}_{\{c_s,-c_t,-c_u\}} \qquad;\qquad \mathpzc{r}_{\{c_s,c_t,c_u\}}=\mathpzc{r}_{\{c_s,c_u,c_t\}}.
\eeq
This is in fact what happens when we
look at the explicit expressions of $\mathpzc{r}^{\pm}_{\vec{p};z}(\hat s,\check s)$.

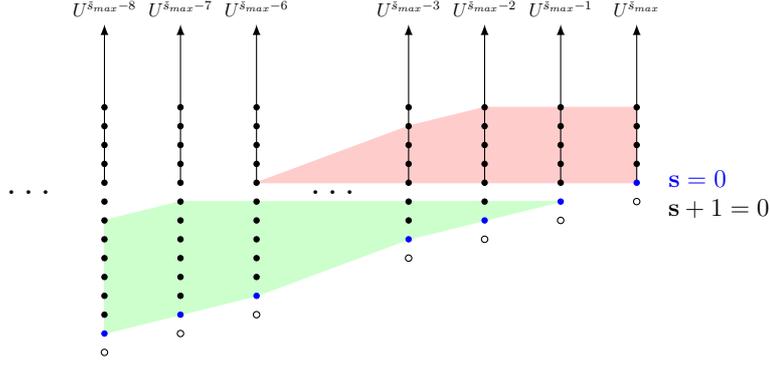
\begin{figure}[t]
\begin{center}
\begin{tikzpicture}

%%%%%%%%%%%% Draw Red %%%%%%%%%%%%%%
\filldraw[ red!20] (-2,0) --  (0,0.75) -- (1,1)  -- (3,1) -- (3,0) -- cycle;
\draw (-1,-0.125) node[scale=1.2] {$\ldots$};
%%%%%%%%%%%% Draw Green %%%%%%%%%%%%%%
\filldraw[ green!20] (-3,-0.25) -- (-4,-0.5) -- (-4,-2)  -- (-2,-1.5)  -- (0,-0.75) -- (2,-0.25) -- cycle;
\draw (-5,-0.125) node[scale=1.2] {$\ldots$};
%%%%%%%%%%%%%%%%%%%%%%%%%%%%%%%%
\draw (3.8,0.05) node[scale=.8,color=blue] {${\bf s}=0$};
\draw (4.08,-0.35) node[scale=.8] {${\bf s}+1=0$};
%%%%%%%%%%%%% Lines and Labels %%%%%%%%%%
\draw[-latex]  (-4,0) --(-4,.5+1.6);  		 
\draw[-latex]  (-3,0) --(-3,.5+1.6);  		 
\draw[-latex]  (-2,0) --(-2,.5+1.6);  		 
\draw[-latex]  (0,0) --(0,.5+1.6);  		 
\draw[-latex]  (1,0) --(1,.5+1.6); 			  
\draw[-latex]  (2,0) --(2,.5+1.6); 			  
\draw[-latex]  (3,0) --(3,.5+1.6); 			  
\draw (3,.5+1.8) node[scale=.6] {$U^{\check s_{max}}$};
\draw (2,.5+1.8) node[scale=.6] {$U^{\check s_{max}-1}$};
\draw (1,.5+1.8) node[scale=.6] {$U^{\check s_{max}-2}$};
\draw (0,.5+1.8) node[scale=.6] {$U^{\check s_{max}-3}$};
\draw (-2,.5+1.8) node[scale=.6] {$U^{\check s_{max}-6}$};
\draw (-3,.5+1.8) node[scale=.6] {$U^{\check s_{max}-7}$};
\draw (-4,.5+1.8) node[scale=.6] {$U^{\check s_{max}-8}$};
%%%%%%%%%%%%%% Dots %%%%%%%%%%%%%%%
%%%%%%
\foreach \y in {-1.75,-1.5,-1.25,-1,-.75,-.5,-.25,0,0.25,0.5,0.75,1}
    \draw[fill=black] (-4,\y) circle (1pt); 
    
\filldraw[blue] (-4,-2) circle (1pt); 
\draw[] (-4,-2.25) circle (1.2pt);  
%%%%%%
\foreach \y in {-1.5,-1.25,-1,-.75,-.5,-.25,0,0.25,0.5,0.75,1}
    \draw[fill=black] (-3,\y) circle (1pt); 
    
\filldraw[blue] (-3,-1.75) circle (1pt); 
\draw[] (-3,-2) circle (1.2pt);  
%%%%%%
\foreach \y in {-1.25,-1,-.75,-.5,-.25,0,0.25,0.5,0.75,1}
    \draw[fill=black] (-2,\y) circle (1pt); 
    
\filldraw[blue] (-2,-1.5) circle (1pt); 
\draw[] (-2,-1.75) circle (1.2pt);  
%%%%%%
\foreach \y in {-.5,-.25,0,0.25,0.5,0.75,1}
    \draw[fill=black] (0,\y) circle (1pt); 
    
\filldraw[blue] (0,-0.75) circle (1pt); 
\draw[] (0,-1.0) circle (1.2pt);  
%%%%%%
\foreach \y in {-.25,0,0.25,0.5,0.75,1}
    \draw[fill=black] (1,\y) circle (1pt); 
    
\filldraw[blue] (1,-0.5) circle (1pt); 
\draw[] (1,-0.75) circle (1.2pt);  
%%%%%%
\foreach \y in {0,0.25,0.5,0.75,1}
    \draw[fill=black] (2,\y) circle (1pt); 
    
\filldraw[blue] (2,-0.25) circle (1pt); 
\draw[] (2,-0.5) circle (1.2pt); 
%%%%%%
\foreach \y in {0.25,0.5,0.75,1}
    \draw[fill=black] (3,\y) circle (1pt); 
    
\filldraw[blue] (3,0) circle (1pt); 
\draw[] (3,-0.25) circle (1.2pt);  

%%%%%%%%%%%%%%%%%%%%%%%%%
\end{tikzpicture}
\end{center}
\caption{The pole structure of ${\cal R}^{(W)}$ in the monomial basis. Red: contributions lying in the window. Green: contributions lying in the Below window. 
\label{fig:win_poles_mon}}
\end{figure}

Two final comments. 
\begin{itemize}
\item[1)] The truncation in the number of poles can be seen to be in one-to-one correspondence with  the bound in the 
degree of the non factorisable polynomial appearing in the numerator of $\mathpzc{r}^{}_{\vec{p};z}$. Under this logic 
we cannot have a numerator past $\mathpzc{r}^{}_{\vec{p};6}$.
\item[2)] Notice also that $\mathpzc{r}^{}_{\vec{p};6}$ and $\mathpzc{r}_{\vec{p};5}$ vanish in the Window precisely because 
of the factors  $(\check s + \tfrac{p_{1}+p_2}{2}-1)$ and $(\check s + \tfrac{p_{1}+p_2}{2})$, thus guaranteeing 
consistency with the five poles picture in Figure~\ref{fig:win_harmonics_poles}. The new picture for the poles in the 
monomial basis is displayed in Figure \ref{fig:win_poles_mon}.
\end{itemize}

%===========================================================================================================
%===========================================================================================================

\subsection{Below Window completion}
\label{below_window_sec}

So far we have focussed on the Window region. We also understood that ${\cal R}_{}^{(W)}$ has 
an immediate analytic continuation to Below Window.  Now we should ask  whether we need or 
not an additional reminder function Below Window? 
The answer is affirmative and in fact we need a contribution ${\cal B}^{(BW)}$ of the following form, 
\be
\mathcal{B}^{(BW)}_{\vec{p}}(\hat{s},\check{s})=\sum_{z= 0}^2 \frac{(\check{s}+\frac{p_{1}+p_2}{2}-z)_{z+1} (\check{s}+\frac{p_{3}+p_4}{2}-z)_{z+1}}{{\bf s} -z} \mathpzc{b}^{}_{\vec{p};z }(\hat{s},\check{s}) \,.
\ee
Then, together, the contribution from $\mathcal{B}^{(BW)}$ and the contributions from both 
${\cal W}^{(AW)}$ and ${\cal R}^{(W)}$ Below Window will reproduce the correct OPE  prediction.

We find that only poles $z=0,1,2$ receive a Below Window completion.  Their explicit form is  
\beq
\!\!\begin{array}{l}
\mathpzc{b}_{\vec{p};2}=\frac{4 (\check s+ \Sigma-1 )}{15}\,,  \\[.2cm]
\mathpzc{b}_{\vec{p};1}=\frac{ (1-2\check s-\Sigma )(4 \check s^2 -4 \check s (1+2 \Sigma )+16- c_s^2 +4 \Sigma -11 \Sigma^2)}{15}\,,\\[.2cm]
\mathpzc{b}_{\vec{p};0}=\frac{ 16 \check s^5 -64\check s^4 \Sigma -8 \check s^3(c_s^2-15 \Sigma^2+4)+16\check s^2 \Sigma (c_s^2+26\Sigma^2  -29)+ \ldots}{60}\,,
\end{array}
\eeq
and can be read in the {\color{red} \tt ancillary file} attached to the arXiv.

Again the truncation to three poles is in one-to-one correspondence with the bound on the degree of the polynomial. 
It is interesting to note the below window region does not exhibit the sphere splitting. For example in 
the ${\tt s}$-channel these functions depend just on $c_s^2$ and thus 
are symmetric under $p_1+p_2\leftrightarrow p_3+p_4$. 

With the knowledge of ${\cal W}^{(AW)}$, ${\cal R}^{(W)}$ and ${\cal B}^{(BW)}$
the bootstrap program for ${\cal M}^{(2,3)}_{}$ is completed, up to the following ambiguities:

%===========================================================================================================
%===========================================================================================================
%\subsection{Ambiguities}
\paragraph{Ambiguities.}

The ambiguities we can add to the one loop function  are tree-like contributions of the form 
of contributions to the Virasoro-Shapiro tree amplitude and do not spoil the one-loop OPE predictions 
in- and below- Window.    For the case of ${\cal M}^{(2,3)}$ we can add the same as the 
functions corresponding to ${\cal M}^{(1,n=3,5,6)}$. The case ${\cal M}^{(1,n=7)}$ is simple 
to exclude since it will contribute to the flat space limit, and there is no such a contribution in flat space.  
Note that by construction the ambiguities listed above will contribute to the $\log^1(U)$ discontinuity, 
Above Window, rather than in the Window, and for the analytic part they will contribute, 
in the Window, rather than Below Window. In other words, they lie 
on top of what we fixed by the OPE data at one-loop, and therefore 
contribute with a free coefficient, as far as the bootstrap program is concerned.

%===========================================================================================================
%===========================================================================================================

\subsection{Large $p$ limit and the flat space amplitude}
\label{sec_largeflatlimit}

In this section we will compute the large $p$ limit of \cite{Aprile:2020luw}, i.e.~the limit of ${\cal M}^{(2,3)}$ when the external charges $p_i$ (and 
thus the Mellin variables) are taken to be large.  By the arguments in \cite{Aprile:2020luw}, this limit 
reduces the amplitude to the full flat space amplitude of IIB supergravity where ${\bf s},{\bf t},{\bf u}$ become 10d Mandelstam invariants. 
The flat space amplitude is quite a simple amplitude,  but  ${\cal M}^{(2,3)}$ involves various pieces 
and therefore it will be quite an interesting computation to show the final result.

 %bs -> 2 sc + \[CapitalSigma], bt -> 2 tc + \[CapitalSigma]
 Let us focus on the ${\tt s}$-channel, and introduce for convenience $\tilde{\bf s}=2{\check s}+{\Sigma}{}$. To take the limit,
we simply need to consider
\beq
\{{\bf s},\tilde{\bf s},c_s\}\ \rightarrow p\, \{ {\bf s},\tilde{\bf s},c_s\}\quad;\quad \Sigma\rightarrow p\Sigma,\qquad;\qquad p\rightarrow \infty
\eeq 
on the various entries in the Mellin amplitude, which we know explicitly as function of $\vec{p}$. %Here we have introduced for convenience $\tilde{\bf s}=2{\check s}+\frac{\Sigma}{2}$. 

The limit on ${\cal W}^{AW}_{\vec{p}}$ is straightforward and reads
\beq\label{flat_alpha3}
 \lim_{p\rightarrow \infty} \frac{ {\cal W}_{\vec{p}}^{AW}}{ p^8} =  \lim_{p\rightarrow \infty} \frac{ \mathpzc{w}_{}(\hat s,\check s,c_s;\Sigma) \psi(-{\bf s})}{ p^8} = \frac{\zeta_3}{90} \Sigma^4 {\bf s}^4 \log(-{\bf s})
\eeq
This is already the corresponding IIB flat space result \cite{Green:2008uj}. %, modulo an ambiguity which we will discuss afterwards. 
We conclude that for the large $p$ limit to hold the contributions from ${\cal R}^{(W)}$ and ${\cal B}^{(BW)}$ must give vanishing contributions in the limit.

One can see that the contribution from ${\cal B}^{(BW)}$ are subleading w.r.t to $p^8$ in the limit. 
However, for each one of the $z$ poles, the contribution from ${\cal R}^{(W)}$ individually are not subleading!  
In fact, we find the following contribution. 
%The highlights are as follows:
%{\color{red} change the notation}
\beq\label{table_largepanc}
\begin{array}{c|cc}
{\rm pole} & & \displaystyle \lim_{p\rightarrow \infty} \frac{{\rm contribution\ from\ pole}\,(z) }{p^8} \\
\hline\\
z=6 &  & \frac{ (\tilde{\bf s}-c_s)^2(\tilde{\bf s}-c_s-2\Sigma)^3((\tilde{\bf s}-\Sigma-c_{tu})(\tilde{\bf s}-\Sigma+c_{tu}))^2}{ 7680 {\bf s} } \\[.2cm] 
\vdots & & \vdots \\[.2cm]
z=0 & &  \frac{(\tilde{\bf s}-c_s)^7(\tilde{\bf s}+c_s)^2  }{7680 {\bf s} }
\end{array}
\eeq
(The full computation can be found in the ancillary {\color{red} {\tt ancillary file}}.) 

By inspection of \eqref{table_largepanc}, we find that the leading term of each pole $z$ is as leading as \eqref{flat_alpha3}: It has degree $8$ in $p$. However, 
when \emph{summing} over all contributions in the second column, the result vanishes! This cancellation is quite remarkable, 
given that non trivial functions of $\tilde{\bf s}$ and charges are involved. We find then that the large $p$ limit of the one-loop amplitude 
${\cal M}^{(2,3)}$, and the flat space IIB S-matrix, match perfectly.  As a byproduct of our analysis here we have given a non-trivial 
confirmation of the geometric picture associated with the large $p$ limit, as described in \cite{Aprile:2020luw}, and shown 
the self-consistency of the one-loop bootstrap program.

%===========================================================================================================
%===========================================================================================================

\section{Conclusions}
In this paper we studied one-loop stringy corrections to the Mellin amplitude of four single-particle operators 
$\langle \mathcal{O}_{p_1} \mathcal{O}_{p_2} \mathcal{O}_{p_3} \mathcal{O}_{p_4} \rangle$. 
We started from the $AdS_5\times S^5$ Mellin representation in terms of variables conjugated to cross ratios, ${\hat s},{\check s},{\hat t},{\check t}$, 
the Mellin amplitude $\mathcal{M}_{\vec{p}}$, and the following kernel of Gamma function, 
\beq%\label{example_Gamma_s}
{\bf \Gamma}=\frac{ \Gamma[\frac{p_1+p_2}{2} -\hat{s}]\Gamma[\frac{p_3+p_4}{2}-\hat{s}]}{  \Gamma[1+\frac{p_1+p_2}{2}+\check{s}]\Gamma[1+\frac{p_3+p_4}{2}+\check{s}]}\Gamma_{\tt t}\Gamma_{\tt u},
\eeq
see discussion around \eqref{ex_gammas}. Then,
we gave explicit formulas for the leading order $(\alpha')^3$ amplitude at one-loop, in the form,
\begin{align}
\label{M13structure_conclusion}
\mathcal{M}^{(2,3)}_{\vec{p}} = \Bigl[ \mathcal{W}^{(AW)}_{\vec{p}}(\hat{s},\check{s}) + \mathcal{R}^{(W)}_{\vec{p}}(\hat{s},\check{s})+ \mathcal{B}^{(BW)}_{\vec{p}}(\hat{s},\check{s}) \Bigr] + {\tt crossing }\,.
\end{align}
The various functions $\mathcal{W}^{(AW)},\mathcal{R}^{(W)},\mathcal{B}^{(BW)}$, introduced in \eqref{M13structure}, 
are described in full details in sections \ref{above_window_sec}, \ref{window_dynamics}, and \ref{below_window_sec}, respectively. 
Each one is bootstrapped from the OPE, and labelled by a corresponding region of 
twists for the exchange of long two-particle operators. We called these regions Above Window, Window and Below Window,
following \cite{Aprile:2019rep}, where the same classification was used to bootstrap the one-loop supergravity amplitude, mainly in position space. The supergravity amplitude is 
still a rather complicated function, due to the fact that the OPE has support for all spins. The $\alpha'$ corrections have instead finite spin supports and therefore are simpler to deal with.  
We believe however that the lessons from the study of $\alpha'$ amplitudes are general, and in fact best expressed in Mellin space.

The advantage of using Mellin space stems from the observation that the 
whole structure of poles in Mellin space can be put in correspondence with the OPE, a result due to Mack  \cite{Mack:2009mi}, which  here we have upgraded to $AdS_5\times S^5$. 
The pole structure takes into account both ${\bf \Gamma}$ and $\mathcal{M}$. 
The poles of $\mathcal{M}$ at tree level were shown to be captured by bold-font variables \cite{Aprile:2020luw}, i.e.~poles given by equations of the 
form ${\bf s}+1=0$, where ${\bf s}=\hat s+ \check s$, similarly for ${\bf t}$, or ${\bf u}$. Quite nicely, this way of parametrising the poles works at one loop, with two important modifications: 
Poles of ${\cal M}$ are now of the form $\psi(-\bf{s})$ or ${\bf s}-z=0$ for $z=0,\ldots, 6$, and the residues, compared to tree level, have generic dependence on the charges $\vec{p}$. 

Considering the various contributions to ${\cal M}^{(2,3)}$ in \eqref{M13structure_conclusion} we have found:

%\begin{itemize}
%\item 
(1) The double logarithmic discontinuity of the one-loop amplitude comes from $\mathcal{W}^{(AW)}$ and is fixed by OPE data Above Window.
It takes the form of $\psi(-{\bf s})$ times a non-trivial polynomial in the Mellin variables, and for many aspects fits into the discussion of the tree level Virasoro-Shapiro amplitude, as given in \cite{Aprile:2020mus}.  
In particular, it can be given as $\Delta^{(8)}$ on a preamplitude. This expectation will be true to all orders in $\alpha'$ and we exemplified the case of $(\alpha')^3$ and $(\alpha')^5$ in the {\tt \color{red} ancillary file}. 

%\item 
(2) The structure of  $\mathcal{R}^{(W)}$ is the main novelty of the one-loop amplitude. This is determined by OPE data in the Window. 
This OPE data is the one responsible for lifting a bonus property of the supergravity amplitude, i.e.~the fact that certain correlators are equal to each other.
Such a degeneracy of amplitudes follows from the special (crossing symmetric) form of ${\bf \Gamma}$, and the peculiar charge 
dependence of the Mellin amplitudes, i.e.~no dependence at all in supergravity, and $(\frac{p_1+p_2+p_3+p_4}{2}-1)_3$ at order $(\alpha')^3$, see section \ref{tree_level_primer} for further explanations.

The first crucial result is that $\mathcal{R}^{(W)}$  takes the form, 
\be
\label{window_splitting1int111}
\mathcal{R}^{(W)}_{\vec{p}}(\hat{s},\check{s})= \sum_{z=0}^6 \frac{1}{{\bf s} -z}\biggl[
(\check{s}+\tfrac{p_{1}+p_2}{2}-z)_{z+1} \,
\mathpzc{r}^{+}_{\vec{p};z} 
(\hat{s},\check{s}) 
+
(\check{s}+\tfrac{p_{3}+p_4}{2}-z)_{z+1}\, \mathpzc{r}^{-}_{\vec{p};z}(\hat{s},\check{s})
\biggr]\,,
\ee
and comes with two separate contributions, singled out by the $z$-dependent Pochhammers. This implies that 
when we look at ${\cal R}^{(W)}_{}$ together with the ${\bf \Gamma}$ functions,  
the total amplitude undergoes the following split,
 \beq%\label{sph_split_intro}
\!\!\!\!\!\!\begin{array}{c}
\begin{tikzpicture}
\draw (-.5,.5) node[scale=.9] {$ \ \ \ \ \ \ \ 
\displaystyle %\frac{1}{{\bf s} -z}
\frac{ {\cal R}^{(W)}_{\vec{p}} }{  \Gamma[1+\frac{p_{1}+p_2}{2}+\check{s}]\Gamma[1+\frac{p_{3}+p_4}{2}+\check{s}]} $};  %\frac{1}{s+\tilde s-z}\Bigg[ \frac{1}{\Gamma[\tilde{s} +1] \Gamma[\tilde{s} +c_s+1]} \Bigg]$};
\draw[-latex] (-0.1,-0.1)--(-1,-1.1);
\draw[-latex] (-0.1,-0.1)--(1.2,-1.1);

\draw (-.5,-1.5) node[scale=.9] {$
\displaystyle
=\frac{1}{ {\bf s} -z}\Bigg[ \frac{\mathpzc{r}_{\vec{p};z}^{+}  }{ \Gamma[-z+\frac{p_{1}+p_2}{2}+\check{s}]\Gamma[1+\frac{p_3+p_4}{2}+\check{s}] }    \,  
+ \frac{\mathpzc{r}_{\vec{p};z}^{-}  }{ \Gamma[1+\frac{p_1+p_2}{2}+\check{s}]\Gamma[-z+\frac{p_3+p_4}{2}+\check{s}] } \Bigg]$};
\end{tikzpicture}
\end{array}
\eeq
The polynomials $\mathpzc{r}^{\pm}$ are related by crossing, $\mathpzc{r}^+_{\vec{p}}  = \mathpzc{r}_{\{-c_s,-c_t,c_u\}}$, $\mathpzc{r}^-_{\vec{p}} = \mathpzc{r}_{\{+c_s,+c_t,c_u\}}$, 
and crucially have generic charge dependence. This structure, which all together we called ``sphere splitting", follows from the OPE.\footnote{From the OPE 
viewpoint this is also true at one-loop in supergravity \cite{Aprile:2019rep}.}
%, that  one-loop predictions}
%in the Window distinguish tree level degenerate correlators (see section \ref{window_split_sec}).} 
Our result for $\mathcal{R}^{(W)}_{}$ shows neatly how the OPE structure translates into a structure in Mellin space. 
% \end{itemize}

The second crucial result is the use of bold font variables to parametrise the poles of ${\cal M}$.
For $\mathcal{W}^{(AW)}$, this implies that $\mathcal{W}^{(AW)}$ not only contributes to the OPE Above Window, where it was bootstrapped, 
but cascades to Window and Below Window regions. Because of this, the OPE predictions in the Window pick the contributions of both $\mathcal{W}^{(AW)}$
and ${\cal R}^{(W)}$. In this sense, ${\cal R}^{(W)}$ is a remainder function, but precisely for this reason, the sum over $z$ in \eqref{window_splitting1int111} truncates
to finitely many poles, $z=0,\ldots 6$, rather than depending on the external charges.  Quite remarkably, the result for ${\cal R}^{(W)}$ can itself be continued Below Window. Thus, with the same logic, 

(3) The function ${\cal B}^{(BW)}$ 
is 
\beq
\mathcal{B}^{(BW)}_{\vec{p}}(\hat{s},\check{s})=\sum_{z\geq 0} \frac{(\check{s}+\frac{p_{1}+p_2}{2}-z)_{z+1} (\check{s}+\frac{p_{3}+p_4}{2}-z)_{z+1}}{{\bf s} -z} \mathpzc{b}^{}_{\vec{p};z }(\hat{s},\check{s}) \,,
\eeq
and is itself a remainder function for $\mathcal{W}^{(AW)}$ and ${\cal R}^{(W)}$ in the Below Window region. In sum, the analytic properties of all these functions are quite spectacular:
they properly continue from the Above Window down to the Below Window region, and correctly switch off outside the physical range of relevance.

Finally, we studied the 10d flat space limit, using the large $p$ limit of \cite{Aprile:2020luw}. The consistency with the 
flat space amplitude of IIB supergravity is again quite remarkable. The limit on 
${\cal W}^{(AW)}$ is already giving the flat amplitude, and 
naively, each pole in $z$ from ${\cal R}^{(W)}$ adds a non vanishing contribution, non trivial in both ${\check s}$ and the charges. 
However, upon summing over $z$ all these extra contributions correctly cancel out!

Let us conclude with an outlook. 
Our main focus has been understanding the structure of the amplitude and the way Mellin space realises the various features of the OPE predictions at one-loop. 
We focused on the $(\alpha')^3$ contribution, but we believe that 
the logic behind the construction of ${\cal M}^{(2,3)}$ is valid to all orders in $(\alpha')^{n+3}$. In particular the sphere splitting is generic, 
and the number of poles in $z$ increases with $n$ but stays finite!
It is only a matter of computational effort to fix the various residues in the Window and Below Window. 
The one-loop bootstrap program in $AdS_5\times S^5$ is thus understood at all orders in $\alpha'$,  but for the usual ambiguities, which needs additional input to be fixed. 
 It would be interesting to find these extra constraints, 
either from localisation \cite{Binder:2019jwn,Chester:2019pvm,Chester:2020dja}, or from sum rules \cite{Alday:2022uxp,Caron-Huot:2022sdy}. In this 
case we would start from a non-perturbative Mellin amplitude \cite{Penedones:2019tng,Caron-Huot:2020adz}, and consider the supergravity expansion in search for constraints.

It would be also interesting to understand the ``sphere splitting" from a diagrammatic point of view. 
Perhaps a 10d master amplitude can be found which undergoes the ``sphere" split onto ${\cal R}^{(W)}$ in a natural way. This
is partially suggested by the remarkable cancellations that take place when we tested the large $p$ flat space limit  in section \eqref{sec_largeflatlimit} and the fact that poles are parametrised by bold font variables. 
Perhaps this point of view would give insight on the way to arrange the one-loop supergravity result in a simple form.  

%{\color{red} MORE?}

\section*{Acknowledgement}
We thank Paul Heslop and Pedro Vieira for many discussions on this and related topics. 
FA is supported by FAPESP bolsa n.2020/16337-8.
JD is supported in part by the ERC Consolidator grant 648630 IQFT. 
MS is supported by a Mayflower studentship from the University of Southampton.
RG is supported by an STFC studentship

%===========================================================================================================
%===========================================================================================================
\appendix

\section{Useful facts about blocks}

In order to make operative the OPE predictions we have to expand the correlator in superconformal blocks. 
Superconformal blocks depend on the type of multiplet exchanged in the OPE, i.e.~protected or long.
In a 4pt function protected multiplets all come from free theory contributions, while unprotected (or long) 
multiplets come from both free theory and interacting contributions. 
In general we will write,
\be\label{blockk}
\langle \mathcal{O}_{p_1} \mathcal{O}_{p_2} \mathcal{O}_{p_3} \mathcal{O}_{p_4} \rangle = \langle \mathcal{O}_{p_1} \mathcal{O}_{p_2} \mathcal{O}_{p_3} \mathcal{O}_{p_4} \rangle_{\rm free}^{\rm prot.} + 
\underbrace{ \langle \mathcal{O}_{p_1} \mathcal{O}_{p_2} \mathcal{O}_{p_3} \mathcal{O}_{p_4} \rangle_{\rm free}^{\rm long} + \langle \mathcal{O}_{p_1} \mathcal{O}_{p_2} \mathcal{O}_{p_3} \mathcal{O}_{p_4} \rangle^{\rm long}_{\rm int}. }_{ LONG }
\ee
Importantly, some multiplets at the unitarity bound, which therefore are semishort in the free theory, 
recombine in the interacting theory to become long. This is explained in great detail in \cite{Doobary:2015gia}.
In this paper we have been concerned with long contributions, and long $\mathcal{N}=4$ superblocks, 
which we will denote by  $\mathbb{L}$, are quite simple to deal with, since they have a factorised form (
and can be made independent of a parameter called $\gamma$  in \cite{Doobary:2015gia} and \cite{Aprile:2021pwd}).

Let us denote the quantum numbers of the exchanged multiplet by $\vec{\tau} = (\tau,l,a,b)$ 
for a superconformal primary of twist $\tau = \Delta-l$, spin $l$, and $su(4)$ representation with Dynkin labels $[aba]$. Then,
\be\label{nn_pert_OPE}
\langle \mathcal{O}_{p_1} \mathcal{O}_{p_2} \mathcal{O}_{p_3} \mathcal{O}_{p_4} \rangle^{\rm long} = \sum_{\vec{\tau}} c^{\vec{p}}_{\vec{\tau}}\, \mathbb{L}^{\vec{p}}_{\vec{\tau}}\,,
 \ee
holds non-perturbatively. Here $\mathbb{L}$ is the superblock for long multiplets defined as %(see e.g. \cite{Doobary:2015gia}), 
\begin{align}
\mathbb{L}^{\vec{p}}_{\vec{\tau}} = %\tilde{\mathcal{I}}\, \frac{\prod_{i<j} g_{ij}^{p_{(ij)}}}{(g_{13} g_{24})^{\Sigma}} \biggl(\frac{V}{\tilde{V}}\biggr)^{p_{(23)}} 
g_{12}^{ \frac{p_1+p_2}{2} } g_{34}^{\frac{p_3+p_4}{2} }\left(\frac{g_{14} }{ g_{24} }\right)^{\!\!\frac{p_1-p_2}{2} }\left(\frac{ g_{14} }{ g_{13} }\right)^{\!\!\frac{p_4-p_3}{2} }\!\!\times\prod_{i,j}(x_i-y_j)\times
\mathcal{B}^{\vec{p}}_{\tau,l}(x,\bar{x}) Y^{\vec{p}}_{[aba]}(y,\bar{y})\,.
\end{align}
The conformal, $\mathcal{B}$, and internal,  $Y$, factors of $\mathbb{L}$ are given by 
\begin{align}
\mathcal{B}^{\vec{p}}_{\tau,l}(x,\bar{x}) &= \frac{F^{(\alpha, \beta)}_{l+\tau/2+2}(x) F^{(\alpha, \beta)}_{\tau/2 + 1}(\bar{x}) - F^{(\alpha, \beta)}_{l+\tau/2+2}(\bar{x}) F^{(\alpha, \beta)}_{\tau/2+1}(x)}{x-\bar{x}} \,, \notag \\
Y^{\vec{p}}_{[aba]}(y,\bar{y}) &= \frac{F^{(-\alpha, -\beta)}_{-b/2}(y) F^{(-\alpha, -\beta)}_{-b/2-a-1}(\bar{y}) - F^{(-\alpha, -\beta)}_{-b/2}(\bar{y}) F^{(-\alpha, -\beta)}_{-b/2-a-1}(y)}{y-\bar{y}}\,,
\end{align}
where
\begin{align}
F^{(\alpha, \beta)}_\rho(x) = x^{\rho-1} {}_2F_1(\rho+\alpha,\rho+\beta,2\rho;x) \,.
\end{align}
and we have introduced the notation $\alpha = \frac{p_2-p_1}{2}$ and $\beta = \frac{p_3-p_4}{2}$.

In the main text, we considered the non perturbative decomposition \eqref{blockk},
and we expanded in $1/N$ and $\alpha'$. Since ${\cal N}=4$ SYM in this regime also has a free theory limit, 
we referred to $\tau$ to the free theory twist, and split the dimension into free plus anomalous. 

In the block decomposition we could assume that $p_1\ge p_2$ and $p_4\ge p_3$. 
This can be done without loss of generality since the OPE does not change if we exchange $1\leftrightarrow 2$ or $3\leftrightarrow 4$. 
In fact, both crossing transformations behave nicely w.r.t.~the blocks. Let us make this statement more precise as follows. From the relation
\eqref{nn_pert_OPE} consider applying crossing  $1\leftrightarrow 2$, then
\beq
\begin{array}{ccc}
\langle \mathcal{O}_{p_2}(x_1) \mathcal{O}_{p_1}(x_2) \mathcal{O}_{p_3}(x_3) \mathcal{O}_{p_4}(x_4) \rangle^{\rm long}  & = &
\sum_{\vec{\tau}} c^{p_1p_2p_3p_4}_{\vec{\tau}}\, \times \mathbb{L}^{p_1p_2p_3p_4}_{\vec{\tau}}(\tfrac{x}{x-1},\tfrac{y}{y-1})\Big|_{g_{1i}\leftrightarrow g_{2i}} \\
 \begin{tikzpicture} \node[rotate=90] (0,0) {$=$};\end{tikzpicture}  & & \\
\sum_{\vec{\tau}} c^{p_2p_1p_3p_4}_{\vec{\tau}}\, \times \mathbb{L}^{p_2p_1p_3p_4}_{\vec{\tau}}(x,y)
\end{array}
\eeq
Now, using (one of) the Pfaff identity ${}_2F_1\bigl(a,b,c;\tfrac{x}{x-1}\bigr) = (1-x)^b {}_2 F_1(c-a,b,c;x)$, the hypergeometric functions 
entering the long blocks can be shown to match up to signs, and we obtain the relation between $c^{p_2p_1p_3p_4}$ and $c^{p_1p_2p_3p_4}$, namely
$c^{p_2p_1p_3p_4}_{\vec{\tau}} = (-1)^{l+a} c^{p_1p_2p_3p_4}_{\vec{\tau}}$ where we used that  $\tau+b$ is always even.
We can repeat the same argument with $3\leftrightarrow 4$ by using (the other one of) the Pfaff identity ${}_2F_1\bigl(a,b,c;\tfrac{x}{x-1}\bigr) =  (1-x)^a {}_2F_1(a,c-b,c;x)$. 
All together we find,
\begin{align}\label{symm_prop_james}
c^{\vec{q}}_{\vec{\tau}} &= (-1)^{l+a} c^{\vec{p}}_{\vec{\tau}} \,  &&\text{for } \vec{q} = (p_2,p_1,p_3,p_4) \text{ or } \vec{q} = (p_1,p_2,p_4,p_3)\,.
\end{align}
Of course, by composing the two, 
\begin{align}
c^{\vec{q}}_{\vec{\tau}} =  c^{\vec{p}}_{\vec{\tau}} \,  &&\text{for } \vec{q} = (p_4,p_3,p_2,p_1)\,.
\end{align}
This one is also obvious because the cross ratios do not change and the blocks are manifestly 
invariant under $p_1\leftrightarrow p_4$ and $p_2\leftrightarrow p_3$.

From the above considerations, we expect that the
CPW coefficients for $(a+l)$ even and the CPW coefficients for $(a+l)$ odd arise from two distinct analytic functions
$c^\pm$, namely,
\be
c^{\vec{p}}_{\vec{\tau}} = 
\begin{cases}
&c^+(\vec{p},\vec{\tau})\quad \text{ $(a+l)$ even,} \\ \notag
&c^-(\vec{p},\vec{\tau}) \quad \text{ $(a+l)$ odd.}
\end{cases}
\ee
which respect the symmetry properties \eqref{symm_prop_james}
for all quantum numbers $\vec{\tau}$ and charges $\vec{p}$.

%===========================================================================================================
%===========================================================================================================
\section{Combining the weight-two contribution}

Note first that we can rewrite the result for ${\cal B}^{(BW)}$ using the same split used for ${\cal R}^{(W)}$. 
In this way we can write
\be
\label{window_splitting1}
\mathcal{W}^{(2)}_{\vec{p}}(\hat{s},\check{s})=
\sum_{z\geq 0} \frac{(\check{s}+p_{12}-z)_{z+1}{\mathcal{P}}^{+}_{\vec{p};z }(\hat{s},\check{s})+(\check{s}+p_{34}-z)_{z+1} {\mathcal{P}}^{-}_{\vec{p};z }(\hat{s},\check{s})}{{\bf s} -z}  \,,
\ee
where $\mathcal{W}^{(2)}_{\vec{p}}(\hat{s},\check{s}):=\mathcal{R}^{(W)}_{\vec{p}}+\mathcal{B}^{(W)}_{\vec{p}}$ and 
\beq
\mathcal{P}^+_{\vec{p}}  = \mathcal{P}_{\{-c_s,-c_t,c_u\}}\qquad;\qquad \mathcal{P}^-_{\vec{p}} = \mathcal{P}_{\{+c_s,+c_t,c_u\}},
\eeq
The list of functions ${\cal P}_{\vec{p};z}$ is given in the {\color{red}\tt ancillary file}. Of course only the poles at $z=0,1,2$ are modified by the addition of ${\cal B}^{(BW)}$ w.r.t.~the contribution of ${\cal R}^{(W)}$. 
%
%by writing
%\be
%\mathcal{P}^{(z)}_{\vec{p} }(\hat{s},\check{s})=\mathcal{R}^{(z)}_{\vec{p}}(\hat{s},\check{s})+\tilde{\mathcal{B}}^{(z)}_{\vec{p} }(\hat{s},\check{s}).
%\ee

Certain analyticity in $z$ can be made manifest for the $\mathcal{P}^{(z)}_{\vec{p} }$. 
Assuming we focus on ${\cal P}^-_z$ for concreteness,
the idea is that the ratio ${\cal P}_z\big/{\Gamma[1+\frac{p_{1}+p_{2}}{2}+\check{s}]\Gamma[-z+\tfrac{p_{3}+p_{4}}{2}+\check{s}]  }$ 
can be further decomposed by writing ${\cal P}_z$ as a polynomial in Pochhammers. 
This is initially suggested by the special form of ${\cal P}_6$ and ${\cal P}_5$ which have a partial factorised form of this sort, i.e.
\beq
\!\!\begin{array}{l}
{\cal P}^{}_{\vec{p};6}=\frac{(\check s + \frac{\Sigma+c_s}{2}-1)(\check s +  \frac{\Sigma+c_s}{2} )}{15}\,,  \\[.2cm]
{\cal P}^{}_{\vec{p};5}=\frac{(\check s +  \frac{\Sigma+c_s}{2} )
(-30+11 c_s+2 c_s^2   +9 \Sigma +3 c_s \Sigma  +\Sigma^2 +c_{tu}^2 + 30 {\check s}-2{\check s} \Sigma-12 {\check s}^2)}{30}\,,\\[.2cm]
\end{array}
\eeq
In this way we can absorb the ${\check s}$ dependence into the $z$-independent $\Gamma$, and get a structure of the form, 
\beq
\frac{ {\cal P}_z({\check s},c_s,c_t,c_u,\Sigma) }{
 \Gamma[1+\frac{p_{1}+p_{2}}{2}+\check{s}]\Gamma[-z+\tfrac{p_{3}+p_{4}}{2}+\check{s}]  }=\sum_{w\ge-1}\frac{ {\cal P}_{z,w}( c_s,c_t,c_u,\Sigma)  }{ 
 \Gamma[-z+\tfrac{p_{3}+p_{4}}{2}+\check{s}] \Gamma[-w+\tfrac{p_{1}+p_{2}}{2}+\check{s}] }
\eeq
for each value of $z$. 
By doing so we find a lattice of points $(w,z)$ of the form
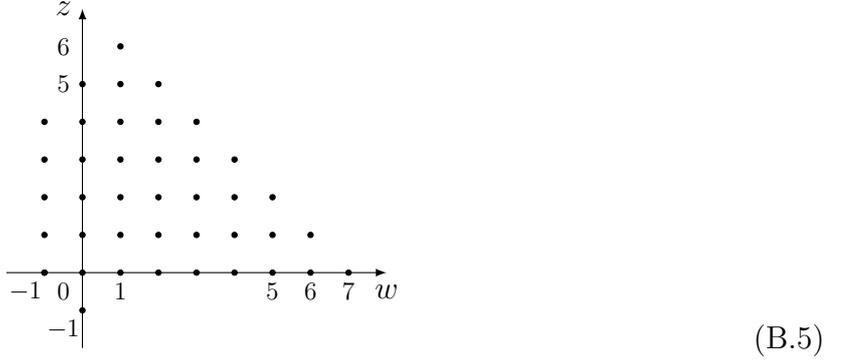
\beq\label{diagonal_Pzw}
\begin{tikzpicture}

\def\scale {.5};

\draw[-latex] (-2*\scale,0)-- (8*\scale,0);
\draw (8*\scale,-.5*\scale) node {$w$};
\draw (7*\scale,-.5*\scale) node[scale=.8] {$7$};
\draw (6*\scale,-.5*\scale) node[scale=.8] {$6$};
\draw (5*\scale,-.5*\scale) node[scale=.8] {$5$};
\draw (1*\scale,-.5*\scale) node[scale=.8] {$1$};
\draw (-0.5*\scale,-.5*\scale) node[scale=.8] {$0$};
\draw (-1.5*\scale,-.5*\scale) node[scale=.8] {$-1$};

\draw (-.5*\scale,7*\scale) node {$z$};
\draw (-.5*\scale,6*\scale) node[scale=.8] {$6$};
\draw (-.5*\scale,5*\scale) node[scale=.8] {$5$};
\draw (-.5*\scale,-1.5*\scale) node[scale=.8] {$-1$};

\draw[-latex] (0,-2*\scale)--(0,7*\scale);

\foreach \xx/\yy in { -1*\scale/0,  0/-1*\scale }
\filldraw[black] (\xx,\yy) circle (1pt);

\foreach \yy in {0,...,4}
\filldraw[black] (-1*\scale,\yy*\scale) circle (1pt);

\foreach \yy in {0,...,5}
\filldraw[black] (0*\scale,\yy*\scale) circle (1pt);

\foreach \yy in {0,...,6}
\filldraw[black] (1*\scale,\yy*\scale) circle (1pt);

\foreach \yy in {0,...,5}
\filldraw[black] (2*\scale,\yy*\scale) circle (1pt);

\foreach \yy in {0,...,4}
\filldraw[black] (3*\scale,\yy*\scale) circle (1pt);

\foreach \yy in {0,...,3}
\filldraw[black] (4*\scale,\yy*\scale) circle (1pt);

\foreach \yy in {0,...,2}
\filldraw[black] (5*\scale,\yy*\scale) circle (1pt);

\foreach \yy in {0,1}
\filldraw[black] (6*\scale,\yy*\scale) circle (1pt);

\foreach \yy in {0}
\filldraw[black] (7*\scale,\yy*\scale) circle (1pt);

\end{tikzpicture}
\eeq
For example, the horizontal axis at $z=0$ corresponds to a polynomial of degree eight in $\tilde{\bf s}$, and therefore there are nine bullet points, the first of which counts
for a degree zero contributions, going with $w=-1$, and the last one for a degree eight contribution, going with $w=7$. 

Rearranging the polynomials on the $-45^\circ$ diagonals, we find simple analytic expressions. For illustration, 
\beq\label{rearrange_Pzw}
\begin{array}{rlc}
{\cal P}_{z,w}&=\tfrac{1}{15}\, {\tt Bin}[^{\,6\,}_{\,z}]&\quad;\quad z+w=7 \\[.2cm]
{\cal P}_{z,w}&=\tfrac{(6c_s+5\Sigma+3(2z-11)}{15}\, {\tt Bin}[^{\,5\,}_{\,z}]&\quad;\quad z+w=6 \\[.2cm]
{\cal P}_{z,w}&=\tfrac{-(c_{tu})^2 + (c_s)^2(31-6z) + \ldots\ }{30}\, {\tt Bin}[^{\,5\,}_{\,z}]&\quad;\quad z+w=5 \\[.2cm]
\vdots\ \ 
\end{array}
%((3 (-11 + 2 ii) + 6 cs + 5 \[CapitalSigma])/15 ) Binomial[5, ii]
\eeq
and so on so forth. The pattern of the binomial ${\tt Bin}$ persists, and becomes ${\tt Bin}[^{-1+\#\bullet}_{\ \ \ \,z}]$ 
where $\#$ counts the number of $\bullet$ on the various diagonals  in \eqref{diagonal_Pzw}. 
This binomial is always singled out by the fact that some of the top degree terms in 
${\cal P}_{z,w}$, as function of $z$, contribute with a constant times such binomial. 
In the above formulae, these top degree terms are $(c_{tu})^0\otimes \{c_s,\Sigma\}$ when $z+w=6$, and $(c_{tu})^2$ 
when $z+w=5$. %Recall that only even powers of $c_{tu}$ can appear at $(\alpha')^3$.

The analytic structure in \eqref{remainder_func_solu2} suggests from the very beginning a sort of analyticity of ${\cal P}_z$ in $z$, 
and the rearrangement into ${\cal P}_{z,w}$ in \eqref{rearrange_Pzw}, is an example.
On the other hand, had we started with a general parametrisation of ${\cal P}_z$ as a polynomial in ${\check s},c_s,c_t,c_u,\Sigma$, and fitted data in- and below- Window, 
we would have found some free coefficients left.  There is indeed a subtlety with a ${\cal P}_z$ so constructed, which is the following: %as a polynomial in $\tilde{\bf s},c_s,c_t,c_u,\Sigma$:
Pieces of ${\cal P}_{z}$ cancel out in the sum
\beq\label{cancell}
\ \ \frac{\bigg[ 
\frac{ \Gamma[{\check s}+\frac{p_1+p_2}{2}+1]  \mathcal{P}_z^{+} }{\Gamma[ {\check s}+\frac{p_1+p_2}{2}-z] } +
 \frac{ \Gamma[{\check s}+\frac{p_3+p_4}{2}+1]  \mathcal{P}_z^{-}  }{\Gamma[ {\check s}+\frac{p_3+p_4}{2}-z] }  \bigg]}{ \Gamma[{\check s}+\frac{p_1+p_2}{2}+1]\Gamma[{\check s}+\frac{p_3+p_4}{2}+1] }. 
\eeq
Using that $\Gamma[X+1]/\Gamma[X-z]=(X-z)_{z+1}$ is polynomial in $X$, and
expanding the whole numerator in \eqref{cancell}, it is simple to see  that contributions of the form $c_s^{2\mathbb{N}+1} f( {\check s},c_s^2,c_t^2,c_u^2,\Sigma)$, for any function $f$, cancel out. 
We haven't encountered this subtlety in our discussion above because the functions in ${\cal R}^{(W)}$, nicely enough, do not have it.

%\bibliographystyle{apsrev4-1}

%\bibliography{refs}

\begin{thebibliography}{99}



%\cite{Heslop:2022qgf}
\bibitem{Heslop:2022qgf}
P.~Heslop,
``The SAGEX Review on Scattering Amplitudes, Chapter 8: Half BPS correlators,''
[arXiv:2203.13019 [hep-th]].
%3 citations counted in INSPIRE as of 18 May 2022

%\cite{Gopakumar:2022kof}
\bibitem{Gopakumar:2022kof}
R.~Gopakumar, E.~Perlmutter, S.~S.~Pufu and X.~Yin,
``Snowmass White Paper: Bootstrapping String Theory,''
[arXiv:2202.07163 [hep-th]].
%6 citations counted in INSPIRE as of 01 Jul 2022



\bibitem{Aprile:2020uxk}
F.~Aprile, J.~M.~Drummond, P.~Heslop, H.~Paul, F.~Sanfilippo, M.~Santagata and A.~Stewart,
``Single particle operators and their correlators in free $ \mathcal{N} $ = 4 SYM,''
JHEP \textbf{11} (2020), 072
%doi:10.1007/JHEP11(2020)072
[arXiv:2007.09395 [hep-th]].

%\cite{Rastelli:2016nze}
\bibitem{Rastelli:2016nze}
L.~Rastelli and X.~Zhou,
``Mellin amplitudes for $AdS_5\times S^5$,''
Phys. Rev. Lett. \textbf{118} (2017) no.9, 091602
%doi:10.1103/PhysRevLett.118.091602
[arXiv:1608.06624 [hep-th]].
%103 citations counted in INSPIRE as of 28 Mar 2021

%\cite{Rastelli:2017udc}
\bibitem{Rastelli:2017udc}
L.~Rastelli and X.~Zhou,
``How to Succeed at Holographic Correlators Without Really Trying,''
JHEP \textbf{04} (2018), 014
%doi:10.1007/JHEP04(2018)014
[arXiv:1710.05923 [hep-th]].
%87 citations counted in INSPIRE as of 28 Mar 2021



%\cite{Alday:2017xua}
\bibitem{Alday:2017xua}
L.~F.~Alday and A.~Bissi,
``Loop Corrections to Supergravity on $AdS_5 \times S^5$,''
Phys. Rev. Lett. \textbf{119} (2017) no.17, 171601
%doi:10.1103/PhysRevLett.119.171601
[arXiv:1706.02388 [hep-th]].
%97 citations counted in INSPIRE as of 18 Mar 2021



%\cite{Aprile:2017bgs}\cite{Aprile:2017qoy}
\bibitem{Aprile:2017bgs}
F.~Aprile, J.~M.~Drummond, P.~Heslop and H.~Paul,
``Quantum Gravity from Conformal Field Theory,''
JHEP \textbf{01} (2018), 035
%doi:10.1007/JHEP01(2018)035
[arXiv:1706.02822 [hep-th]].
%84 citations counted in INSPIRE as of 18 Mar 2021


%\cite{Alday:2017vkk}
\bibitem{Alday:2017vkk}
L.~F.~Alday and S.~Caron-Huot,
``Gravitational S-matrix from CFT dispersion relations,''
JHEP \textbf{12} (2018), 017
doi:10.1007/JHEP12(2018)017
[arXiv:1711.02031 [hep-th]].
%110 citations counted in INSPIRE as of 20 Sep 2021



%\cite{Aprile:2017qoy}
\bibitem{Aprile:2017qoy}
F.~Aprile, J.~M.~Drummond, P.~Heslop and H.~Paul,
``Loop corrections for Kaluza-Klein AdS amplitudes,''
JHEP \textbf{05} (2018), 056
%doi:10.1007/JHEP05(2018)056
[arXiv:1711.03903 [hep-th]].
%44 citations counted in INSPIRE as of 18 Mar 2021


%\cite{Aprile:2019rep}
\bibitem{Aprile:2019rep}
F.~Aprile, J.~Drummond, P.~Heslop and H.~Paul,
``One-loop amplitudes in AdS$_{5}\times{}$S$^{5}$ supergravity from $ \mathcal{N} $ = 4 SYM at strong coupling,''
JHEP \textbf{03} (2020), 190
doi:10.1007/JHEP03(2020)190
[arXiv:1912.01047 [hep-th]].






%
%%\cite{Alday:2018pdi}Alday:2018kkw
%\bibitem{Alday:2018pdi}
%L.~F.~Alday, A.~Bissi and E.~Perlmutter,
%``Genus-One String Amplitudes from Conformal Field Theory,''
%JHEP \textbf{06} (2019), 010
%%doi:10.1007/JHEP06(2019)010
%[arXiv:1809.10670 [hep-th]].
%%52 citations counted in INSPIRE as of 24 Mar 2021


%\cite{Alday:2018kkw}\cite{Alday:2019nin}
\bibitem{Alday:2018kkw}
L.~F.~Alday,
``On genus-one string amplitudes on $AdS_5 \times S^5$,''
JHEP \textbf{04} (2021), 005
doi:10.1007/JHEP04(2021)005
[arXiv:1812.11783 [hep-th]].
%41 citations counted in INSPIRE as of 24 Aug 2021






%\cite{Aprile:2019rep}
\bibitem{Aprile:2019rep}
F.~Aprile, J.~Drummond, P.~Heslop and H.~Paul,
``One-loop amplitudes in AdS$_{5}\times{}$S$^{5}$ supergravity from $ \mathcal{N} $ = 4 SYM at strong coupling,''
JHEP \textbf{03} (2020), 190
%doi:10.1007/JHEP03(2020)190
[arXiv:1912.01047 [hep-th]].
%22 citations counted in INSPIRE as of 18 Mar 2021


%==========================================================================================



%\cite{Alday:2019nin}
\bibitem{Alday:2019nin}
L.~F.~Alday and X.~Zhou,
``Simplicity of AdS Supergravity at One Loop,''
JHEP \textbf{09} (2020), 008
doi:10.1007/JHEP09(2020)008
[arXiv:1912.02663 [hep-th]].
%32 citations counted in INSPIRE as of 24 Aug 2021



%\cite{Bissi:2020wtv}
%\bibitem{Bissi:2020wtv}
\bibitem{Bissi:2020woe}
A.~Bissi, G.~Fardelli and A.~Georgoudis,
``Towards All Loop Supergravity Amplitudes on $AdS_5 \times S^5$,''
[arXiv:2002.04604 [hep-th]].
%7 citations counted in INSPIRE as of 28 Mar 2021
%
%\cite{Bissi:2020woe}
%\bibitem{Bissi:2020woe}
A.~Bissi, G.~Fardelli and A.~Georgoudis,
``All loop structures in Supergravity Amplitudes on $AdS_5 \times S^5$ from CFT,''
[arXiv:2010.12557 [hep-th]].
%0 citations counted in INSPIRE as of 26 Mar 2021



%==========================================================================================






%===============================================================================




%\cite{Aprile:2017xsp}\cite{Aprile:2018efk}
\bibitem{Aprile:2017xsp}
F.~Aprile, J.~M.~Drummond, P.~Heslop and H.~Paul,
``Unmixing Supergravity,''
JHEP \textbf{02} (2018), 133
%doi:10.1007/JHEP02(2018)133
[arXiv:1706.08456 [hep-th]].
%50 citations counted in INSPIRE as of 18 Mar 2021


%\cite{Aprile:2018efk}
\bibitem{Aprile:2018efk}
F.~Aprile, J.~Drummond, P.~Heslop and H.~Paul,
``Double-trace spectrum of $N=4$ supersymmetric Yang-Mills theory at strong coupling,''
Phys. Rev. D \textbf{98} (2018) no.12, 126008
%doi:10.1103/PhysRevD.98.126008
[arXiv:1802.06889 [hep-th]].
%43 citations counted in INSPIRE as of 18 Mar 2021



%===============================================================================




%\cite{Caron-Huot:2018kta}
\bibitem{Caron-Huot:2018kta}
S.~Caron-Huot and A.~K.~Trinh,
``All tree-level correlators in AdS$_{5}\times{}$S$^{5}$ supergravity: hidden ten-dimensional conformal symmetry,''
JHEP \textbf{01} (2019), 196
%doi:10.1007/JHEP01(2019)196
[arXiv:1809.09173 [hep-th]].
%58 citations counted in INSPIRE as of 18 Mar 2021



%===============================================================================



%\cite{Binder:2019jwn}
\bibitem{Binder:2019jwn}
D.~J.~Binder, S.~M.~Chester, S.~S.~Pufu and Y.~Wang,
``$ \mathcal{N} $ = 4 Super-Yang-Mills correlators at strong coupling from string theory and localization,''
JHEP \textbf{12} (2019), 119
%doi:10.1007/JHEP12(2019)119
[arXiv:1902.06263 [hep-th]].
%38 citations counted in INSPIRE as of 24 Mar 2021



%\cite{Chester:2019pvm}\cite{Binder:2019jwn}
\bibitem{Chester:2019pvm}
S.~M.~Chester,
``Genus-2 holographic correlator on AdS$_{5}\times{}$S$^{5}$ from localization,''
JHEP \textbf{04} (2020), 193
%doi:10.1007/JHEP04(2020)193
[arXiv:1908.05247 [hep-th]].
%27 citations counted in INSPIRE as of 18 Mar 2021







%\cite{Chester:2020dja}
\bibitem{Chester:2020dja}
S.~M.~Chester and S.~S.~Pufu,
``Far beyond the planar limit in strongly-coupled $ \mathcal{N} $ = 4 SYM,''
JHEP \textbf{01} (2021), 103
doi:10.1007/JHEP01(2021)103
[arXiv:2003.08412 [hep-th]].
%44 citations counted in INSPIRE as of 05 Jul 2022

%===============================================================================



%===============================================================================

%\cite{Mack:2009mi}
\bibitem{Mack:2009mi}
G.~Mack,
``D-independent representation of Conformal Field Theories in D dimensions via transformation to auxiliary Dual Resonance Models. Scalar amplitudes,''
[arXiv:0907.2407 [hep-th]].
%214 citations counted in INSPIRE as of 19 Oct 2021


%\cite{Penedones:2010ue}
\bibitem{Penedones:2010ue}
J.~Penedones,
``Writing CFT correlation functions as AdS scattering amplitudes,''
JHEP \textbf{03} (2011), 025
doi:10.1007/JHEP03(2011)025
[arXiv:1011.1485 [hep-th]].
%366 citations counted in INSPIRE as of 20 Sep 2021


%\cite{Aprile:2020luw}
\bibitem{Aprile:2020luw}
F.~Aprile and P.~Vieira,
``Large $p$ explorations. From SUGRA to big STRINGS in Mellin space,''
JHEP \textbf{12} (2020), 206
%doi:10.1007/JHEP12(2020)206
[arXiv:2007.09176 [hep-th]].
%7 citations counted in INSPIRE as of 18 Mar 2021

%\cite{Goncalves:2014ffa}
\bibitem{Goncalves:2014ffa}
V.~Gon\c{c}alves,
``Four point function of $\mathcal{N}=4$ stress-tensor multiplet at strong coupling,''
JHEP \textbf{04} (2015), 150
doi:10.1007/JHEP04(2015)150
[arXiv:1411.1675 [hep-th]].
%67 citations counted in INSPIRE as of 11 May 2022


%\cite{Alday:2018pdi}
\bibitem{Alday:2018pdi}
L.~F.~Alday, A.~Bissi and E.~Perlmutter,
``Genus-One String Amplitudes from Conformal Field Theory,''
JHEP \textbf{06} (2019), 010
doi:10.1007/JHEP06(2019)010
[arXiv:1809.10670 [hep-th]].
%62 citations counted in INSPIRE as of 23 Sep 2021




%\cite{Drummond:2019hel}Drummond:2020uni
\bibitem{Drummond:2019hel}
J.~M.~Drummond and H.~Paul,
``One-loop string corrections to AdS amplitudes from CFT,''
JHEP \textbf{03} (2021), 038
doi:10.1007/JHEP03(2021)038
[arXiv:1912.07632 [hep-th]].
%14 citations counted in INSPIRE as of 27 Mar 202

%\cite{Drummond:2020uni}
\bibitem{Drummond:2020uni}
J.~M.~Drummond, R.~Glew and H.~Paul,
``One-loop string corrections for AdS Kaluza-Klein amplitudes,''
[arXiv:2008.01109 [hep-th]].
%4 citations counted in INSPIRE as of 27 Mar 2021


%\cite{Drummond:2019odu}Drummond:2020dwr
\bibitem{Drummond:2019odu}
J.~M.~Drummond, D.~Nandan, H.~Paul and K.~S.~Rigatos,
``String corrections to AdS amplitudes and the double-trace spectrum of $ \mathcal{N} $ = 4 SYM,''
JHEP \textbf{12} (2019), 173
doi:10.1007/JHEP12(2019)173
[arXiv:1907.00992 [hep-th]].
%15 citations counted in INSPIRE as of 27 Mar 2021


%\cite{Drummond:2020dwr}\cite{Aprile:2020mus}
\bibitem{Drummond:2020dwr}
J.~M.~Drummond, H.~Paul and M.~Santagata,
``Bootstrapping string theory on AdS$_5 \times S^5$,''
[arXiv:2004.07282 [hep-th]].~
%7 citations counted in INSPIRE as of 23 Mar 2021
%
%\cite{Aprile:2020mus}Abl:2020dbx
\bibitem{Aprile:2020mus}
F.~Aprile, J.~M.~Drummond, H.~Paul and M.~Santagata,
``The Virasoro-Shapiro amplitude in AdS$_5\times$S$^5$ and level splitting of 10d conformal symmetry,''
[arXiv:2012.12092 [hep-th]].
%0 citations counted in INSPIRE as of 18 Mar 2021


%\cite{Abl:2020dbx}
\bibitem{Abl:2020dbx}
T.~Abl, P.~Heslop and A.~E.~Lipstein,
``Towards the Virasoro-Shapiro Amplitude in AdS5xS5,''
[arXiv:2012.12091 [hep-th]].
%0 citations counted in INSPIRE as of 22 Mar 2021


%\cite{Alday:2022uxp}
\bibitem{Alday:2022uxp}
L.~F.~Alday, T.~Hansen and J.~A.~Silva,
``AdS Virasoro-Shapiro from dispersive sum rules,''
[arXiv:2204.07542 [hep-th]].
%2 citations counted in INSPIRE as of 04 Jul 2022

%\cite{Caron-Huot:2022sdy}
\bibitem{Caron-Huot:2022sdy}
S.~Caron-Huot, F.~Coronado, A.~K.~Trinh and Z.~Zahraee,
``Bootstrapping $\mathcal{N}=4$ sYM correlators using integrability,''
[arXiv:2207.01615 [hep-th]].
%0 citations counted in INSPIRE as of 26 Jul 2022



%\cite{Dolan:2006ec}
\bibitem{Dolan:2006ec}
F.~A.~Dolan, M.~Nirschl and H.~Osborn,
``Conjectures for large N superconformal N=4 chiral primary four point functions,''
Nucl. Phys. B \textbf{749} (2006), 109-152
doi:10.1016/j.nuclphysb.2006.05.009
[arXiv:hep-th/0601148 [hep-th]].
%50 citations counted in INSPIRE as of 10 Jan 2022

%%\cite{Uruchurtu:2008kp}
%\bibitem{Uruchurtu:2008kp}
%L.~I.~Uruchurtu,
%``Four-point correlators with higher weight superconformal primaries in the AdS/CFT Correspondence,''
%JHEP \textbf{03} (2009), 133
%doi:10.1088/1126-6708/2009/03/133
%[arXiv:0811.2320 [hep-th]].
%%35 citations counted in INSPIRE as of 10 Jan 2022
%
%%\cite{Uruchurtu:2011wh}
%\bibitem{Uruchurtu:2011wh}
%L.~I.~Uruchurtu,
%``Next-next-to-extremal Four Point Functions of N=4 1/2 BPS Operators in the AdS/CFT Correspondence,''
%JHEP \textbf{08} (2011), 133
%doi:10.1007/JHEP08(2011)133
%[arXiv:1106.0630 [hep-th]].
%%37 citations counted in INSPIRE as of 10 Jan 2022


%\cite{Arutyunov:2017dti}
\bibitem{Arutyunov:2017dti}
G.~Arutyunov, S.~Frolov, R.~Klabbers and S.~Savin,
``Towards 4-point correlation functions of any $ \frac{1}{2} $ -BPS operators from supergravity,''
JHEP \textbf{04} (2017), 005
doi:10.1007/JHEP04(2017)005
[arXiv:1701.00998 [hep-th]].
%23 citations counted in INSPIRE as of 29 Jun 2022



%\cite{Eden:2000bk}
\bibitem{Eden:2000bk}
B.~Eden, A.~C.~Petkou, C.~Schubert and E.~Sokatchev,
``Partial nonrenormalization of the stress tensor four point function in N=4 SYM and AdS / CFT,''
Nucl. Phys. B \textbf{607} (2001), 191-212
doi:10.1016/S0550-3213(01)00151-1
[arXiv:hep-th/0009106 [hep-th]].
%147 citations counted in INSPIRE as of 13 Jan 2022


%===============================================================================

\bibitem{Doobary:2015gia}
R.~Doobary and P.~Heslop,
``Superconformal partial waves in Grassmannian field theories,''
JHEP \textbf{12} (2015), 159
doi:10.1007/JHEP12(2015)159
[arXiv:1508.03611 [hep-th]].


%\cite{Aprile:2021pwd}
\bibitem{Aprile:2021pwd}
F.~Aprile and P.~Heslop,
``Superconformal blocks in diverse dimensions and $BC$ symmetric functions,''
[arXiv:2112.12169 [hep-th]].
%3 citations counted in INSPIRE as of 04 Jul 2022

%===============================================================================




%\cite{Aprile:2021mvq}Drummond:2022dxd
\bibitem{Aprile:2021mvq}
F.~Aprile and M.~Santagata,
``Two particle spectrum of tensor multiplets coupled to AdS$_{3}\times{}$S$^{3}$gravity,''
Phys. Rev. D \textbf{104} (2021) no.12, 126022
doi:10.1103/PhysRevD.104.126022
[arXiv:2104.00036 [hep-th]].
%9 citations counted in INSPIRE as of 31 Mar 2022

%\cite{Drummond:2022dxd}
\bibitem{Drummond:2022dxd}
J.~M.~Drummond, R.~Glew and M.~Santagata,
``BCJ relations in ${AdS}_5 \times S^3$ and the double-trace spectrum of super gluons,''
[arXiv:2202.09837 [hep-th]].
%1 citations counted in INSPIRE as of 31 Mar 2022


\bibitem{unpublished}
F.~Aprile, J.~M.~Drummond, R.~Glew, unpublished.


%\cite{Green:2008uj}
\bibitem{Green:2008uj}
M.~B.~Green, J.~G.~Russo and P.~Vanhove,
``Low energy expansion of the four-particle genus-one amplitude in type II superstring theory,''
JHEP \textbf{02} (2008), 020
doi:10.1088/1126-6708/2008/02/020
[arXiv:0801.0322 [hep-th]].
%119 citations counted in INSPIRE as of 19 May 2022


%\cite{Penedones:2019tng}
\bibitem{Penedones:2019tng}
J.~Penedones, J.~A.~Silva and A.~Zhiboedov,
``Nonperturbative Mellin Amplitudes: Existence, Properties, Applications,''
JHEP \textbf{08} (2020), 031
doi:10.1007/JHEP08(2020)031
[arXiv:1912.11100 [hep-th]].
%59 citations counted in INSPIRE as of 05 Jul 2022


%\cite{Caron-Huot:2020adz}
\bibitem{Caron-Huot:2020adz}
S.~Caron-Huot, D.~Mazac, L.~Rastelli and D.~Simmons-Duffin,
``Dispersive CFT Sum Rules,''
JHEP \textbf{05} (2021), 243
doi:10.1007/JHEP05(2021)243
[arXiv:2008.04931 [hep-th]].
%60 citations counted in INSPIRE as of 05 Jul 2022


\end{thebibliography}

\end{document}